%% file: main-file_instrumentation-paper.tex
\documentclass[instruments,article,moreauthors,pdftex]{mdpi} 

\firstpage{1} 
\pubvolume{xx}
\issuenum{1}
\articlenumber{5}
\pubyear{2021}
\copyrightyear{2021}
\history{Received: date; Accepted: date; Published: date}
\setitemize{parsep=6pt,itemsep=0pt,leftmargin=*,labelsep=5.5mm}
\setenumerate{parsep=6pt,itemsep=0pt,leftmargin=*,labelsep=5.5mm}
\setlist[description]{itemsep=0mm}

\include{packagesAndDefinitions}
\Title{Commissioning of a High Pressure Time Projection Chamber with Optical Readout}
\Author{
  A.~Deisting $^{8}$ * \orcidA{},
  A.~V.~Waldron $^{3}$ * \orcidJ{}, 
  E.~Atkin $^{3}$ \orcidE{}, 
  G.~J.~Barker $^{11}$, 
  A.~Basharina-Freshville $^{10}$,
  C.~Betancourt $^{12}$,
  S.~B.~Boyd $^{11}$, 
  D.~Brailsford $^{5}$ \orcidG{}, 
  Z.~Chen-Wishart $^{8}$, 
  L.~Cremonesi $^{10,\alpha}$ \orcidL{}, 
  A.~Dias $^{8}$ \orcidB{},
  P.~Dunne $^{3}$ \orcidF{},  
  J.~Haigh $^{11}$, 
  P.~Hamacher-Baumann $^{9}$ \orcidI{}, 
  S.B.~Jones $^{10}$ \orcidH{},
  A.~Kaboth $^{8}$,   
  A.~Korzenev $^{2}$,
  W.~Ma $^{9}$,
  P.~Mermod $^{2}$ $\dagger$,
  M.~Mironova $^{3,6}$,
  J.~Monroe $^{8}$ \orcidC{}, 
  R.~Nichol $^{10}$,
  T.S.~Nonnenmacher $^{3}$,     
  J.~Nowak $^{5}$ \orcidD{}, 
  W.~Parker $^{8,\beta}$, 
  H.~Ritchie-Yates~$^{8}$,
  S.~Roth~$^{9}$,
  R.~Saakyan $^{10}$, 
  N.~Serra $^{12}$,
  Y.~Shitov $^{3,4}$,  
  J.~Steinmann $^{9}$, 
  A.~Tarrant $^{8,\gamma}$ \orcidK{}, 
  M.~A.~Uchida $^{3,1}$  
  S.~Valder $^{11}$, 
  M.~Ward $^{8,7}$,  
  M.O.~Wascko $^{3}$ \orcidM{}
}

\AuthorNames{Alexander Deisting; Abigail Victoria Waldron; Edward Atkin; Gary Barker; Anastasia Basharina-Freshville; Christopher Betancourt; Steven Boyd; Dominic Brailsford; Zachary Chen-Wishart; Linda Cremonesi; Adriana Dias; Patrick Dunne; Jennifer Haigh; Philip Hamacher-Baumann; Sebastian Jones; Asher Kaboth; Alexander Korzenev; William Ma; Philippe Mermod; Maria Mironova; Jocelyn Monroe; Ryan Nichol; Toby Nonnenmacher; Jaroslaw Nowak; William Parker; Harrison Ritchie-Yates; Stefan Roth; Ruben Saakyan; Nicola Serra; Yuri Shitov; Jochen Steinmann; Adam Tarrant; Melissa Uchida; Sammy Valder; Mark Ward and Morgan Wascko}

\address{%
	$^{1}$ \quad Cavendish Laboratory, Cambridge CB3 0HE, UK; m.a.uchida@imperial.ac.uk

	$^{2}$ \quad DPNC Universit\'e de Gen\`eve, 1205 Genf, Switzerland; korzenev@mail.cern.ch (A.K.); Philippe.Mermod@cern.ch (P.M.);
	
	$^{3}$ \quad The Blackett Laboratory, Imperial College London, London SW7 2BW, UK; e.atkin17@imperial.ac.uk (E.A.); p.dunne12@imperial.ac.uk (P.D.); maria.mironova@physics.ox.ac.uk (M.M.); toby.nonnenmacher14@imperial.ac.uk (T.N.); Shitov@JINR.ru (Y.S.); m.wascko@imperial.ac.uk (M.W.)
	
	$^{4}$ \quad JINR, 141980 Dubna, Russia
	
	$^{5}$ \quad Lancaster University, Bailrigg, Lancaster LA1 4YW, UK; d.brailsford@lancaster.ac.uk (D.B.); j.nowak@lancaster.ac.uk (J.N.)
	
	$^{6}$ \quad Department of Physics, Oxford University, Oxford OX1 3PU, UK
	
  $^{7}$ \quad Queen’s University, Kingston, ON K7L 3N6, Canada; mark.ward@snolab.ca

	$^{8}$ \quad Royal Holloway, University of London, Egham Hill, Egham, TW20 0EX, UK; Zachary.Chen-Wishart.2016@live.rhul.ac.uk (Z.C.-W.); Adriana.Dias.2011@live.rhul.ac.uk (A.D.); Asher.Kaboth@rhul.ac.uk (A.K.); jocelyn.monroe@rhul.ac.uk~(J.M.); william.parker@physics.ox.ac.uk (W.P.); Harrison.Ritchie-Yates.2013@live.rhul.ac.uk (H.R.-Y.); A.Tarrant@liverpool.ac.uk (A.T.)
	
  $^{9}$ \quad III. Physikalisches Institut, RWTH Aachen University, 52056 Aachen, Germany; hamacher.baumann@physik.rwth-aachen.de (P.H.-B.); ma@physik.rwth-aachen.de (W.M.); stefan.roth@physik.rwth-aachen.de (S.R.); jochen.steinmann@physik.rwth-aachen.de (J.S.)
	
	$^{10}$ \quad University College London, Gower St, Kings Cross, London WC1E 6BT, UK; anastasia.freshville@ucl.ac.uk~(A.B.-F.); l.cremonesi@qmul.ac.uk (L.C.); sebastian.jones.17@ucl.ac.uk~(S.B.J.); r.nichol@ucl.ac.uk~(R.N.); r.saakyan@ucl.ac.uk~(R.S.)
	
	$^{11}$ \quad University of Warwick, Coventry CV4 7AL, UK; g.j.barker@warwick.ac.uk (G.J.B.); S.B.Boyd@warwick.ac.uk (S.B.B.); J.Haigh.2@warwick.ac.uk (J.H.); s.valder@warwick.ac.uk (S.V.)
	
	$^{12}$ \quad Physik-Institut, Universit\"at Z\"uriche, R\"amistrasse 71, 8006 Z\"urich, Switzerland; christopher.betancourt@cern.ch (C.B.); nicola.serra@cern.ch (N.S.)
}

\corres{Correspondence: alexander.deisting@cern.ch (A.D), a.waldron@imperial.ac.uk (A.V.W.)\\
$\alpha$ \quad Now at: Queen Mary University of London, Mile End Road, London E1 4NS, UK\\
$\beta$ \quad Now at: University of Oxford, Clarendon Laboratory, Parks Road, Oxford OX1 3PU, UK\\
$\gamma$\quad Now at: University of Liverpool, The Oliver Lodge Laboratory, Liverpool L69 7ZE, UK\\
$\dagger$ \quad Passed away on 20 August 2020}


\begin{document}

% Abstract (Do not insert blank lines, i.e. \\) 
%\abstract{
\input{sec_0_abstract}
%}

%\maketitle
%%\linenumbers

% Introduction and motivation 
\input{sec_1_introduction}

% Justification of the gas choice 
\input{sec_2_gasMixtures}

% HPTPC description
\input{sec_3_hptpc.tex}

% Calibration, analysis and performance of the TPC ccd readout
\input{sec_5_ccdAnalysis.tex}

% Calibration, analysis and performance of the TPC charge readout
\input{sec_6_chargeAnalysis.tex}

% combined quantities from the charge and ccd analysis
\input{sec_7_combinedCCDandChargeAnalysis.tex}

% Outlook and Summary
\input{sec_8_summary.tex}

\authorcontributions{Conceptualization, Anastasia Basharina-Freshville, Steven Boyd, Dominic Brailsford, Linda Cremonesi, Patrick Dunne, Jennifer Haigh, Asher Kaboth, Jocelyn Monroe, Ryan Nichol, Jaroslaw Nowak, Ruben Saakyan, Nicola Serra, Yuri Shitov and Morgan Wascko; 

Data curation, Alexander Deisting, Abigail Waldron, Edward Atkin, Dominic Brailsford, Zachary Chen-Wishart, Linda Cremonesi, Adriana Dias, Patrick Dunne, Jennifer Haigh, Philip Hamacher-Baumann, Sebastian Jones, Asher Kaboth, Alexander Korzenev, Maria Mironova, Jocelyn Monroe, Ryan Nichol, Toby Nonnenmacher, Jaroslaw Nowak, William Parker, Harrison Ritchie-Yates, Stefan Roth, Adam Tarrant, Melissa Uchida, Sammy Valder, Mark Ward and Morgan Wascko; 

Formal analysis, Alexander Deisting, Abigail Waldron, Edward Atkin, Dominic Brailsford, Zachary Chen-Wishart, Adriana Dias, Patrick Dunne, Jennifer Haigh, Philip Hamacher-Baumann, Sebastian Jones, Maria Mironova, Toby Nonnenmacher, William Parker, Harrison Ritchie-Yates, Adam Tarrant and Mark Ward; 

Funding acquisition, Gary Barker, Anastasia Basharina-Freshville, Steven Boyd, Asher Kaboth, Jocelyn Monroe, Ryan Nichol, Jaroslaw Nowak, Stefan Roth, Ruben Saakyan and Morgan Wascko;

Investigation, Alexander Deisting, Abigail Waldron, Edward Atkin, Gary Barker, Anastasia Basharina-Freshville, Christopher Betancourt, Dominic Brailsford, Zachary Chen-Wishart, Linda Cremonesi, Adriana Dias, Patrick Dunne, Jennifer Haigh, Philip Hamacher-Baumann, Sebastian Jones, Asher Kaboth, Alexander Korzenev, William Ma, Philippe Mermod, Maria Mironova, Jocelyn Monroe, Ryan Nichol, Toby Nonnenmacher, Jaroslaw Nowak, William Parker, Harrison Ritchie-Yates, Stefan Roth, Ruben Saakyan, Nicola Serra, Yuri Shitov, Jochen Steinmann, Adam Tarrant, Melissa Uchida, Sammy Valder, Mark Ward and Morgan Wascko;

Methodology, Alexander Deisting, Abigail Waldron, Gary Barker, Anastasia Basharina-Freshville, Christopher Betancourt, Steven Boyd, Dominic Brailsford, Zachary Chen-Wishart, Linda Cremonesi, Adriana Dias, Patrick Dunne, Jennifer Haigh, Philip Hamacher-Baumann, Sebastian Jones, Asher Kaboth, Alexander Korzenev, William Ma, Philippe Mermod, Maria Mironova, Jocelyn Monroe, Ryan Nichol, Toby Nonnenmacher, Jaroslaw Nowak, William Parker, Harrison Ritchie-Yates, Stefan Roth, Ruben Saakyan, Nicola Serra, Yuri Shitov, Jochen Steinmann, Adam Tarrant, Melissa Uchida, Sammy Valder, Mark Ward and Morgan Wascko;

Project administration, Alexander Deisting, Gary Barker, Steven Boyd, Linda Cremonesi, Asher Kaboth, Jocelyn Monroe, Ryan Nichol, Jaroslaw Nowak and Morgan Wascko;

Resources, Steven Boyd, Dominic Brailsford, Linda Cremonesi, Asher Kaboth, Jocelyn Monroe, Ryan Nichol, Jaroslaw Nowak, Stefan Roth and Morgan Wascko;

Software, Alexander Deisting, Abigail Waldron, Edward Atkin, Patrick Dunne, Jennifer Haigh, Philip Hamacher-Baumann, Sebastian Jones, Toby Nonnenmacher, William Parker, Harrison Ritchie-Yates, Adam Tarrant and Sammy Valder;

Supervision, Alexander Deisting, Abigail Waldron, Gary Barker, Anastasia Basharina-Freshville, Steven Boyd, Linda Cremonesi, Patrick Dunne, Asher Kaboth, Jocelyn Monroe, Ryan Nichol, Jaroslaw Nowak, Ruben Saakyan and Morgan Wascko;

Validation, Alexander Deisting, Abigail Waldron, Edward Atkin, Dominic Brailsford, Zachary Chen-Wishart, Linda Cremonesi, Patrick Dunne, Sebastian Jones, Asher Kaboth, Maria Mironova, Jocelyn Monroe, Ryan Nichol, Toby Nonnenmacher, Harrison Ritchie-Yates and Adam Tarrant;

Visualization, Alexander Deisting, Abigail Waldron, Edward Atkin, Zachary Chen-Wishart, Adriana Dias, Patrick Dunne, Philip Hamacher-Baumann, William Parker and Harrison Ritchie-Yates; 

Writing – original draft, Alexander Deisting, Abigail Waldron, Zachary Chen-Wishart, Patrick Dunne, Asher Kaboth, Jocelyn Monroe, William Parker, Harrison Ritchie-Yates, Adam Tarrant and Morgan Wascko;

Writing – review \& editing, Alexander Deisting, Abigail Waldron, Edward Atkin, Dominic Brailsford, Zachary Chen-Wishart, Linda Cremonesi, Adriana Dias, Patrick Dunne, Philip Hamacher-Baumann, Asher Kaboth, Jocelyn Monroe, Jaroslaw Nowak, Harrison Ritchie-Yates and Morgan Wascko.}

%For research articles with several authors, a short paragraph specifying their individual contributions must be provided. We copied the contributions from submission system. Please check and confirm. 

% The following statements should be used ``Conceptualization, X.X. and Y.Y.; methodology, X.X.; software, X.X.; validation, X.X., Y.Y. and Z.Z.; formal analysis, X.X.; investigation, X.X.; resources, X.X.; data curation, X.X.; writing--original draft preparation, X.X.; writing--review and editing, X.X.; visualization, X.X.; supervision, X.X.; project administration, X.X.; funding acquisition, Y.Y. All authors have read and agreed to the published version of the manuscript.'', please turn to the  \href{http://img.mdpi.org/data/contributor-role-instruction.pdf}{CRediT taxonomy} for the term explanation. Authorship must be limited to those who have contributed substantially to the work reported.

%%%%%%%%%%%%%%%%%%%%%%%%%%%%%%%%%%%%%%%%%%
\funding{This research was funded in part by Science and Technology Facilities Council grant number ST/N003233/.}
% or ``This research was funded by NAME OF FUNDER grant number XXX.'' and  and ``The APC was funded by XXX''. Check carefully that the details given are accurate and use the standard spelling of funding agency names at \url{https://search.crossref.org/funding}, any errors may affect your future funding.

\acknowledgments{We wish to acknowledge support for summer students from the Ogden Trust and St. Andrews University, and outstanding support during the beam test from Johannes Bernhard of CERN as well as Johan Borg, Rebecca Conybeare, Nicole Cullen, Kate Gould, Maria Khaleeq, Veera Mikola, Duncan Parker, Christopher Thorpe and Simon Williams.}

\conflictsofinterest{The authors declare no conflict of interest.}
% Authors must identify and declare any personal circumstances or interest that may be perceived as inappropriately influencing the representation or interpretation of reported research results. Any role of the funders in the design of the study; in the collection, analyses or interpretation of data; in the writing of the manuscript, or in the decision to publish the results must be declared in this section. If there is no role, please state ``The funders had no role in the design of the study; in the collection, analyses, or interpretation of data; in the writing of the manuscript, or in the decision to publish the results''.

\appendix

% Bibliography
%\AtNextBibliography{\small}
%\printbibliography[heading=bibintoc]

%=====================================
% References, variant B: external bibliography
%=====================================
\section*{References}
%\todo[inline]{\large if there are more than 4 authors make it first et al. or collaboration name}

\externalbibliography{yes}
\bibliography{hptpc-instrumentation-paper.bib}

\end{document}

%% file: packagesAndDefinitions.tex
\usepackage{amsfonts}
\usepackage{amsmath}           %allows the use of align
\usepackage{amssymb}
\usepackage{amstext}
\usepackage[affil-it]{authblk}
\usepackage[UKenglish]{babel}
\usepackage{bm}                %bold in math mode
\usepackage{booktabs}
\usepackage{xcolor}
\usepackage{enumitem}
\usepackage{fancyhdr}
\usepackage{float}
\usepackage{graphicx}
\usepackage[space]{grffile}    %allows spaces in picture names
\usepackage[utf8]{inputenc}
\usepackage{longtable}
\usepackage{marvosym}
\usepackage{multicol}
\usepackage{setspace}
\usepackage[separate-uncertainty,multi-part-units = single, allow-number-unit-breaks]{siunitx}
\usepackage{subfig}
\usepackage{tikz-timing}
\usepackage[colorinlistoftodos,prependcaption,textsize=tiny]{todonotes} % Add TODO notes to the text
\usepackage{wasysym} % permille

\usepackage{url}
\def\UrlBreaks{\do\/\do-}
\usepackage{csquotes}

\newcommand*\patchAmsMathEnvironmentForLineno[1]{%
  \expandafter\let\csname old#1\expandafter\endcsname\csname #1\endcsname
  \expandafter\let\csname oldend#1\expandafter\endcsname\csname end#1\endcsname
  \renewenvironment{#1}%
     {\linenomath\csname old#1\endcsname}%
     {\csname oldend#1\endcsname\endlinenomath}}% 
\newcommand*\patchBothAmsMathEnvironmentsForLineno[1]{%
  \patchAmsMathEnvironmentForLineno{#1}%
  \patchAmsMathEnvironmentForLineno{#1*}}%
\AtBeginDocument{%
\patchBothAmsMathEnvironmentsForLineno{equation}%
\patchBothAmsMathEnvironmentsForLineno{align}%
\patchBothAmsMathEnvironmentsForLineno{flalign}%
\patchBothAmsMathEnvironmentsForLineno{alignat}%
\patchBothAmsMathEnvironmentsForLineno{gather}%
\patchBothAmsMathEnvironmentsForLineno{multline}%
}

\newcommand{\ar}{$\textrm{Ar}$}
\newcommand{\cf}{$\textrm{CF}_{4}$}
\newcommand{\co}{$\textrm{CO}_{2}$}
\newcommand{\n}{$\textrm{N}_{2}$}
\newcommand{\arco}{$\textrm{Ar}/\textrm{CO}_2$}
\newcommand{\arcois}[1]{$\textrm{Ar}/\textrm{CO}_2$ (#1)}
\newcommand{\arcf}{$\textrm{Ar}/\textrm{CF}_4$}

\newcommand{\arn}{$\textrm{Ar}/\textrm{N}_2$}
\newcommand{\nen}{$\textrm{Ne}/\textrm{N}_2$}
\newcommand{\arnis}[1]{$\textrm{Ar}/\textrm{N}_2$ (#1)}
\newcommand{\arcon}{$\textrm{Ar}/\textrm{CO}_2/\textrm{N}_2$}
\newcommand{\arconis}[1]{$\textrm{Ar}/\textrm{CO}_2/\textrm{N}_2$ (#1)}

\newcommand{\figref}[1]{Fig.~\ref{#1}}    % Standard
\newcommand{\figrefbra}[1]{Fig.~\ref{#1}}   % In brackets 
\newcommand{\Figref}[1]{Figure~\ref{#1}}    % Beginning of a sentence
\newcommand{\tabref}[1]{Tab.~\ref{#1}}      % Standard
\newcommand{\tabrefbra}[1]{Tab.~\ref{#1}}   % In brackets 
\newcommand{\Tabref}[1]{Table~\ref{#1}}     % Beginning of a sentence
\newcommand{\secref}[1]{Sec.~\ref{#1}}      % Standard
\newcommand{\secrefbra}[1]{Sec.~\ref{#1}}   % In brackets
\newcommand{\Secref}[1]{Section~\ref{#1}}   % Beginning of a sentence

%% file: sec_0_abstract.tex
\begin{abstract}
\noindent Measurements of proton-nucleus scattering and high resolution neutrino-nucleus interaction imaging are key to reduce neutrino oscillation systematic uncertainties in future experiments. A High Pressure Time Projection Chamber (HPTPC) prototype has been constructed and operated at Royal Holloway University of London and CERN as a first step in the development of a HPTPC capable of performing these measurements as part of a future long-baseline neutrino oscillation experiment such as the Deep Underground Neutrino Experiment. In this paper we describe the design and operation of the prototype HPTPC with an argon based gas mixture. We report on the successful hybrid charge and optical readout, using four CCD cameras, of signals from $^{241}\textrm{Am}$ sources.
\end{abstract}

%% file: sec_1_introduction.tex
\section{Introduction}
\label{hptpcPaper:sec:Introduction}

High Pressure Time Projection Chambers (HPTPCs) are an area of growing international interest. The Deep Underground Neutrino Experiment (DUNE) envisions the use of an HPTPC as part of its near detector and European groups have held a series of workshops on HPTPC development over the last five years. Given the recent indication of non-zero CP violation in the Tokai to Kamioka (T2K) experiment's data \cite{Abe:2019vii}, it is timely to quantify the potential impact of HPTPC neutrino ($\nu$) detector technology on mitigation of the dominant neutrino-interaction cross-section uncertainties for the future long-baseline neutrino oscillation programme.\\
Final State Interactions (FSIs) of nucleons produced in neutrino interactions are among the leading sources of systematic uncertainties in neutrino oscillation experiments \cite{Alvarez-Ruso:2017oui}. Gas TPCs are ideal for precisely characterizing FSI effects because of their high track reconstruction efficiency, low momentum threshold and $4\!\:\pi$ angular coverage of final state particles, which are all key to distinguishing between interaction models. For example, the proton multiplicity and momentum distributions for neutrino charged current interactions on argon calculated by the neutrino interaction Monte Carlo generators NEUT \cite{Hayato:2002sd} and GENIE \cite{Andreopoulos:2015wxa} are highly discrepant in the fraction of events with few ejected protons, and at low proton momentum, below \SI{250}{\mega\electronvolt\per c} \cite{NEUETandGENIE}. This is below the proton detection threshold in water Cherenkov detectors (\SI{1100}{\mega\electronvolt\per c}) and below that of liquid argon TPCs, around \SI{400}{\mega\electronvolt\per c} \cite{MicroBooNE-1024}. A gas-filled HPTPC however has a low enough momentum threshold to resolve FSI model discrepancies, and therefore an HPTPC has unique capability to address the dominant systematic uncertainty in neutrino oscillation measurements.\\
This paper describes the design, commissioning and calibration of a prototype HPTPC detector. \Secref{hptpcPaper:sec:HPTPCDesignConsiderations} describes the prototype detector and readout design, \secref{hptpcPaper:sec:gasChoice} motivates the choice of gas target, \secref{hptpcPaper:sec:highPressureVessel} describes the high pressure vessel and the gas system, \secref{hptpcPaper:sec:hptpc} details the TPC hardware including its high voltage supply and data acquisition system. In \secref{hptpcPaper:sec:ccdAnalysis} and \secref{hptpcPaper:sec:chargeAnalysis} the analysis of camera images and charge signal waveforms, respectively, are explained and results of the commissioning measurements are presented. \Secref{hptpcPaper:sec:ChargeLight} contains a combined analysis of the optical and charge readout signals.

\subsection{Design considerations}
\label{hptpcPaper:sec:HPTPCDesignConsiderations}

The need for lower momentum measurements \cite{NEUETandGENIE} motivates the choice of a gas-filled detector for the task of measuring neutrino-nucleus scattering. Another key consideration for the detector is that it has sufficient target mass to achieve a low statistical error on measured final state kinematic distributions. This requirement drives the choice of a high pressure gas as it has higher density and therefore higher mass.\\
The momentum threshold goal for our HPTPC prototype is designed to probe the discrepant low-momentum region of parameter space \cite{NEUETandGENIE}. The threshold goal for a well-reconstructed proton in argon at \SI{5}{barA} (\SI{10}{barA}) is $\sim\!\!\SI{50}{\mega\electronvolt\per c}$ (about \SI{70}{\mega\electronvolt\per c}).
This drives choices in the readout design, such that at $\sim\!\!\SI{50}{\mega\electronvolt\per c}$ a proton track is sampled by $\sim\!\!10$ measurements in the readout plane. We also aim to cover the momentum range above \SI{320}{\mega\electronvolt\per c} (\SI{50}{\mega\electronvolt} kinetic energy) where no measurements currently exist \cite{NEUETandGENIE}.

The track length of a \SI{50}{\mega\electronvolt\per c} proton in a \SI{5}{barA} argon target is $\sim\!\!\SI{10}{\milli\meter}$. To achieve 10 samples along such a track, a readout plane with a granularity of order \SI{1}{\milli\meter\squared} is needed. Conventional segmented pad planes of current experiments (e.g. T2K) have a pad size of order \SI{1}{\centi\meter\squared} at a cost of about 8 EUR per channel. Given that an area of \SI{20}{\meter\squared} is realistic for the readout plane of a future HPTPC near detector at a long-baseline neutrino oscillation experiment, a solution with a lower cost per channel is attractive.

The transverse diffusion in pure \ar{} at \SI{5}{barA} is too large to allow for drift lengths of several \SI{10}{\centi\meter} whilst permitting \SI{1}{\milli\meter} track sampling. When adding a quencher such as \co{} to the argon gas, the diffusion is reduced, allowing for \SI{1}{\milli\meter} track sampling and a \SI{50}{\centi\meter} drift length (\secrefbra{hptpcPaper:sec:gasChoice}).

\subsection{Optical readout}
\label{hptpcPaper:sec:opticalReadout}

A relatively new development in TPC readout technology that offers a low cost per channel is optical readout. TPCs have been in use since the late 1970s, typically with direct readout of the drifted charge. CCD optical readout of time projection chambers was first demonstrated in $\sim$1990 by~\cite{Breskin:1988sd},~\cite{Buckland:1994gc} and~\cite{Fonte:1989658}, and more recently has been developed by the DMTPC project for direction-sensitive dark matter searches~\cite{Dujmic:2007bd}, by the CYGNO collaboration~\cite{Baracchini_2020}, by an optical TPC for precision nuclear physics cross section measurements \cite{Gai:2011yb}	, for X-ray imaging~\cite{FUJIWARA20177}, for proton imaging~\cite{Klyachko:2011} and by the CERN gas detectors group for radiography using x-rays \cite{Brunbauer:2018nyz}. DMTPC demonstrated that a TPC with optical readout can realise a sub-\si{\milli\meter\squared} segmentation over a readout plane with an area larger than \SI{1}{\meter\squared} \cite{Leyton:2016nit}. For a recent review, we refer the reader to \cite{BATTAT20161}.

\indent An optical TPC is instrumented with a cathode and (several) anode electrodes which define its signal collection and amplification regions. Ionisation electrons from charged particles propagating through the TPC move in the drift field to the amplification region where avalanche charge multiplication and scintillation photon production occurs (\figrefbra{fig:hptpcPaper:sec:HPTPCOverview:sketch:xz}). A schematic of how our optical TPC operates is shown in \figref{fig:hptpcPaper:sec:HPTPCOverview:sketch}. The anodes may also be equipped with charge readout to provide high resolution tracking in the drift direction, as in ~\cite{Leyton:2016nit}. CCD or CMOS cameras view the amplification plane through lenses from outside of the pressure vessel containing the TPC and target gas, collecting the scintillation light and subsequently providing tracking information in the amplification plane. The design considerations for optical TPCs are described in detail in~\cite{BATTAT20161}.\\
In an optical TPC, the track reconstruction resolution in the amplification plane depends on the optical plate scale. This scale is determined by the requirement that the object be in focus, which sets a minimum object distance given an image distance and focal length of the lens, and on the optical system demagnification, which is the ratio of the object to image distances. Typical demagnification values are 5-10. The area of the amplification region imaged by each CCD pixel (a ‘vixel’), determines the smallest unit of track segment measurement possible with a given optical system and detector geometry. We define a vixel to be a box with an area of $A_{\text{vixel}}$ for the sides parallel to the readout plane and a height corresponding to the length an electron drifts during one CCD exposure time.\\
The minimum sensible vixel size is determined by the transverse diffusion of the ionization electrons from a particle track in the TPC, as they drift to the amplification region. The track reconstruction resolution in the drift direction is determined by the number of samples along the track, which depends on the track length, drift velocity, and readout rate.\\
The momentum threshold for track reconstruction depends on the minimum deposited energy at which a cluster of vixels can be identified as a track. This depends primarily on the signal-to-noise ($\mathcal{S}:\mathcal{N}$) ratio per vixel. In general, vixels with $\mathcal{S}:\mathcal{N}>5$ can be identified as part of a particle track~\cite{Deaconu:2017vam}.\\
The expected signal size, that is the number of photons collected per pixel, is given by
\begin{equation}
\begin{split}
	N_{\mathcal{S}} \ = \ \biggl[ \frac{\varepsilon_{\text{particle}}}{W} \times G \times (\gamma/e^-) \biggr] \times  \biggl[ T_{\text{anode}} \times T_{\text{cathode}} \times T_{\text{window}} \times T_{\text{lens}} \biggr] \\
	\times  \biggl[ \frac{1}{16 \times f_{\text{stop}}^2 \times (1+m_{\text{d}})^2} \biggr] \times QE^* \quad ,
\end{split}
\label{eqn:signal_contributions}
\end{equation}
where the first term in brackets is the number of photons produced in the amplification region, which depends on the ionization energy deposited per vixel by a particle with energy $\varepsilon_{\text{particle}}$, the energy to liberate one electron-ion pair in the gas $W$, the gas amplification factor (gain) $G$, and the scintillation photon-to-electron ratio $(\gamma/e^-)$ of the gas. The second term in brackets is the total photon transmission of the system, which depends on the transmittance of the lens ($T_{\text{lens}}$), the pressure vessel window ($T_{\text{window}}$), and the cathode ($ T_{\text{cathode}}$) and anode meshes ($T_{\text{anode}}$) through which the CCD views the amplification region, averaged over the scintillation emission spectrum. The third term in brackets is the geometric acceptance of the optical system, which depends on the lens aperture to focal length ratio ($f_{\text{stop}}$) and the demagnification ($m_{\text{d}}$). The last term $QE^*$ is the CCD quantum efficiency averaged over the scintillation emission spectrum. Other elements which enhance (\textit{e.g.} reflections) or reduce the signal are not taken into account.\\
The noise per vixel depends on the quadrature sum of the shot noise which is $\sqrt{N_{\text{signal}}}$, the read noise $N_{\text{read}}$, and the dark rate of the camera times the exposure time ($N_{\text{pixels}} \cdot R(T) \cdot t_{\text{exposure}}$):
\begin{equation}
N_{\mathcal{N}} \ = \ \sqrt{N_{\text{signal}} + N_{\text{read}}^2 + N_{\text{pixels}} \cdot R(T) \cdot t_{\text{exposure}} } \quad.
\label{eqn:noise_contributions}
\end{equation}
In the dark noise term, $N_{\text{pixels}}$ is the number of CCD pixels grouped into a readout bin, $t_{\text{exposure}}$ is the exposure time of a pixel, and $R(T)$ is the dark rate which is a function of temperature $T$. Here, a readout bin is a group of camera pixels which is grouped together and read out as one. Typically a cooled CCD can suppress the dark current to $<\SI{0.1}{electrons\per pixel \per\second}$, whilst the read noise is of order 10 electrons RMS, so for exposure times of order seconds the read noise dominates. The area determined by $N_{\text{pixels}} \times A_{\text{vixel}}$ can be thought of as an effective pad size of the readout, where $A_{\text{vixel}}$ is the vixel area imaged by one CCD pixel.\\
In the prototype detector described here, the vixel size is $\sim\!\!236\times\SI{236}{\micro\meter\squared}$ in the readout plane, and the readout binning operated was $4\times4$ ($N_{\text{pixels}}$ = 16) and $8\times8$ ($N_{\text{pixels}}$ = 64), producing an effective pad size after readout binning of $\sim\!\!\SI{0.86}{\milli\meter\squared}$ and $\sim\!\!\SI{3.46}{\milli\meter\squared}$ respectively. In this way, a \SI{10}{\milli\meter} long track, corresponding to a \SI{50}{\mega\electronvolt\per c} proton, is sampled at 5-10 points, as the vixel area in the readout plane is a square. The area $A_{\text{vixel}}$ is calculated by dividing the area imaged by one camera ($\sim\!\!71\times\SI{71}{\centi\meter\squared}$) by the camera's pixel layout of $3056\times\SI{3056}{pixel\squared}$ and accounting for the readout binning. The height of a vixel corresponds to the full drift length, since we operated the cameras with an exposure times of \SI{0.5}{\second} to \SI{1}{\second}.

\subsection{HPTPC prototype overview}
\label{hptpcPaper:sec:HPTPCOverview}

\begin{figure}
\centering
\subfloat[longitudinal view]{\label{fig:hptpcPaper:sec:HPTPCOverview:sketch:xz}
\includegraphics[width=0.39\columnwidth]{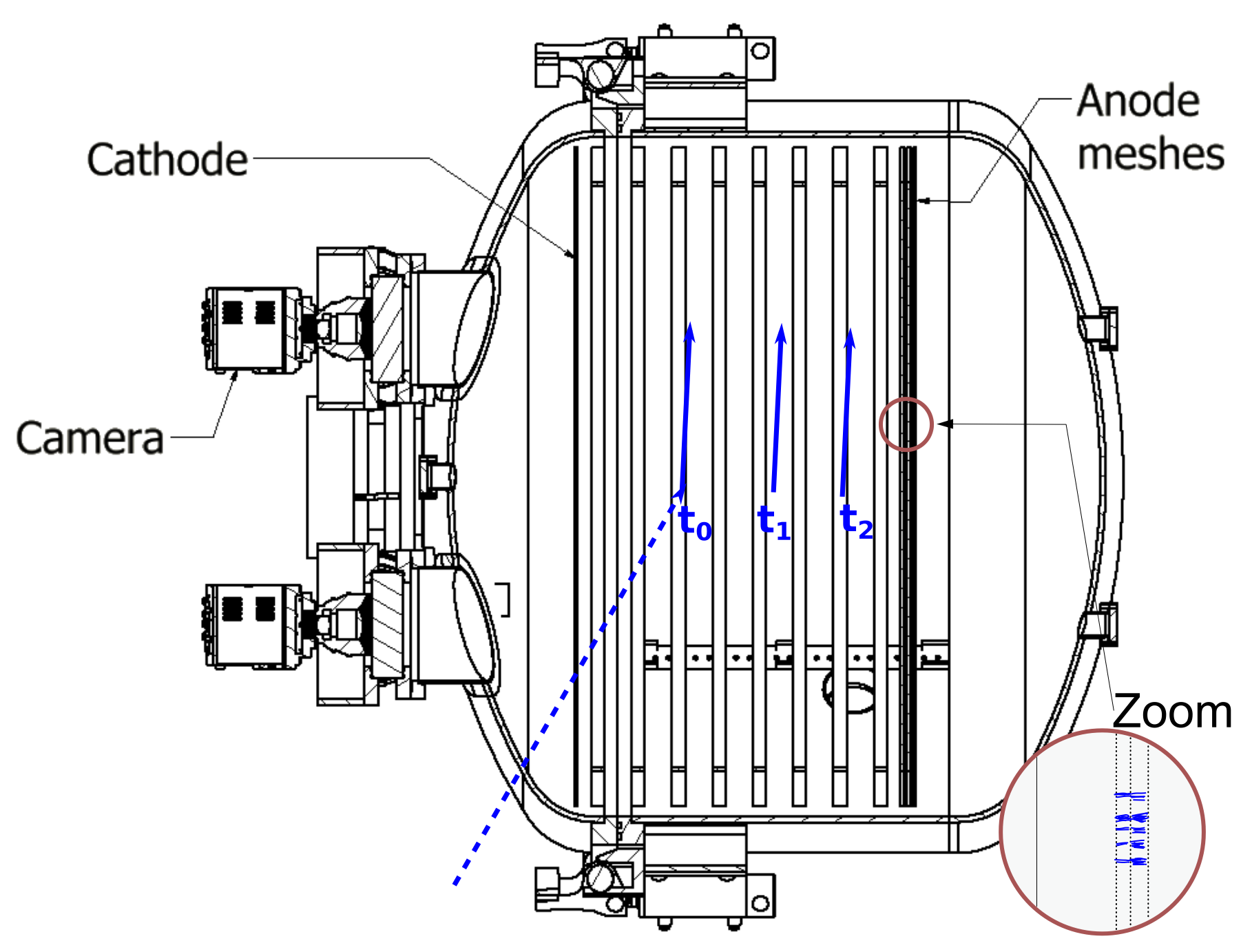}}
\subfloat[camera view]{\label{fig:hptpcPaper:sec:HPTPCOverview:sketch:xy}
\includegraphics[width=0.3\columnwidth]{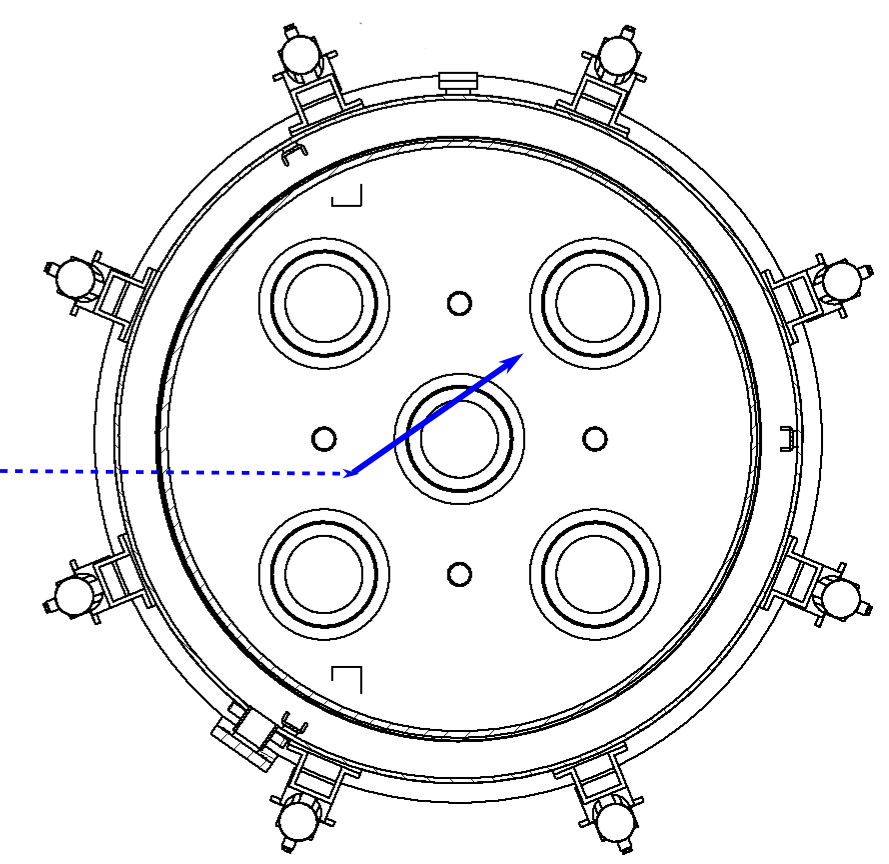}}
\caption{\label{fig:hptpcPaper:sec:HPTPCOverview:sketch}Cross-sectional view of the HPTPC through the \protect\subref{fig:hptpcPaper:sec:HPTPCOverview:sketch:xz} plane parallel to the drift field $E$ and \protect\subref{fig:hptpcPaper:sec:HPTPCOverview:sketch:xy} the plane perpendicular to $E$. A particle (dotted line) scatters on an atom or molecule in the gas at the time $t_0$, ejects a charged particle from the nucleus which in turn ionises gas atoms along its trajectory (arrow, Fig. \protect\subref{fig:hptpcPaper:sec:HPTPCOverview:sketch:xz}). These ionisation electrons are moved by $E$ towards the anode meshes and are eventually amplified. The positions of these ionisation electrons as they drift are labelled $t_1$ and $t_2$. Photons produced during the amplification are then imaged by cameras and provide the 2D projection of the interaction (Fig. \protect\subref{fig:hptpcPaper:sec:HPTPCOverview:sketch:xy}), the zoomed inlet in \protect\subref{fig:hptpcPaper:sec:HPTPCOverview:sketch:xz} illustrates where avalanches form and the photons are emitted.}  
\end{figure}
The prototype HPTPC detector described here is housed in a stainless steel (type 304L) vessel (\secrefbra{hptpcPaper:sec:highPressureVessel}) of volume \SI{1472}{\liter} which is rated to \SI{5}{barG}. We use \si{barA} to denote absolute pressure and \si{barG} for pressure measurements relative to ambient pressure. The vessel features feed-throughs for high voltage and instrumentation, optical windows and camera mounting hardware. The detector layout is sketched in \figref{fig:hptpcPaper:sec:HPTPCOverview:sketch}. The vessel's internal rail system supports a TPC, which has \SI{44.7}{\centi\meter} drift length and \SI{111}{\centi\meter} diameter (\secrefbra{hptpcPaper:sec:hptpc}).  
The TPC drift region is enclosed by the cathode mesh at negative voltage and the first anode mesh. Two more anode meshes at increasing positive voltage follow in order to amplify primary ionisations.

The working principle of the detector is illustrated in \figref{fig:hptpcPaper:sec:HPTPCOverview:sketch}. A particle entering the drift volume (\textit{e.g.} a neutrino) scatters at a time $t_0$ on an atom or molecule, thereby ejecting protons from the struck nucleus. These final state particles ionise gas atoms and molecules along their path (indicated schematically with an arrow in \figrefbra{fig:hptpcPaper:sec:HPTPCOverview:sketch:xz}). The resulting primary ionisation electrons drift in the electric field $E$ towards the anode meshes and are eventually amplified in the high electric field close to the meshes' wires and between the meshes. In the avalanche, electrons and photons are produced and the latter can then be recorded by the cameras, which provide an image of the interaction (\figrefbra{fig:hptpcPaper:sec:HPTPCOverview:sketch:xy}) with the locations as well as the intensity, where the latter is proportional to the energy deposited in the drift volume. Reading out the charge signals induced by the avalanches at the anode meshes provides additional time information.  The duration of these charge signals in the anodes will be proportional to the track length projected into the drift direction.  The advantages of using this charge readout include the ability to calibrate gas mixtures that emit very little light and the ability to correlate light and charge signals.

The optical readout system for the HPTPC prototype described here uses four CCD cameras, which are mounted onto the high pressure vessel and image the amplification stage from the cathode side, through the windows of the pressure vessel, as well as through the cathode and anode meshes. Each camera views one quadrant of the amplification region, through lenses focussed on the amplification plane (\secrefbra{hptpcPaper:subsec:hptpc:ccds}). 
The HPTPC's charge readout system reads the charge induced on the whole (un-segmented) plane of each of the three anodes.  The signals are decoupled, amplified and shaped by commercial front end electronics, and subsequently digitized synchronously in time with the CCD data acquisition.

Throughout the paper we use a Cartesian coordinate system in which all electrodes are $x/y$ planes at a constant $z$ and where the $z$ axis is parallel to the electric field direction. The origin is located in the centre of the anode 1 mesh and $z$ increases towards the cathode. In the $x/y$ planes we occasionally use polar coordinates where $r$ points from the centre to the edge of the TPC.

%% file: sec_2_gasMixtures.tex
\section{Gas Requirements}
\label{hptpcPaper:sec:gasChoice}

The typical wavelength sensitivity range of CCD cameras is \SI{350}{\nano\meter} to \SI{850}{\nano\meter} (\secrefbra{hptpcPaper:subsec:hptpc:ccds}), and therefore the gas is required to have a high photon (or electro-luminescence) yield in this wavelength range. A noble gas is the obvious choice for the dominant part of the gas mixture, since it lacks the rotational and vibrational degrees of freedom which absorb photons.\\ 
Gaseous argon has been shown to emit not only light in the Vacuum Ultra Violet (VUV), but also in the near infra-red (NIR) wavelengths~\cite{Buzulutskov2011}. Scintillation light measurements at pressures higher than \SI{1}{barA} show that the NIR light yield normalised to the number of amplification electrons decreases with increasing pressure \cite{Fraga2001}. This can, however, be compensated by a larger gain of the amplification stage. In \cite{Fraga2001} the authors show that additions of \cf{} leads to a high photon yield in the visible (VIS) and NIR: In \ar{} gas with a small (\SI{5}{\%}) admixture of \cf{}, the scintillation photon yield in optical wavelengths is 0.1-0.3 per avalanche electron, and is a weak function of the reduced electric field. Neon, on the other hand, emits in the NIR region as well \cite{LINDBLOM1988204}. Admixtures of nitrogen have been shown to result in a higher intensity electro-luminescence in the VIS, as compared to the NIR neon electro-luminescence. A \nen{} mixture is therefore also a good candidate for a TPC with optical readout.\\
We chose argon as the principle component of our gas mixture because an \ar{} based mixture is foreseen for the HPTPC of DUNE's near detector. Demonstrating the technological readiness of an HPTPC with this gas mixture makes a strong case for using this technology as part of a near detector in a long-baseline neutrino oscillation experiment with far detectors with identical targets. Argon has already been proven to emit light at high pressure in the wavelength range to which our cameras are sensitive \cite{Fraga2000}. Furthermore argon is considerably less expensive than neon gas. \\
\begin{figure}
\centering
\includegraphics[trim =0 0 0 0, clip = true, width=0.6\textwidth]{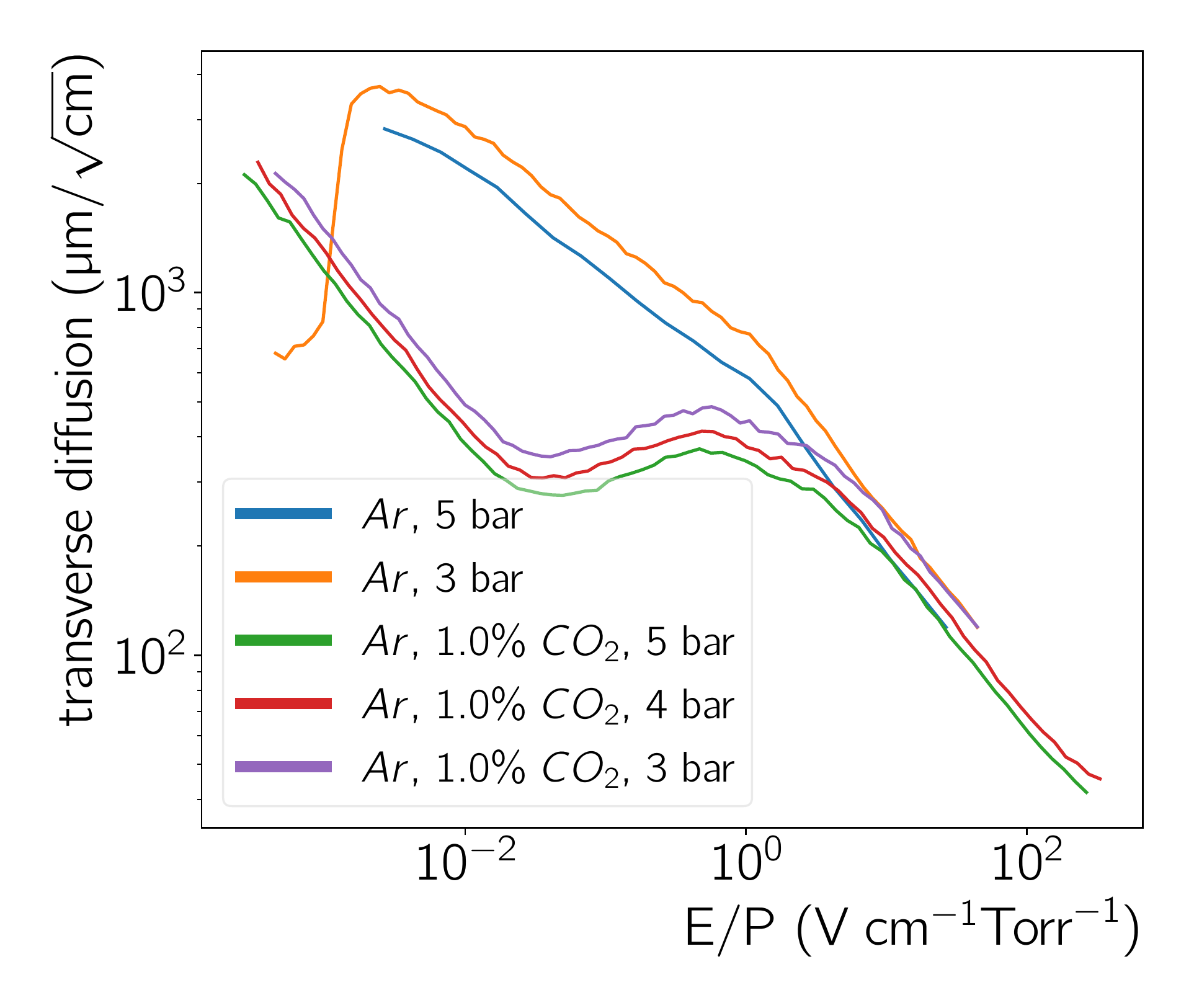}
\caption{\label{hptpcPaper:sec:gasChoice:fig:diff}Transverse diffusion for pure argon and different \arco{} mixtures simulated using \textsc{Magboltz} \cite{Biagi1018382}.}
\end{figure}
Operating a TPC with pure argon comes with the disadvantage that stable operation is notoriously difficult at high gains, and that the transverse diffusion is high. For a drift field of $\sim\!\!\SI{200}{\volt\per\centi\meter}$ the transverse diffusion in pure \ar{} at \SI{5}{barA} ($E/P\sim\!\SI{0.05}{\volt\per\centi\meter\per Torr}$) is about $\SI{1000}{\micro\meter\per\sqrt{\textrm{cm}}}$ \cite{Biagi1018382} as can be seen in \figref{hptpcPaper:sec:gasChoice:fig:diff}. An optical readout with cameras provides an effective segmentation of the readout plane into segments of less than a \SI{1}{\milli\meter\squared}, as discussed in \secref{hptpcPaper:sec:HPTPCDesignConsiderations}. The diffusion in pure argon for drift lengths of several 10s of \si{\centi\meter} is too large to exploit the advantages of a fine segmentation. Adding a quencher reduces the diffusion and enables higher gains under more stable operating conditions. For example, in \arcois{99/1} the diffusion is reduced by an order of magnitude as compared to pure argon (\figrefbra{hptpcPaper:sec:gasChoice:fig:diff}). This allows drift lengths of up to \SI{50}{\centi\meter} whilst retaining the requirement that the transverse diffusion not exceed twice the readout segment length of \SI{1}{\milli\meter}.\\
The typical quencher for an \ar{} mixture is carbon-dioxide, however \co{} has been shown to lower the light yield \cite{Fraga2000}. \n{} on the other hand is not a good quencher in \ar{}, but provides small reduction in the light yield \cite{Fraga2000}. In this paper we experimented with pure \ar{} and different \arco{}, \arn{} and \arcon{} mixtures to establish which gas provides the largest light gain in the NIR (\secrefbra{hptpcPaper:subsec:hptpcPerfomrance:lightGain}). During the operation of the high pressure TPC periods of sparking occurred (\secrefbra{hptpcPaper:subsec:readoutAndDAQ:CCDSparkDetect}), which had a large influence on the gas eventually used in the measurements presented in this paper. Other gases and admixtures are also interesting to study, however, these studies are not part of the measurements for this paper.

%% file: sec_3_hptpc.tex
\section{High pressure vessel design}
\label{hptpcPaper:sec:highPressureVessel}

\begin{figure}
\centering
\includegraphics[width=\columnwidth]{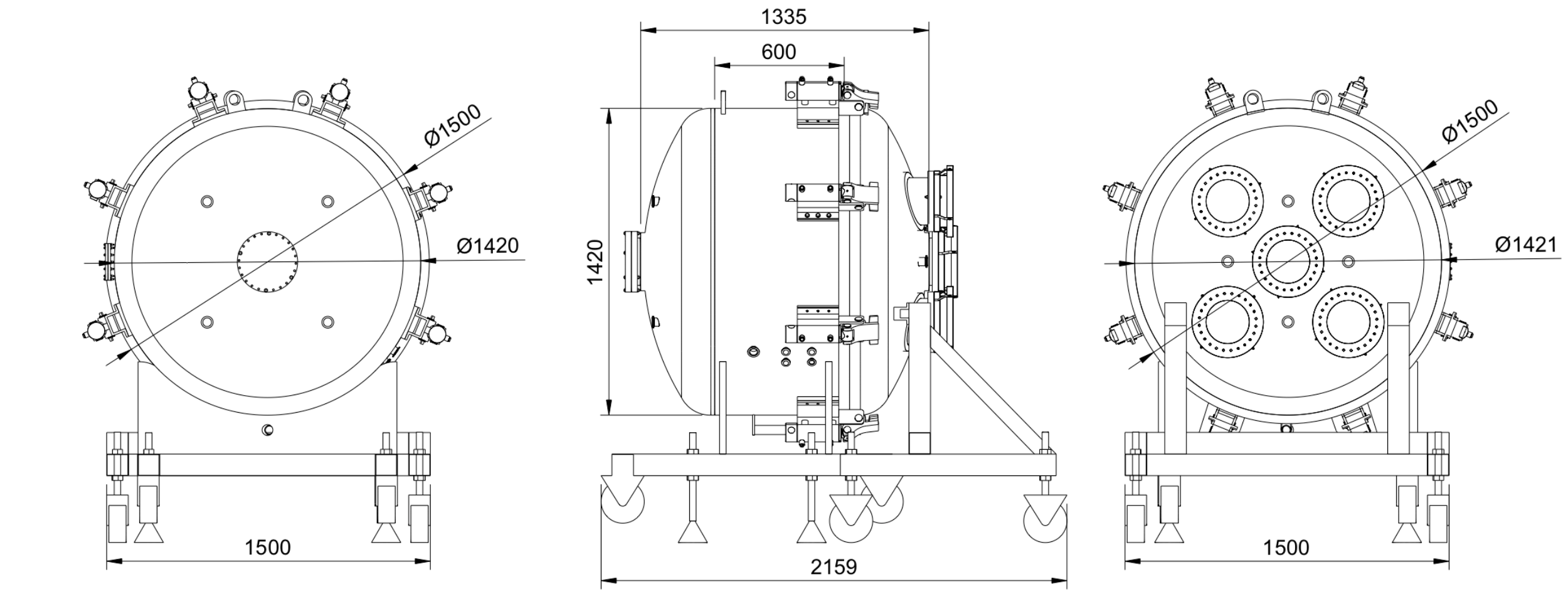}
\caption{\label{hptpcPaper:sec:highPressureVessel:fig:Vesse1} Schematic drawings of the pressure vessel: end view of the back side (left), side view with the vessel door to the left (middle), and end view of the door side (right).}
\end{figure}
The pressure vessel design is shown in \figrefbra{hptpcPaper:sec:highPressureVessel:fig:Vesse1}. The vessel is 304L stainless steel, with an inner (outer) diameter of \SI{140}{\centi\meter} (\SI{142}{\centi\meter}). The total length of the vessel, including the domed ends, is \SI{138.6}{\centi\meter}; the length of the domed sections is \SI{32.5}{\centi\meter} each, leaving a length of \SI{73.6}{\centi\meter} in the cylindrical straight section which hosts the TPC. The weight of the vessel (empty) is \SI{2370}{\kilogram}. 

One of the domed ends of the cylinder is fully detachable to gain access to the vessel's interior, \textit{e.g} for the TPC installation. The detachable door is connected to the body of the vessel via a large DN 1500 flange. The door and the body of the vessel are mounted to separate steel frames with wheels, both with adjustable feet for elevation adjustment. A double O-ring seal of viton and a rectangular silicone layer between the door and body flange sides provide gas tightness. The door and body flanges are clamped together with 8 hydraulic pistons and 8 screwable clamps, with a force up to \SI{50}{\newton\meter}. The helium leak tightness specification is \SI{2.5e-9}{\milli\bar\liter\per\second}.
 
The vessel flanging is indicated in \figref{hptpcPaper:sec:highPressureVessel:fig:Vesse1}. The door is equipped with five DN200 and four KF40 flanges (\figrefbra{hptpcPaper:sec:highPressureVessel:fig:Vesse1}, left), while the body features one DN200 and four KF40 flanges on the side opposite of the door (\figrefbra{hptpcPaper:sec:highPressureVessel:fig:Vesse1}, right), four KF25 flanges and one KF40 flange on the left side of the body (\figrefbra{hptpcPaper:sec:highPressureVessel:fig:Vesse1}, middle), and four KF40 flanges on the right side. The KF25 and KF40 flanges are used for High Voltage (HV), gas and vacuum system feed-throughs. The 5 DN200 flanges on the door are each equipped with a custom optical window flange and camera mount incorporating a \SI{60}{\milli\meter} thick quartz optical window. The body flanges host two independent pressure relief systems. The first is a \SI{5}{barG} burst disk backed by a \SI{5}{barG} pressure relief valve. The second, on an independent body flange, is a \SI{6}{barG} burst disk.

The interior of the vessel houses three steel rails that run longitudinally along the walls, separated at approximately $\SI{120}{^\circ}$, to allow mounting of equipment inside the chamber. The interior surfaces are shot blasted.

The vessel is rated to a \SI{6}{barA}. To verify this after construction the vessel was filled with water and subjected to the test pressure of \SI{7.2}{barG} for 10 minutes. No evidence of leaks or material deformation was observed. The hydrostatic pressure was subsequently decreased to the working pressure of \SI{5}{bar} absolute pressure and maintained for 90 minutes to verify the tightness of the pressurized vessel. After the test the vessel was emptied and dried with nitrogen gas. All optical windows were installed during this test, validating the design of the custom DN200 optical flanges. Given the \SI{1472}{\liter} volume, the vessel is a category IV pressure vessel. This hydrostatic test was used to follow conformity assessment procedure MAT-17-CE-G-CRTO02/17 to obtain the declaration of conformity with pressure vessel directive 97/23/CE.

\subsection{Gas system}
\label{hptpcPaper:subsec:highPressureVessel:gasSystem}

\begin{figure}
\centering
\includegraphics[width=0.6\columnwidth]{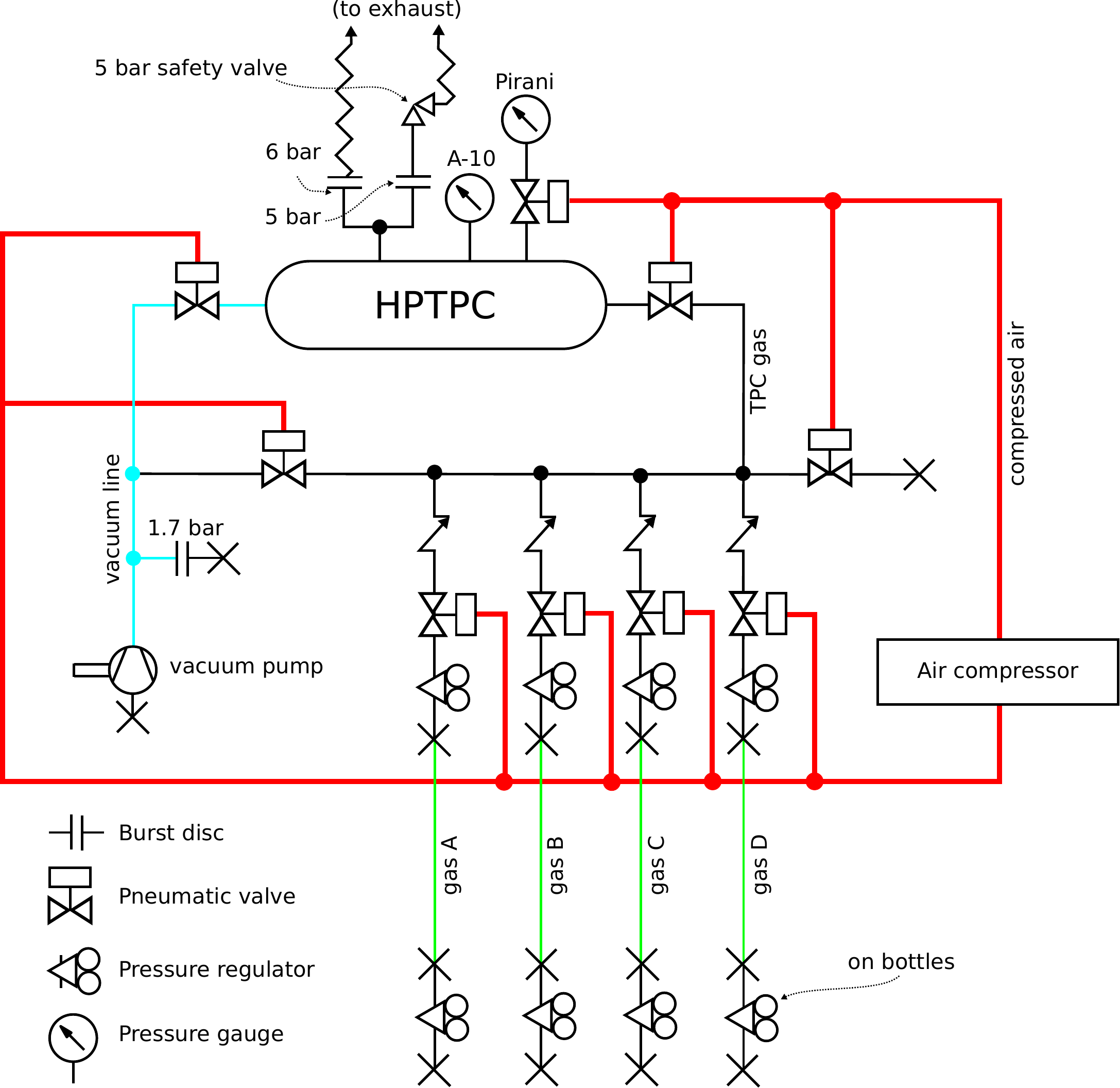}
\caption{\label{hptpcPaper:sec:highPressureVessel:fig:PNID} Diagram of the gas fill and evacuation system for the HPTPC vessel.}
\end{figure}

The gas and evacuation system for the HPTPC prototype detector described here is shown in \figrefbra{hptpcPaper:sec:highPressureVessel:fig:PNID}.

The gas filling strategy for the HPTPC foresees to evacuate (and purge) the vessel prior to the target gas fill. An Agilent Triscroll 800 dry vacuum pump is used to pump down the vessel to a pressure of approximately \SI{1e-6}{barA} before gas operations. The same pump is used to evacuate the fill line from the gas system to the vessel to reduce contamination, during the filling procedure or in case a gas fill is topped up to a higher pressure. The system enables mixing of gases from four different inputs, using eight Aura gas pressure regulators with manometers and threaded connections (four in the primary $200-\SI{10}{\bar}$ stage and four in the secondary $10-\SI{0}{\bar}$ stage). Mixtures are achieved by filling with different gases in turn, while the respective proportions are adjusted by partial pressure. The lines from gas bottle to the gas system are purged using gas from the bottle whenever a new bottle is connected.

All valves can be opened and closed remotely using the slow control system. The system consists of eight \SI{0.5}{in} Swagelok solenoid valves and one \SI{1.5}{in} Carten solenoid valve, all pneumatically activated. The gas pressure is monitored by a Wika A-10 digital pressure gauge (from \SI{0.8}{barA} to \SI{6}{barA} absolute pressure), and by an Inficon PGC550 combined capacitance-Pirani vacuum gauge for pressures between \SI{5e-8}{barA} to just below atmospheric pressure. Since the Pirani gauge is not suitable for over-pressure it is protected by an electronically controlled valve when the pressure exceeds \SI{0.8}{barA}. The slow control system logs the gas pressure from the two gauges as well as the ambient laboratory temperature, measured with sensor SynAccess TS-0300, for later use in the analysis.

\section{Time Projection Chamber}
\label{hptpcPaper:sec:hptpc}

\begin{figure}
\centering
\subfloat[]{\includegraphics[trim=240 480 160 160, clip=true, width=0.59\columnwidth]{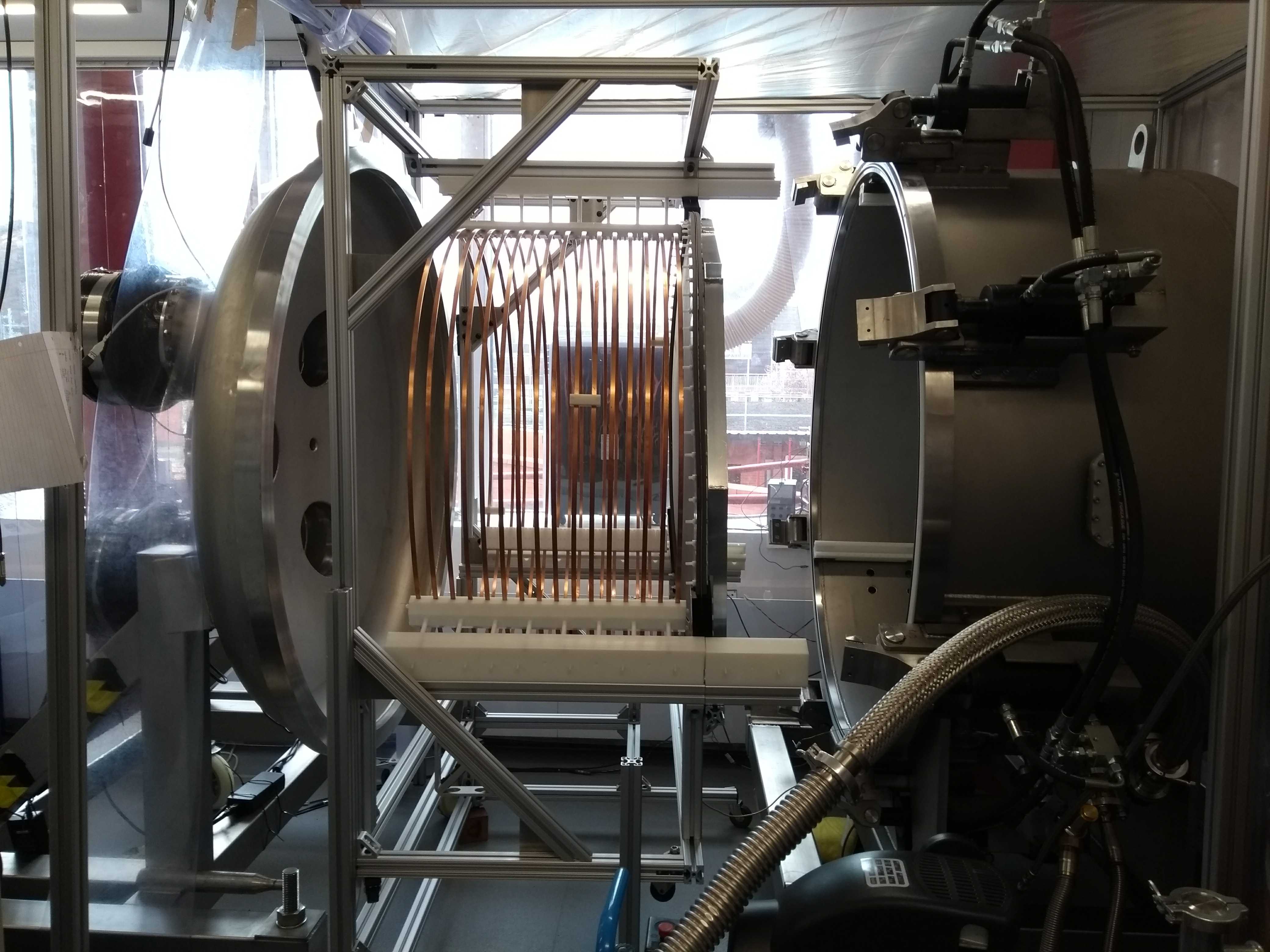}
\label{hptpcPaper:sec:hptpc:fig:fc:FCoutside}}
\subfloat[]{\includegraphics[width=0.39\columnwidth]{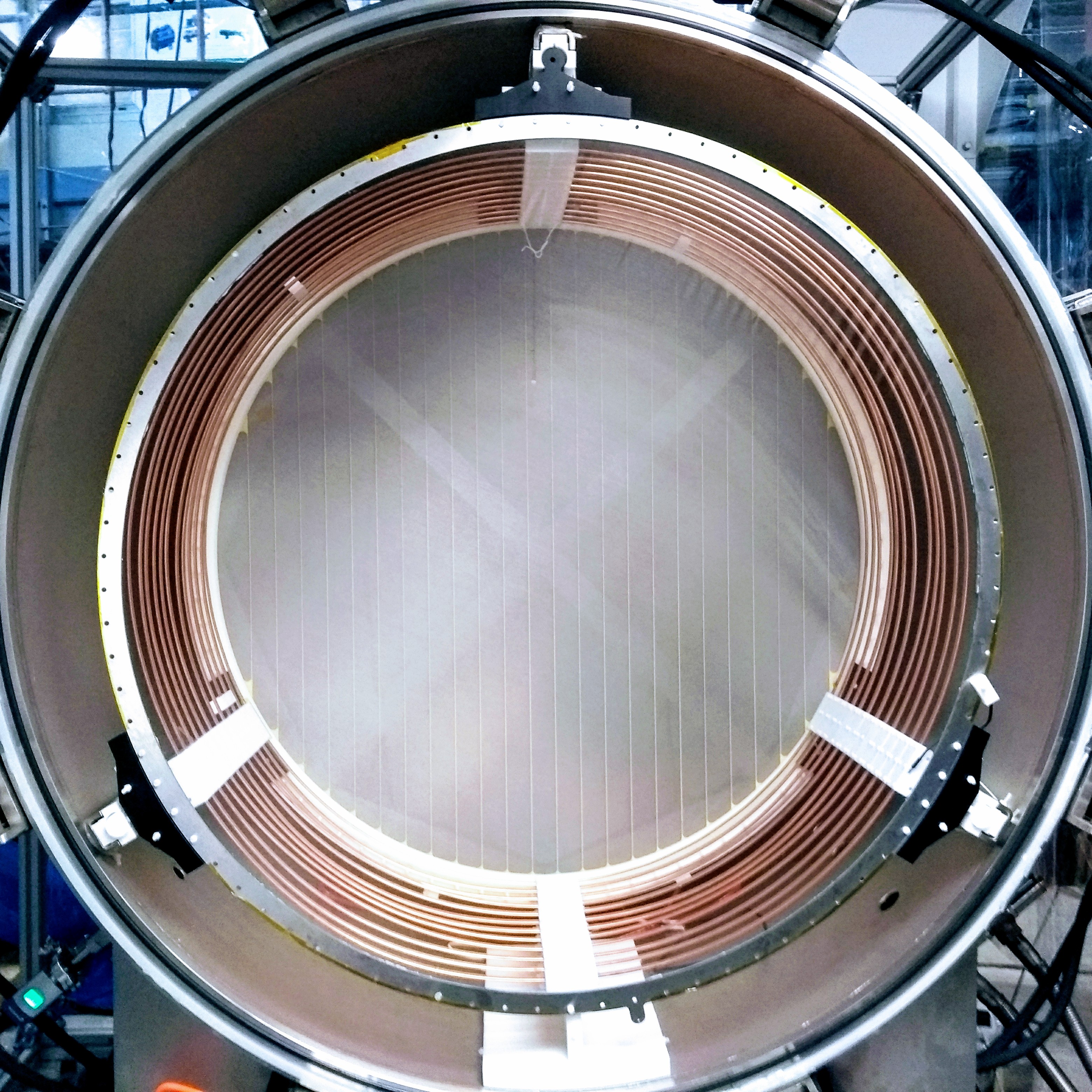}
\label{hptpcPaper:sec:hptpc:fig:fc:FCinside}}
\caption{\label{hptpcPaper:sec:hptpc:fig:fc}\protect\subref{hptpcPaper:sec:hptpc:fig:fc:FCoutside} The field cage before insertion into the pressure vessel and \protect\subref{hptpcPaper:sec:hptpc:fig:fc:FCinside} after insertion. The latter picture is photographed through the high-transparency cathode towards the amplification region and shows the full TPC.}
\end{figure}

The principal components of the time projection chamber are the field cage and the electrodes that define the drift and amplification regions. \Figref{hptpcPaper:sec:hptpc:fig:fc} shows the field cage ring structure, and amplification region before the assembly is inserted into the pressure vessel (left), and \emph{in-situ}---including the cathode---before the pressure vessel is closed (right).

\subsection{Field cage}

\indent\label{hptpcPaper:subsec:hptpc:FieldCage} The field cage (\figrefbra{hptpcPaper:sec:hptpc:fig:fc:FCoutside}) is constructed of 12 copper rings with an inner diameter of \SI{111}{\centi\meter}, and length of \SI{1.0}{\centi\meter} in $z$ and \SI{0.6}{\centi\meter} in $r$. The distance between two neighbouring rings is \SI{2.5}{\centi\meter}. Each ring is supplied with HV via the cathode in series with \SI{3}{\mega\ohm} resistors held in place with compression fittings between subsequent rings. The last ring on the field cage facing the amplification region is connected to ground via a resistor who's value is chosen depending on the spacing between the final ring and the amplification region to maintain field uniformity. The total length of the field cage is \SI{42.4}{\centi\meter}, resulting in a \SI{44.7}{\centi\meter} drift distance between the cathode and the amplification region.

The field cage assembly is supported from the three internal rails on the pressure vessel by machined Delrin parts. One set of these supports houses the resistor chain. The size of the support between the vessel rails and the field cage is adjustable.

\subsection{Cathode electrode}

The cathode electrode is a \SI{25}{lpi} (lines per inch) steel mesh made from \SI{27}{\micro\meter} diameter wires. Due to its low wire density the mesh has a calculated transparency of $\sim\!\!\SI{97}{\%}$, which allows for camera imaging of the amplification region through the cathode mesh (\figrefbra{hptpcPaper:sec:hptpc:fig:fc:FCinside}).

A \SI{122}{\centi\meter}$\times$\SI{122}{\centi\meter} square of this mesh was stretched to a tension of \SI{6.4}{\newton\per\centi\meter} on a Gr\"{u}nig G-STRETCH 210 mesh stretching machine. After stretching, the mesh was epoxied using DP460 epoxy to a circular stainless steel ring, with \SI{118}{\centi\meter} outer diameter, \SI{112}{\centi\meter} inner diameter, and \SI{0.3}{\centi\meter} thickness.

The tension measurement employs a Teren Instruments HT-6510N tension meter; measurements are made at 9 points on the mesh. The average tension reported here is the average of measurements at these 9 locations, after the stretching and relaxation procedure is completed. The standard deviation of repeated tension measurements across the 9 spatial locations is measured to be $0.4-\SI{0.8}{\newton\per\centi\meter}$ \cite{maRitchieYates2018}.

The cathode assembly is supported on the three internal rails of the vessel with machined Delrin pieces. The spacing of the cathode to the first field cage ring is constrained by the Delrin supports of the cathode and the mating support parts of the closest field cage ring.

\subsection{Gas amplification stage}
\label{hptpcPaper:subsec:hptpc:AmplificationStage}

The amplification stage is constructed from three electrodes (anodes), separated by two resistive spacers. The resistive spacers are \SI{121}{\centi\meter} outer diameter, \SI{112}{\centi\meter} inner diameter rings, with 24 wide beams, \SI{0.1}{\centi\meter} each, crossing them, laser cut from polyester shim stock. The spacer beams are visible as the vertical lines in the anode plane in \figref{hptpcPaper:sec:hptpc:fig:fc:FCinside}.
\label{hptpcPaper:subsubsec:hptpc:meshes}\\
The amplification region flatness is constrained by its support frame, which consists of two ring-shaped frames made of Nylon bolted together, which sandwich the anodes and resistive separators. The frame dimensions are \SI{118}{\centi\meter} outer diameter, \SI{112}{\centi\meter} inner diameter, with thickness \SI{1.6}{\centi\meter}. The two rings, and each amplification mesh and resistive spacer ring have 88 aligning drilled holes. A stack is formed with one support frame on the bottom, followed by alternating the three amplification meshes with the two resistive spacers and finished with the second support frame. Nylon bolts are passed through the 88 drilled holes in the stack. Finally, on the side facing away from the camera readout, a hexagonal aluminium stiffener is bolted to the framed assembly, attached to the nylon bolts at 16 of the 88 drilled holes points.\\
The three anodes are constructed from steel meshes with \SI{121}{\centi\meter} diameter. Anodes 1 and 2 are made from \SI{100}{lpi} meshes with a wire diameter of \SI{27}{\micro\meter}. The optical transparency of the anode 1 and 2 mesh is \SI{89}{\%}. The third anode is made from a \SI{250}{lpi} mesh with \SI{40}{\micro\meter} diameter. We chose the meshes with the smaller wire diameters for the two meshes closest to the field cage in order to achieve the highest gas amplification in the first stages, and minimize the loss of light because of imaging the amplification region through the cathode, anode 1 and anode 2 meshes. The meshes are epoxied to stainless steel rings (outer diameter \SI{118}{\centi\meter}, inner-diameter \SI{112}{\centi\meter}, thickness \SI{0.1}{\centi\meter}) after stretching the meshes as described for the cathode. The procedure for stretching the anode meshes takes approximately a week of successive stretching and relaxation of the mesh. Following this procedure, the average tension force on the anode 1 and 2 meshes is \SI{16.8}{\newton\per\centi\meter}. The measured tension is uniform over the plane of the anode mesh to better than \SI{5}{\%}. The average tension force on the anode 3 mesh is \SI{38}{\newton\per\centi\meter}. After stretching, the meshes are epoxied to the stainless steel support ring in the same way as described for the cathode \cite{maRitchieYates2018}. The goal for the distance between the anode 1 and anode 2 (anode 2 and anode 3) meshes is \SI{0.5}{\milli\meter} (\SI{1}{\milli\meter}). A measurement of the capacitance of the amplification region is described in \secref{hptpcPaper:sec:dataAnalysis:subsec:chargeCalib}. The capacitance measurement implies the distances achieved were approximately \SI{1}{\milli\meter} (\SI{2}{\milli\meter}) spacing. This is likely due to the epoxy and spacer thickness tolerances as well as flatness variation.\\
Like the cathode, the amplification region assembly is supported on the three internal rails on the pressure vessel using machined Delrin parts. These supports constrain the amplification region distance to the closest field cage ring.

\subsection{High-voltage distribution system}
\label{hptpcPaper:subsubsec:hptpc:HV}

The anode meshes are provided with positive high voltage (HV) by either a CAEN NDT1470 or a CAEN N1470 multi-channel Power Supply (PS), which is controlled through a serial link over USB. The cathode power supply is a Spellman SL 30 PS with a maximal output voltage of \SI{30}{\kilo\volt}. The resulting limit on the electric field in the field cage is over \SI{600}{\volt\per\centi\meter}. The cathode PS voltage is controlled by varying an analogue input from $0$ to $\SI{10}{\volt}$, which results in an output voltage from the PS of \SI{0}{\volt} up to its maximum voltage. This analogue signal is generated by the slow control system using a LabJack U3-HV USB Data AcQuisition (DAQ) device which is connected to the Spellman PS control input.

\begin{figure}
\centering
\includegraphics[width=0.65\columnwidth]{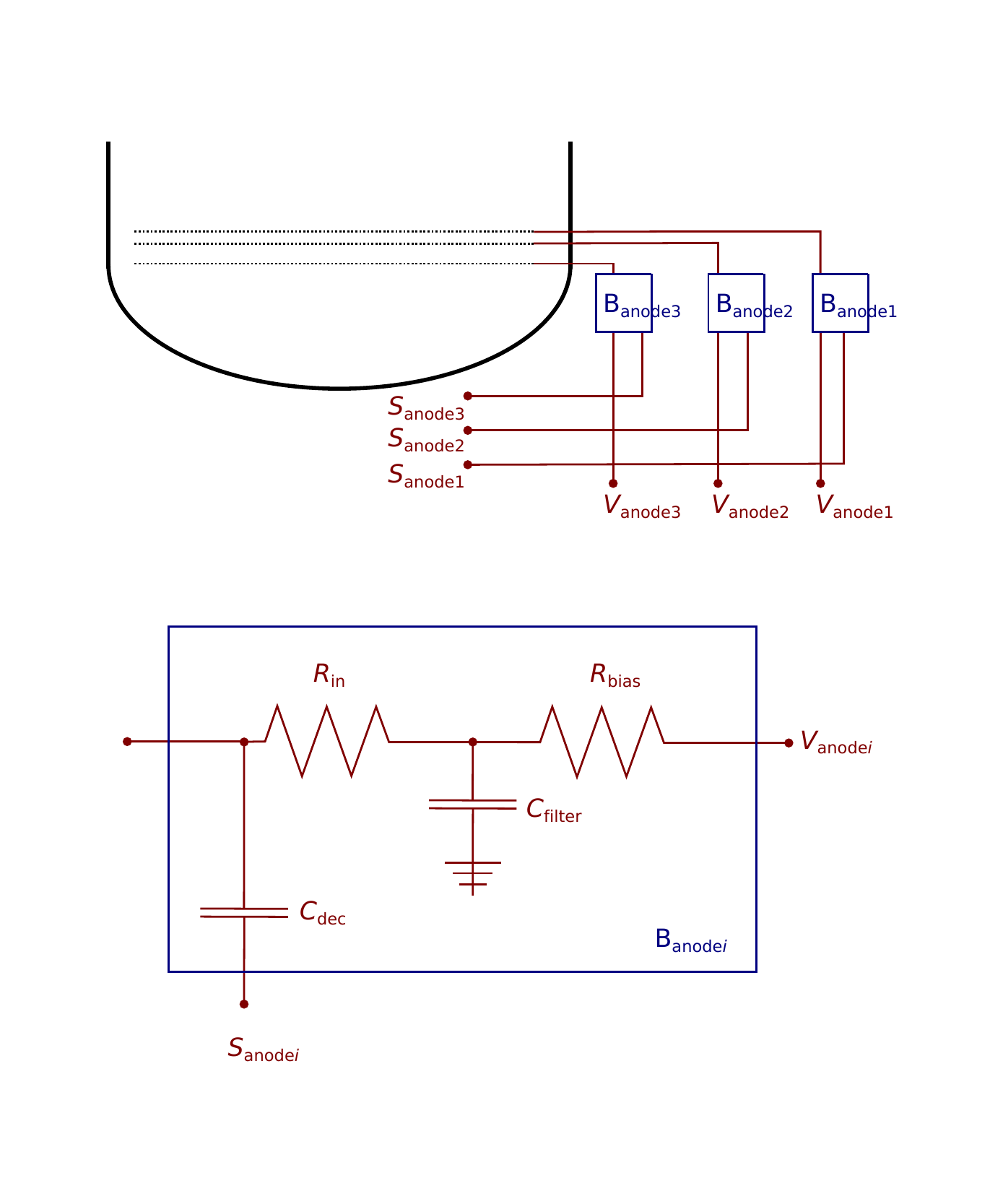}
\caption{\label{hptpcPaper:subsubsec:hptpc:HV:fig:HV}Schematic of the circuit to bring high voltage ($V_{\textrm{anode}i}$, $i\in1,2,3$) to the anode meshes and to decouple the signal from the high voltage lines. The signals are decoupled in bias boxes via a $\SI{10}{\nano\farad}$ decoupling capacitor ($C_{\textrm{dec}}$) and are then fed to the signal line ($S_{\textrm{anode}i}$). These bias boxes $B_{\textrm{anode}i}$ feature also a protection and filtering circuit consisting of a bias resistor ($R_{\textrm{bias}}=\SI{200}{\mega\ohm}$), filter capacitor ($C_{\textrm{filter}}=\SI{10}{\nano\farad}$), and input resistor at the detector input ($R_{\textrm{in}}$=\SI{10}{\mega\ohm}).}
\end{figure}

The various meshes are connected to the power supplies through the following chain: Inside the pressure vessel all meshes are connected to Kapton coated copper wires which in turn are connected to the HV feed-throughs that pass through the pressure vessel wall. To provide extra insulation, these wires have ceramic beads threaded along their entire length, and the resulting assembly is also surrounded by a fibreglass sheath. In the case of the anodes the HV feed-throughs are rated to \SI{10}{\kilo\volt}; in the case of the cathode the feed-through is rated to \SI{20}{\kilo\volt}.
Outside the pressure vessel, each anode's feed-through connects via coaxial cables to a custom 'bias box'. These bias boxes decouple the charge signals from the constant current HV as shown in \figref{hptpcPaper:subsubsec:hptpc:HV:fig:HV}. Therefore each bias box connects to the respective PS and each box has a signal output which is fed into the TPCs charge readout system. Signals are routed from these signal outputs through a preamplifier, described in \secrefbra{hptpcPaper:subsubsec:hptpc:amplifiers}. The RC constant of the $R_{\textrm{in}}$ resistor and the respective mesh capacitance of $\sim\!\!\SI{5}{\nano\farad}$ as well as the RC constant of the filter circuit limit the charge-up speed of the meshes and in turn help to quench discharges. The output from these preamplifiers is fed into the detector's DAQ system which is described in \secrefbra{hptpcPaper:sec:readoutAndDAQ}.

The cathode feed-through is connected to a coaxial power supply cable using a custom Delrin assembly which separates the grounded outer conductor of the cable from the voltage carrying inner conductor. The grounds of the power supplies (both anode and cathode) are connected together in a grounding circuit which is coupled to the pressure vessel.

The voltages and currents supplied by each power supply channel are recorded by the detector's slow control system, for use in later analysis.

\subsection{Charge signal measurement}
\label{hptpcPaper:subsubsec:hptpc:amplifiers}

The pre-amplifiers used for the detector's charge readout are charge-sensitive CREMAT CR-113 (or CR-112) hosted in CR-150-R5 evaluation boards. The specified gains of the pre-amplifiers are \SI{1.3}{\milli\volt\per\pico\coulomb} (or \SI{13}{\milli\volt\per\pico\coulomb} respectively). A measurement of the agreement of our preamplifiers with this value can be found in \secrefbra{hptpcPaper:sec:dataAnalysis:subsec:chargeCalib}. The output signals from the preamplifiers are digitised by a CAEN N6730 8-channel digitizer, with \SI{2}{\volt} dynamic range and \SI{500}{\mega\hertz} sampling frequency.

\subsection{Optical signal measurement}
\label{hptpcPaper:subsec:hptpc:ccds}

\begin{figure}
\centering
\subfloat[]{\label{hptpcPaper:subsec:hptpc:ccds:cameraMountDrawing}
\includegraphics[trim=10 0 3 0, clip=true, width=0.39\columnwidth]{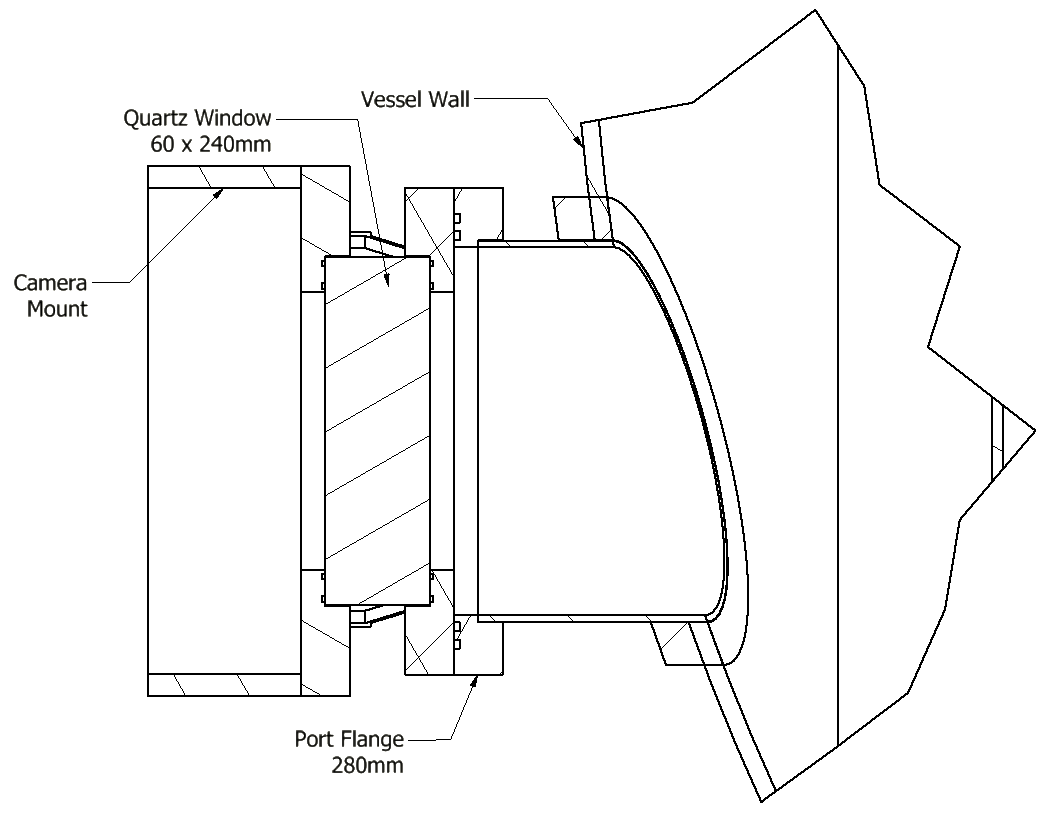}}
\subfloat[]{\label{hptpcPaper:subsec:hptpc:ccds:cameraMountPhoto}
\includegraphics[width=0.4\columnwidth]{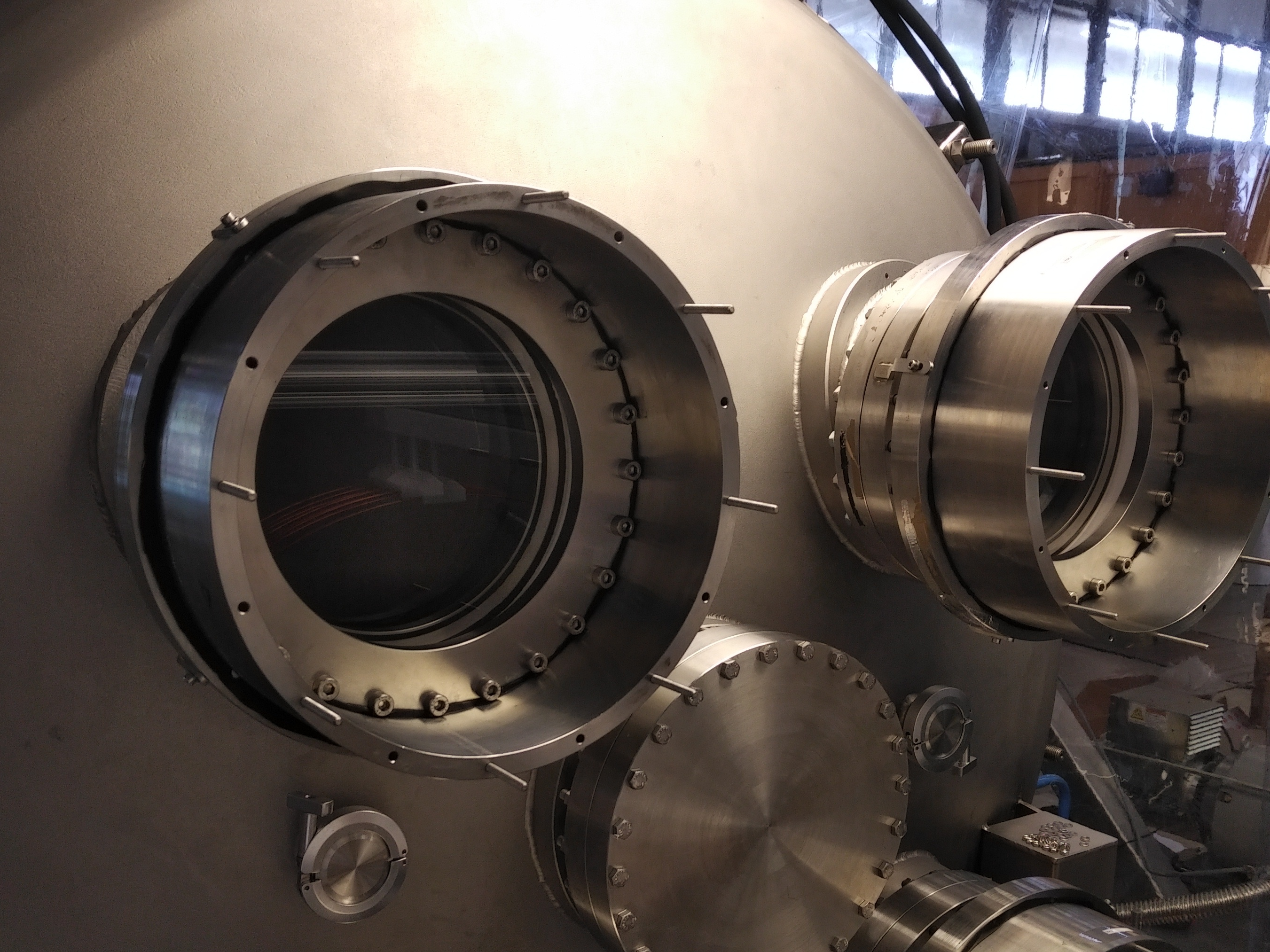}}
\caption{\label{hptpcPaper:subsec:hptpc:ccds:cameraMount}\protect\subref{hptpcPaper:subsec:hptpc:ccds:cameraMountDrawing} Drawing of the optical flange with the camera mount. The thick quartz is necessary to ensure that the assembly can withstand the pressure difference between the vessel pressure and ambient pressure. \protect\subref{hptpcPaper:subsec:hptpc:ccds:cameraMountPhoto} A photograph of the assembly with the camera removed.}
\end{figure}

The optical readout system uses four FLI Proline PL09000 CCDs, each of which contains a front-illuminated Kodak KAF-09000 chip with $3056\times3056$ active pixels (9.3 \si{Mp}), and a pixel size of $12\times\SI{12}{\micro\meter\squared}$. The chip has a quantum efficiency (QE) in the range of $50-\SI{70}{\%}$ for photons with a wavelength between \SI{475}{\nano\meter} and \SI{750}{\nano\meter}. In the wavelength range from \SI{350}{\nano\meter} to \SI{925}{\nano\meter} the QE is always larger than than \SI{20}{\%}. These wavelength ranges cover the full VIS part of a spectrum and extend towards the NIR and UV, which makes the TPC sensitive in the regions of the spectrum discussed in \secref{hptpcPaper:sec:gasChoice}.

Each camera's field of view is centred on a quadrant of the amplification plane. Each camera is coupled to a Nikon f/1.2 \SI{50}{\milli\meter} focal length lens with a $\SI{54.8}{^\circ}$ angle of view. The cameras are mounted to optical flanges as shown in \figref{hptpcPaper:subsec:hptpc:ccds:cameraMount}. Quartz windows of \SI{6}{\centi\meter} thickness are used to allow for the desired overpressure in the vessel. The transmission of the optical flanges is measured to be $97^{+3}_{-4}\;\%$ for red light. The camera lenses have a transmission of \SI{70}{\%} (\SI{90}{\%}) at \SI{420}{\nano\meter} (\SI{750}{\nano\meter}) wavelengths, where the lens transmission includes all photons lost between the 7 elements of this compound lens.

Adding up the TPC drift distance, the non-active area between the cathode and the vessel door as well as the path through the camera assembly, the total object distance is approximately \SI{102}{\centi\meter} which is larger than the minimum focal distance of our camera lenses. At this distance the system images a $71\times\SI{71}{\centi\meter}$ field of view with a vixel size of $\sim\!\!\SI{40}{\micro\meter}$, when no extra readout binning is applied. Considering the full optical path including quartz window and lens, we estimate a geometric acceptance of the optical system -- the third term in brackets in Equation~\eqref{eqn:signal_contributions} -- of approximately $1.1\times10^{-4}$. Achieving a high enough gain in the amplification region to produce enough photons for signals to be detected above the noise, given this acceptance, is key.

To achieve optimal noise performance the CCDs are cooled to $-\SI{25}{^{\circ}\textrm{C}}$ to $-\SI{30}{^{\circ}\textrm{C}}$. The cameras are equipped with an internal thermoelectric cooler which can cool the CCD to approximately $\SI{50}{^{\circ}\textrm{C}}$ below the ambient camera temperature. This is supplemented by a water cooling system attached to each camera to reduce its ambient temperature by $\SI{15}{^{\circ}\textrm{C}}$. At $-\SI{25}{^{\circ}\textrm{C}}$ operating temperature, the read noise per pixel is in the range of 9.6 to $\SI{11.3}{e^-}$, and the dark rate is $0.006-\SI{0.025}{\textrm{e}^{-}\per\textrm{pixel}\per\second}$ (the range of variation is across the four cameras).

The CCDs digitize the number of electrons collected in each pixel in each exposure. For scale, the typical conversion gains of the cameras are $1.52-\SI{1.55}{\text{e}^{-}\per ADU}$, where ADU is \textit{analogue-to-digital units}.
To mitigate the dominant effect of readout noise, pixels are grouped prior to digitization. This grouping reduces the noise per pixel in the group by approximately $1/\sqrt{N_{\text{pixels}}}$ given the relative scales of readout noise and dark current rate in a \SI{1}{\second} exposure. Typically we use $8\times8$ groupings ($N_{\text{pixels}}=64$) as this gives an acceptable balance between readout noise and readout pixel size, with the effective vixel width (pixel width in the amplification plane) being approximately \SI{2}{\milli\meter}.

\subsection{Slow control}
\label{hptpcPaper:sec:readoutAndDAQ}

The slow control software sets and monitors the detector voltages, gas pressure and ambient temperature. The software has a web based user interface, and uses java and C++ software to interface with an SQL database. The database contains the values of the monitored variables as well as the desired set points for these variables. The detector control code reads the set points from this database and communicates with the high voltage power supplies to set the required voltage and read out the measured voltage and current into the database. Control of the gas system is also achieved through the same web interface, which is able to launch code communicating with the valve control hardware to perform filling, venting and evacuation automatically.

\subsection{Data Acquisition}
\label{hptpcPaper:subsec:readoutAndDAQ:ReadoutStrategy} 

The DAQ system triggers and acquires data from the charge and optical readout hardware. DAQ commands are sent from the same web interface used for slow control to a DAQ PC which communicates with the cameras and the CAEN N6730 digitiser used for the charge signals to initiate each run. A run consists of a user-specified number of camera exposures (data frames), which are acquired simultaneously from the four cameras as well as the charge waveforms digitized during the exposure time. Additionally, at the start of the run a specified number of frames are acquired while the camera shutters are closed. The use of these frames is to subtract off the baseline behaviour of each pixel in the CCD chip when it is not exposed to light (\secrefbra{hptpcPaper:subsec:readoutAndDAQ:CCDnoise}). After these empty frames the data frames are taken with the camera shutter open. The detector can run in two triggering modes. In the first mode the data frames are taken immediately after each other, separated only by the CCD readout time. In the second mode the data frames are taken based on an external trigger signal. During the CCD exposure time the charge waveform digitiser (see \secrefbra{hptpcPaper:subsubsec:hptpc:amplifiers}) is triggered by signals larger or smaller than a user-configurable threshold amount above the baseline on each channel, and then records waveforms of typical duration \SI{100}{\micro\second} around each trigger, including a configurable period of time before the trigger event. The digitiser can also be triggered externally. In both triggering modes any trigger causes all eight channels of the digitiser to be read out simultaneously. The DAQ system stores the configuration of all of the parameters described in this subsection for each run.

%% file: sec_5_ccdAnalysis.tex
\section{Optical Readout Analysis and Performance}
\label{hptpcPaper:sec:ccdAnalysis}

In this section we report on the results of data taking with $^{241}\textrm{Am}$ sources mounted in the pressure vessel. The HPTPC was also tested using a beam at CERN, the analysis of which is ongoing and not presented here. After explaining the CCD calibration (\secref{hptpcPaper:subsec:readoutAndDAQ:CCDSparkDetect} and \secref{hptpcPaper:subsec:readoutAndDAQ:CCDnoise}) we show a first scan of various gas mixtures (\secrefbra{hptpcPaper:sec:hptpcPerfomrance:opticalRO}) to establish the most promising mixture for a more comprehensive light gain measurement. This in-depth measurement with a single mixture and its analysis is then shown in \secref{hptpcPaper:subsec:hptpcPerfomrance:lightGain}.\\
Am-241 predominantly emits either a \SI{5.486}{\mega\electronvolt} (\SI{84.8}{\%}) or \SI{5.443}{\mega\electronvolt} (\SI{13.1}{\%}) alpha particle ($\alpha$) and different energy gamma-rays ($\gamma$), where the most probable ones have an energy of \SI{59.54}{\kilo\electronvolt} or \SI{26.34}{\kilo\electronvolt} \cite{IEAE2019}. Furthermore there is a substantial amount of x-ray radiation in the range from \SI{10}{\kilo\electronvolt} to $\sim\!\!\SI{20}{\kilo\electronvolt}$. The $\alpha$ particles pass through a foil before they enter the gas volume, therefore their energy is reduced by about \SI{860}{\kilo\electronvolt} to $\sim\!\!\SI{4.56}{\mega\electronvolt}$ \cite{Henderson2016}. Such $\alpha$ particles are stopped inside the gas volume and deposit their full remaining energy. The $\gamma$-rays have high enough energy to escape the active TPC volume. According to a \textsc{heed} \cite{SMIRNOV2005474} and \textsc{Garfield++} \cite{garfieldpp} simulation which takes the HPTPC's geometry into account only \SI{1.2}{\%} of all $\gamma$-rays interact in the counting gas. The lower energy x-rays are more likely to interact; when integrating over all x-ray energies we find that \SI{58}{\%} are absorbed in the active gas volume. Their overall contribution is still not large, since the ratio of the x-ray count over $\gamma$-ray count is about \SI{12}{\%}. The emission distribution of the $^{241}\textrm{Am}$ in the forward hemisphere is roughly isotropic for the different kinds of radiation. Furthermore there is a contribution from cosmic rays.\\
\begin{figure}
\centering
\includegraphics[width=0.5\columnwidth, trim = 0 0 0 0, clip=true]{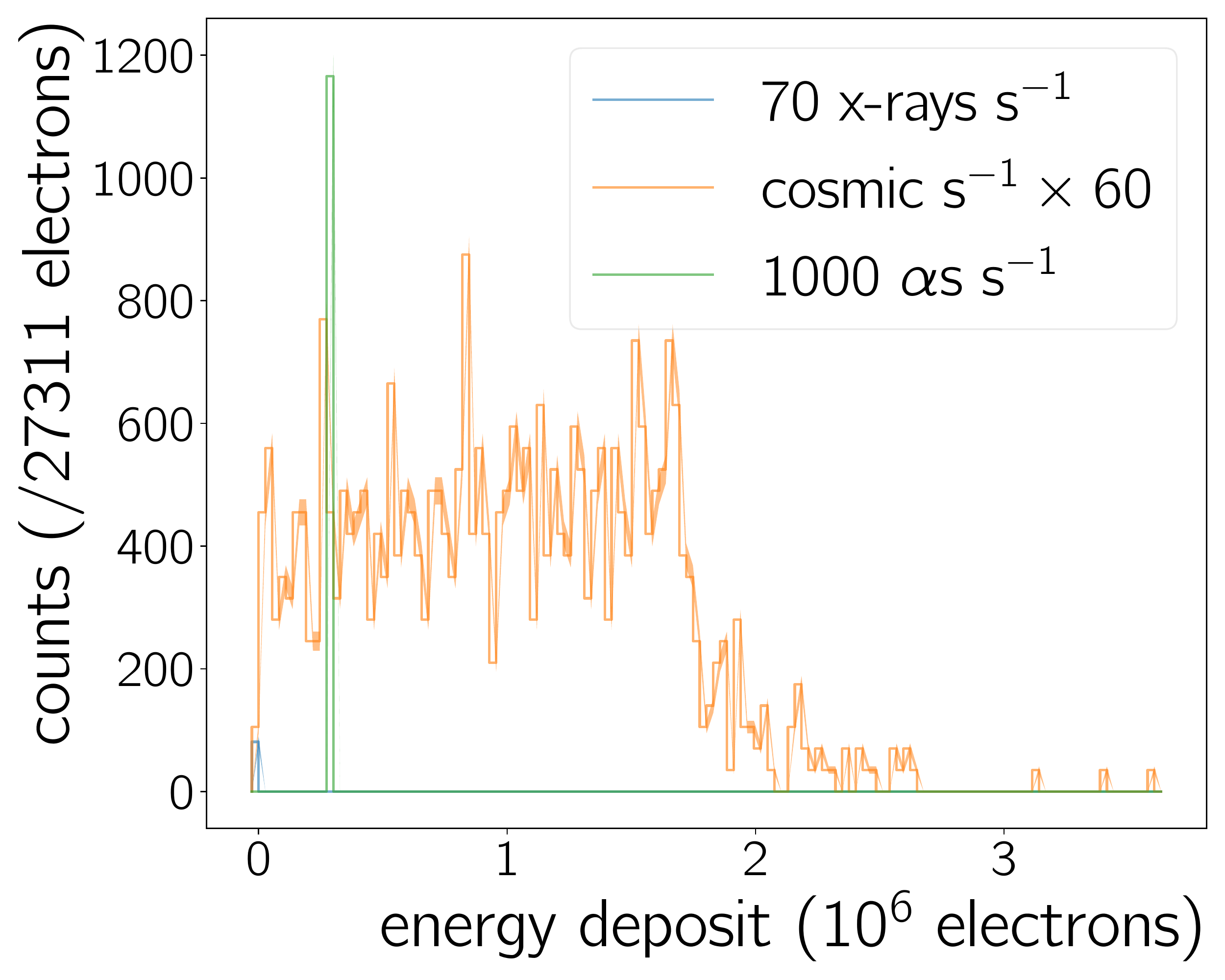}
\caption{\label{hptpcPaper:sec:hptpcPerfomrance:fig:energyDepostiInDet}Simulated energy deposits of $^{241}\textrm{Am}$ decay radiation and cosmic muons inside a gas volume filled with \arcois{90/10}. Energy deposits are measured in the number of liberated electrons during the energy deposit. This is the result of a \textsc{heed} \cite{SMIRNOV2005474} and \textsc{Garfield++} \cite{garfieldpp} study taking into account the approximate layout of the HPTPC and the information in \cite{IEAE2019,Henderson2016}.}
\end{figure}
\label{hptpcPaper:sec:ccdAnalysis:subsec:energyDeposit}
\Figref{hptpcPaper:sec:hptpcPerfomrance:fig:energyDepostiInDet} shows the result of a \textsc{heed} and \textsc{Garfield++} simulation of the expected energy deposits by these different sources of radiation, which does not take any trigger effects, electronic noise, gas gain or an amplifier response into account. For the simulation we assume a quadrant of the HPTPC's volume with a source location similar to the location in the experiment. The normalisation of the three different kinds of radiation in \figref{hptpcPaper:sec:hptpcPerfomrance:fig:energyDepostiInDet} is given by the result of the simulation: For \SI{1000}{Bq} of $^{241}\textrm{Am}$ decays, all 1000 $\alpha$ particles interact in the active volume every second as do 70 x-rays and $\gamma$-rays. Note that the x-rays and $\gamma$-rays contribute only at the low energy end of the spectrum. The contribution from the cosmic rays per second is scaled up by a factor of 60 to make the shape of the cosmic ray spectrum better visible. The most distinct feature of the spectrum is the $\alpha$-peak from the $^{241}\textrm{Am}$ decay at $\sim\!\!\SI{175E+3}{electrons}$.\footnote{It turned out to be not feasible to simulate stopping of $\alpha$ particles in \textsc{heed}. Therefore we ultimately simulated \SI{11.8}{\mega\electronvolt} $\alpha$ particles, evaluated their most probable energy loss and scaled this energy loss to \SI{4.56}{\mega\electronvolt}.} For a gas pressure of \SI{1}{atm} these ionisations are created along a \SI{5}{\centi\meter} to \SI{10}{\centi\meter} long trajectory, yielding a high ionisation density along the track. For larger pressures, the track length decreases and ionisation density increases. When amplified, this high ionisation density will result in many photons produced in a small area. An $\alpha$ particles' energy deposit in the detector is thus more easy to image with cameras than less ionising forms of radiation. Furthermore, a gain measurement is possible since the total energy deposited in the detector is known.\\
In the amplitude spectrum of the charge readout (\textit{cf}. \secrefbra{hptpcPaper:subsec:hptpcPerfomrance:chargeGain}), we expect to see something qualitatively similar to the spectrum in \figref{hptpcPaper:sec:hptpcPerfomrance:fig:energyDepostiInDet}. However, the simulation does not take into account the energy resolution of the amplification plane, hence the actual measured quantity -- amplitudes or light intensity -- will exhibit a spread larger than what is shown in the plot. Furthermore electronic noise is not included, which is a substantial contribution at threshold.\\
Two different configurations were used in our measurements: one using five $^{241}\textrm{Am}$ sources and one using a single $^{241}\textrm{Am}$ source. In the single source configuration the source is either visible in the overlap region of the top two cameras or the bottom two cameras. In the five source configuration the sources are arranged in a cross configuration and are distributed such that there are always two sources in the overlap region of two cameras and that the central source can be seen by all cameras. Occasional sparks can be used to map these positions in the recorded frames (\figrefbra{hptpcPaper:sec:hptpcPerfomrance:fig:sparkEvent}).

\subsection{Spark Detection}
\label{hptpcPaper:subsec:readoutAndDAQ:CCDSparkDetect}

\begin{figure}
\centering
\subfloat[]{\label{hptpcPaper:sec:hptpcPerfomrance:fig:sparkEvent}
\includegraphics[height=0.27\textheight, trim = 0 0 0 0, clip=true]{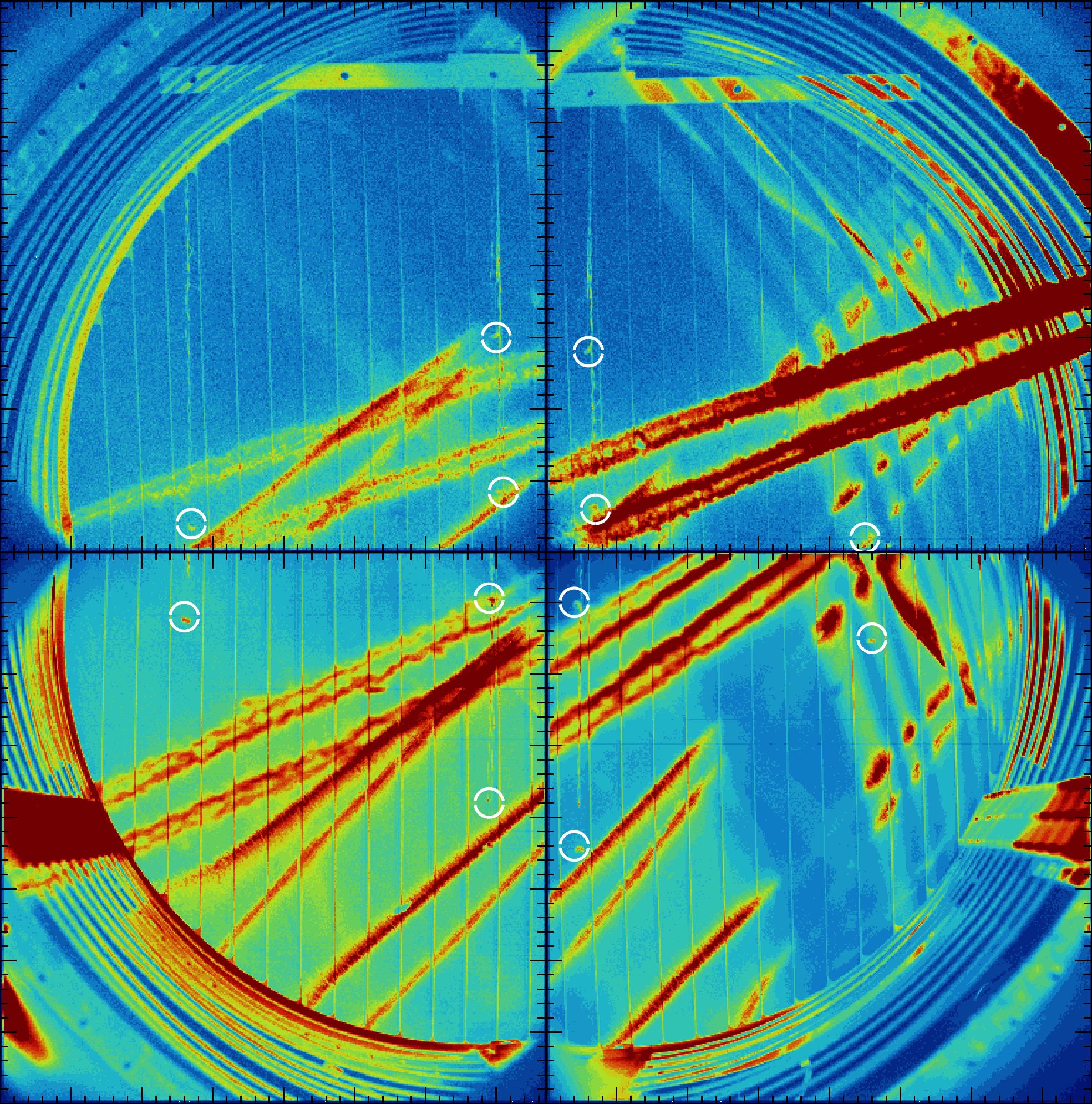}}
\subfloat[]{\label{hptpcPaper:sec:hptpcPerfomrance:fig:calibSources:light}
\includegraphics[height=0.27\textheight, trim = 26 26 26 26, clip=true]{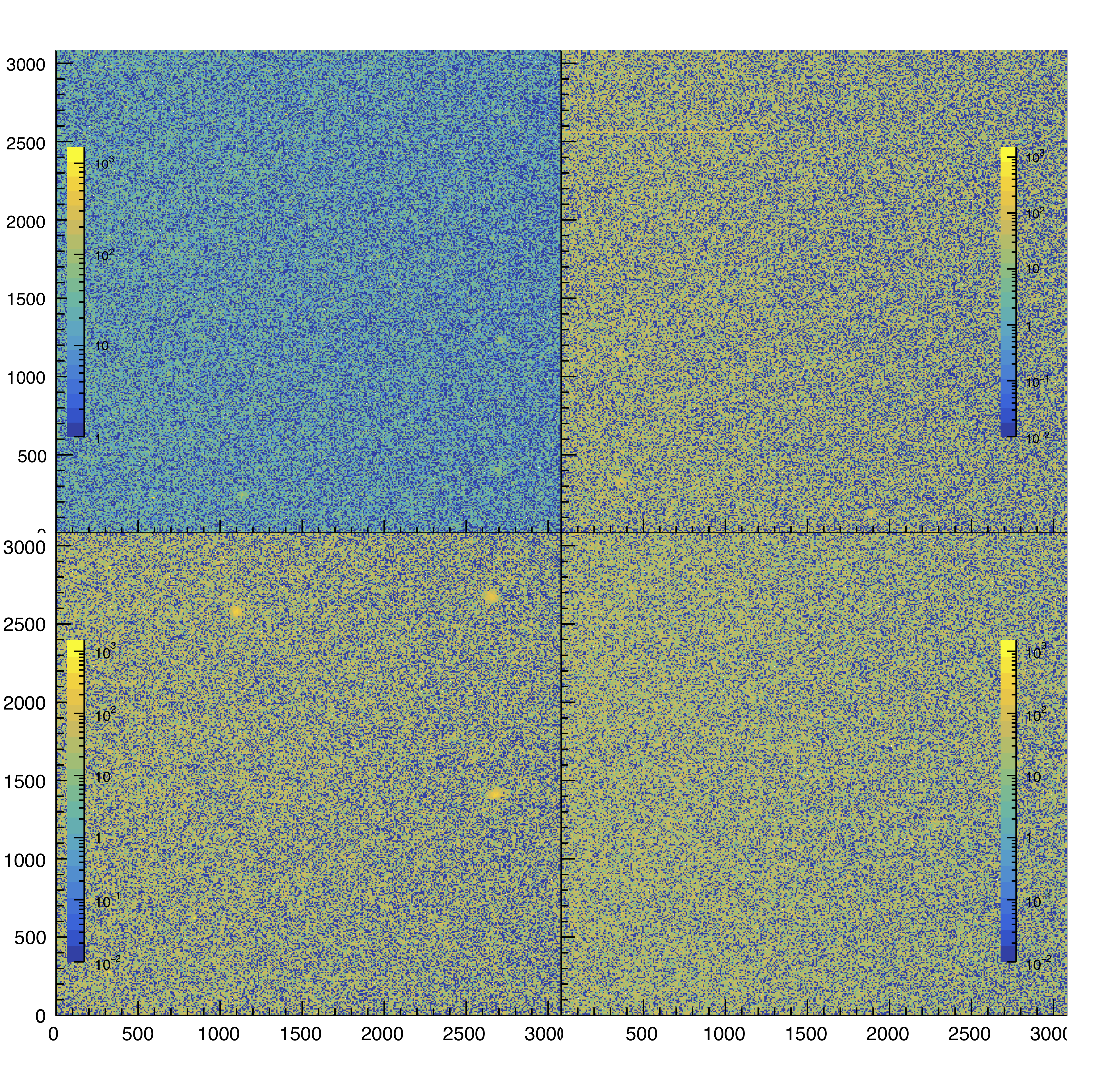}}
\caption{\label{hptpcPaper:sec:hptpcPerfomrance:fig:sparkEventAndCalibSources}CCD images showing the readout plane of the HPTPC; the vertical (horizontal) image axis points along the $y$ ($x$) direction. The colour encodes the light intensity in arbitrary units. \protect\subref{hptpcPaper:sec:hptpcPerfomrance:fig:sparkEvent} Simultaneously recorded frames during a spark event. The locations of the $^{241}\textrm{Am}$ sources (marked by circles) inside the TPC are visible during the spark event as well as the field cage rings and the anode support, \textit{cf.} \figref{hptpcPaper:sec:hptpc:fig:fc:FCinside}. \protect\subref{hptpcPaper:sec:hptpcPerfomrance:fig:calibSources:light} Light yield from the calibration sources for \SI{200}{\second} exposure time in pure Argon at \SI{3}{bar} absolute pressure. The intensity of the image in the top left frame differs from the other three frames, because the corresponding camera has a different conversion gain.}
%$V_1=\SI{1500}{\volt}$, $V_2=\SI{2900}{\volt}$, $V_3=\SI{3750}{\volt}$, $V_{\textrm{C}}=\SI{-8500}{\volt}$
\end{figure}
A major source of noise comes from sparking in the chamber. These sparks mostly originated along the boltholes of the amplification region and from the cathode feedthrough. The frequency of these events increased with the anode and cathode voltages and ultimately limited the maximum voltages we were able to reach. Our gas mixture choices were thus driven by finding mixtures which allowed to operate the detector without many discharges at large enough voltages to see charge and light signals. The gas mixtures listed in \secref{hptpcPaper:sec:hptpcPerfomrance:opticalRO} allowed us to operate the detector in a stable manner. Other gas mixtures as \arcf{} were also tried out during the initial testing, but resulted in too many discharges to perform a light gain measurement and are hence not reported on.\\
Sparks cause a large fraction of pixels in an image to become significantly brighter, an example is shown in \figref{hptpcPaper:sec:hptpcPerfomrance:fig:sparkEvent}. The camera pixels measure charge in \si{ADU}. Images with sparking are rejected from the analysis as follows. First, events in which one of the CCD images has a pixel-value RMS above $\sim\!\!\SI{100}{ADU}$ to \SI{300}{ADU} are rejected as sparks. Of the remaining images, those with events in which one or more of the CCD images have 100,000 pixels above \SI{100000}{ADU} are also removed from the analysis. The exact thresholds depend on the actual CCD camera and detector settings, \textit{e.g.} the RMS thresholds vary from \SI{133}{ADU} to \SI{300}{ADU} between the four cameras. The exact values for each threshold have been identified by comparing the properties of spark images selected by eye to images without sparks.

\subsection{CCD Camera Calibration}
\label{hptpcPaper:subsec:readoutAndDAQ:CCDnoise}

\begin{figure}
\centering
\subfloat[]{\label{hptpcPaper:subsec:readoutAndDAQ:CCDnoise:fig:ccdnoise:raw}
\includegraphics[width=0.49\columnwidth, trim = 0 0 0 30, clip=true]{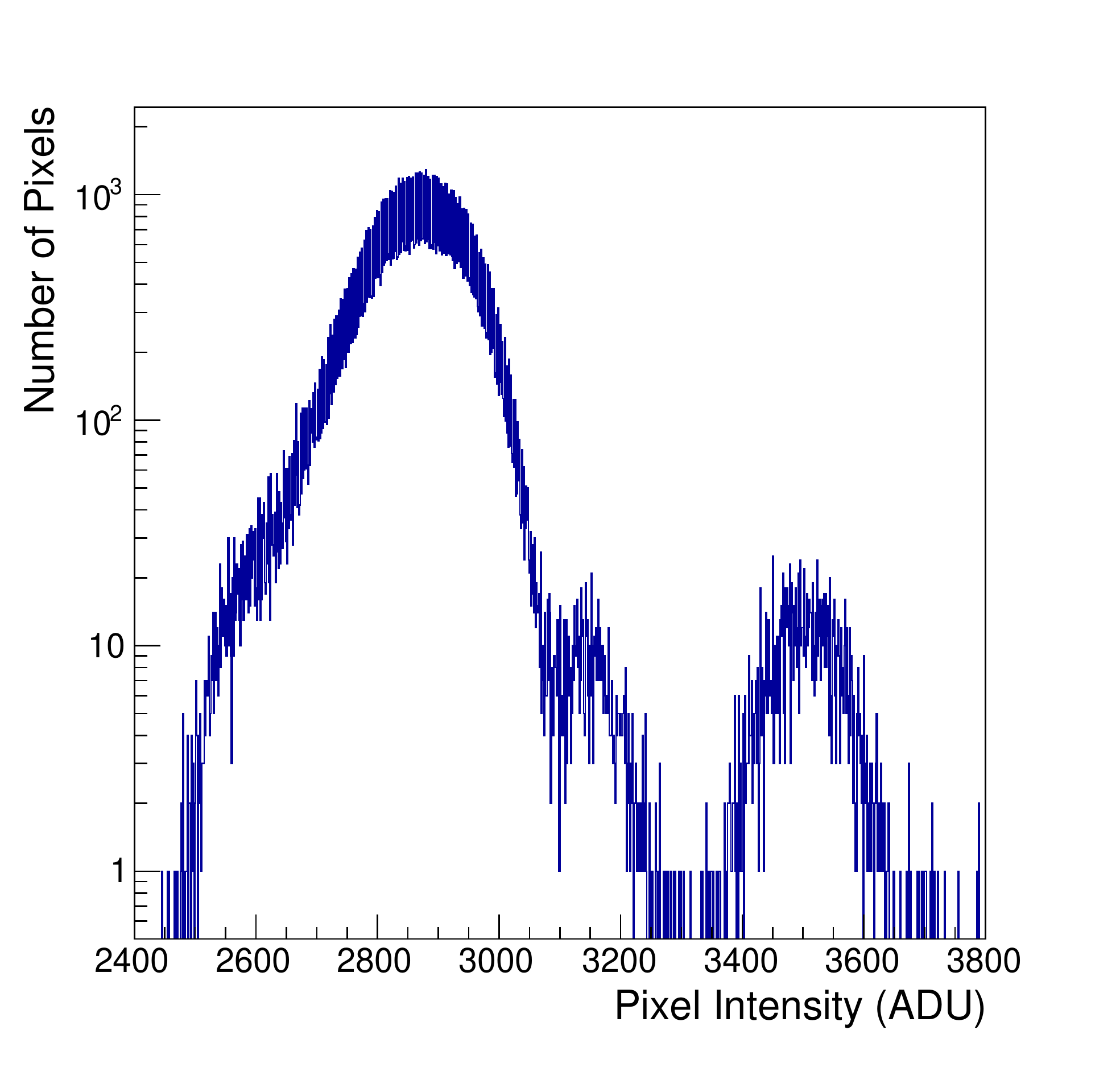}}
\subfloat[]{\label{hptpcPaper:subsec:readoutAndDAQ:CCDnoise:fig:ccdnoise:biasSub}
\includegraphics[width=0.49\columnwidth, trim = 0 0 0 30, clip=true]{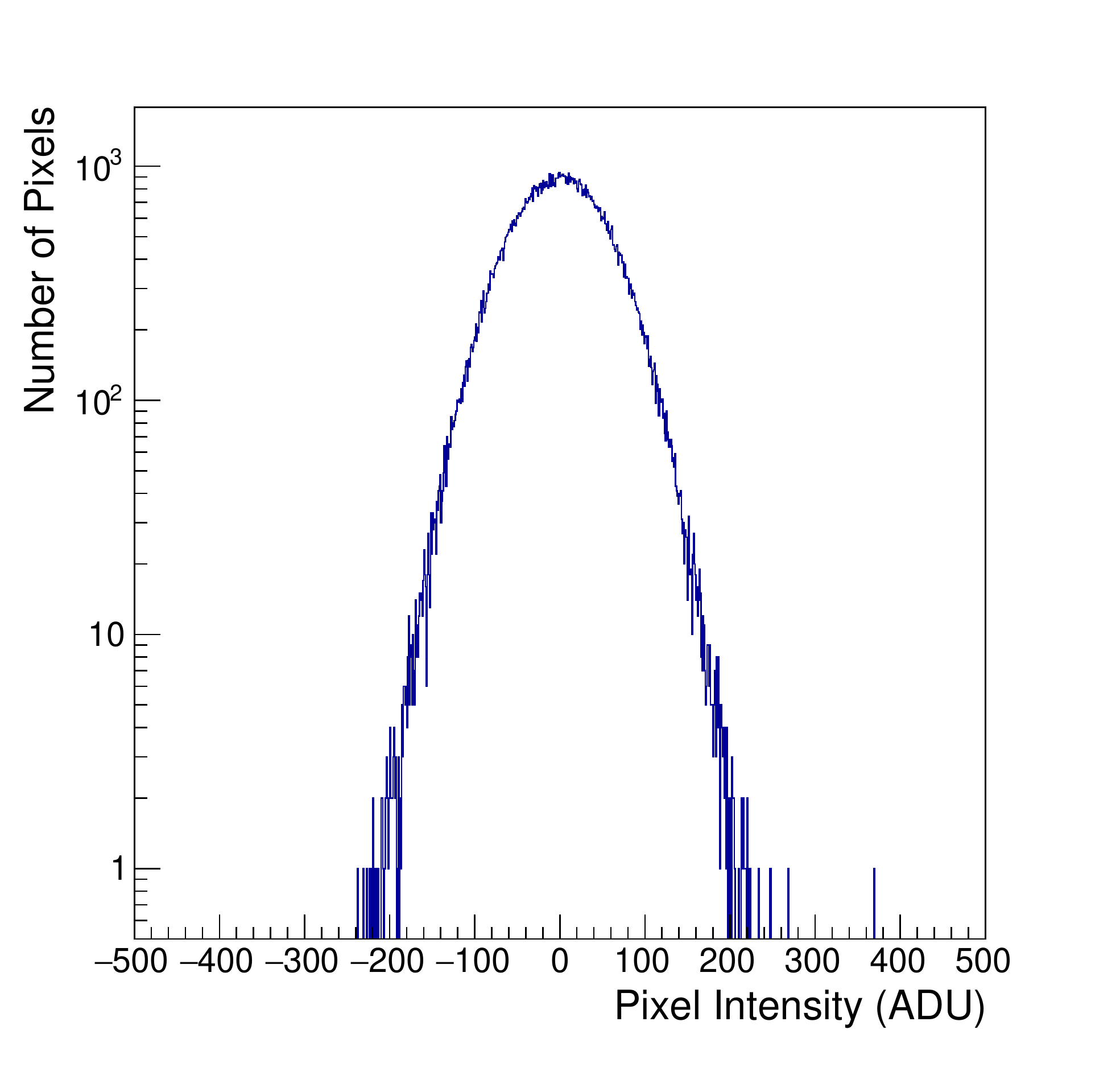}}
\caption{\label{hptpcPaper:subsec:readoutAndDAQ:CCDnoise:fig:ccdnoise}Analogue-to-Digital Unit (ADU) distribution of all pixels of an exposure frame before \protect\subref{hptpcPaper:subsec:readoutAndDAQ:CCDnoise:fig:ccdnoise:raw} and after \protect\subref{hptpcPaper:subsec:readoutAndDAQ:CCDnoise:fig:ccdnoise:biasSub} bias subtraction.}
\end{figure}

The CCD camera calibration removes variations in pixel gain, transient phenomena, and time dependent noise sources. The first step of the CCD calibration is the subtraction of \textit{bias frames}, which deals with persistent features and noise sources, and accounts for variations in pixel gain. At the start of each run we take 5--10 bias frames with the shutter closed. These are averaged and then subtracted from all \textit{exposure frames} with shutter open in the same run, where a typical run consists of 20--100 images (per camera) with an exposure time of \SI{2}{\second} per frame.\\
A source of transient noise is \textit{hot pixels}, created \textit{e.g.} by cosmic muons passing through the camera chip and saturating pixels. These hot pixels are usually confined to individual frames but can remain saturated over several exposure lengths. If they occur in the bias frames they must be corrected before the bias frame subtraction from the exposure frames. Each pixel value of each bias frame is compared to the values of the same pixel in the other bias frames, and if its \si{ADU} reading has changed by more than five standard deviations of its mean \si{ADU} value, the value of the pixel is set to that of the previous bias frame.\\
The next step of the CCD calibration is the temperature dependent image mean correction. The temperature of the CCDs is seen to increase with the number of events taken in a run. This results in a natural upwards drift in the pixels' intensities with time which contributes to the noise. This effect is corrected for by calculating each CCD frame's average pixel value and then subtracting that value from every pixel within that frame. This process is applied to every frame in all runs.\\
The impact of these calibration steps is to reduce the pixel intensity variance. The distribution of pixel values before and after bias subtraction is shown in \figref{hptpcPaper:subsec:readoutAndDAQ:CCDnoise:fig:ccdnoise}.

\subsubsection{Calibration without closed-shutter bias frames}
\label{hptpcPaper:subsec:hptpcPerfomrance:lightGain:subsubsect:lightgaincalibration}

The measurements in this paper are grouped into two data taking periods: One, where we explore different gas mixtures to find the most promising gas for an in depth measurement campaign (\secrefbra{hptpcPaper:sec:hptpcPerfomrance:opticalRO}), and the second period where only the gas identified in the first period is studied (\secrefbra{hptpcPaper:subsec:hptpcPerfomrance:lightGain}). During the beginning of the second period it was discovered that camera 2 (which was set up to take the light gain data) had a stuck open shutter. Due to time constraints we continued with data taking despite this and have adjusted our calibration accordingly as detailed in this section.\\
To address mechanical shutter failure, a procedure was developed to acquire bias frames for calibration with the shutter open. To avoid stray light from the sources or sparks, 1000 \SI{2}{\second} shutter-open frames were acquired daily with the TPC voltages switched off. The anode meshes need to be slowly brought up to the desired voltages in order to reduce the probability of sparking and the subsequent need to reduce the voltages for some time. Reaching the target voltages in a gas mixture with low or no quencher content can thus take on the order of hours, when starting from zero. For this reason we decided to take these shutter-open bias frames not before every run. These frames are then used to produce a single, low noise \textit{super bias} frame to be subtracted from each event taken that same day.

\paragraph{\textbf{Super bias frame creation}}

The method used to create each super bias is to first remove any anomalous pixels by the method described in \secref{hptpcPaper:subsec:readoutAndDAQ:CCDnoise}. Next, a 1D distribution for each pixel in the super bias is created and filled with the $N_{\text{bias}}=1000$ \si{ADU} values measured by that pixel in all 1000 bias frames. The mean and standard deviation ($\sigma_{\text{pixel}}$) of that distribution is calculated and any \si{ADU} value above $3\;\sigma_{\text{pixel}}$ of the mean are removed. A Gaussian is fitted to the remaining 1D distribution of each pixel. The centre of the Gaussian gives the ADU value of that pixel in the super bias. 
As mentioned in \secref{hptpcPaper:subsec:readoutAndDAQ:CCDnoise}, bias subtraction using a bias frame, taken close in time to the event frame can help to reduce temperature (and therefore time) dependent noise. Due to the significant time difference between bias and event frames additional corrections of temperature/time dependent effects need to be implemented before the super bias frame can be used as a bias frame for exposure frames.\\
\begin{figure}
\centering
\subfloat[]{\label{hptpcPaper:sec:opticalgain:fig:Row_Corrcetion_Not_RC}
\includegraphics[width=0.49\columnwidth, trim = 0 0 0 0, clip=true]{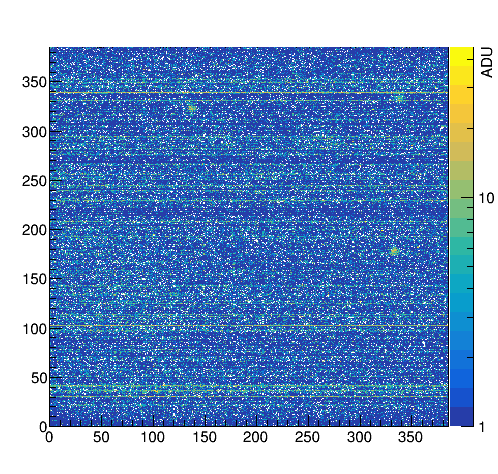}}
\subfloat[]{\label{hptpcPaper:sec:opticalgain:fig:Row_Corrcetion_RC}
\includegraphics[width=0.49\columnwidth, trim = 0 0 0 0, clip=true]{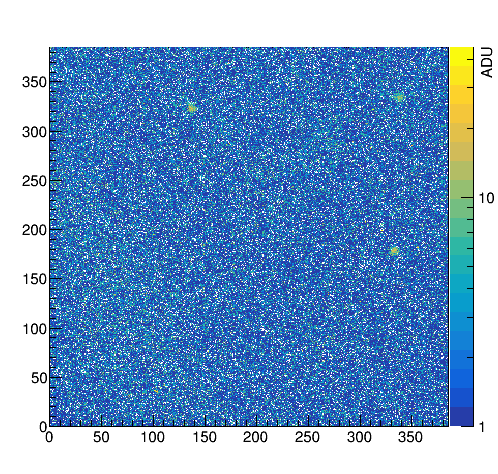}}
\caption{\label{hptpcPaper:sec:opticalgain:fig:Row_Corrcetion}Example of the average of 100 bias subtracted events with event and bias frame taken days apart \protect\subref{hptpcPaper:sec:opticalgain:fig:Row_Corrcetion_Not_RC} before row correction (demonstrating row CCD artefacts) \protect\subref{hptpcPaper:sec:opticalgain:fig:Row_Corrcetion_RC} after row correction (demonstration correction of row CCD artefacts). The colour in both plots encodes the \si{ADU} value at the position of a pixel, while the horizontal and vertical axis shows the $y$ and $x$ coordinate, respectively.}
\end{figure}
\Figref{hptpcPaper:sec:opticalgain:fig:Row_Corrcetion_Not_RC} shows an example of row pedestal artefacts. The scale of the effect has been artificially increased for demonstration purposes by using exposure and bias frames taken days apart. This effect occurs when the pedestal of each pixel within a row changes by some amount between taking the bias and exposure frame. These effects are not observed when the bias frames are recorded directly before the exposure frames as part of the same run, since the pedestal value shift occurs only between runs. In order to use the super bias frames we apply a row correction to every row in the super bias subtracted image. For this correction the average \si{ADU} value of a row is calculated whilst omitting any anomalous pixels or any pixels located within the region of interested for the analysis, \textit{i.e.} the source locations. This average is then subtracted from each pixel in that row. \Figref{hptpcPaper:sec:opticalgain:fig:Row_Corrcetion_Not_RC} shows the same image as \figref{hptpcPaper:sec:opticalgain:fig:Row_Corrcetion_RC} after row correction has been applied.\\
\begin{figure}
\centering
\subfloat[]{\label{hptpcPaper:sec:opticalgain:fig:Mean_vs_Event_Not_RC}
\includegraphics[width=0.49\columnwidth, trim = 0 0 0 0, clip=true]{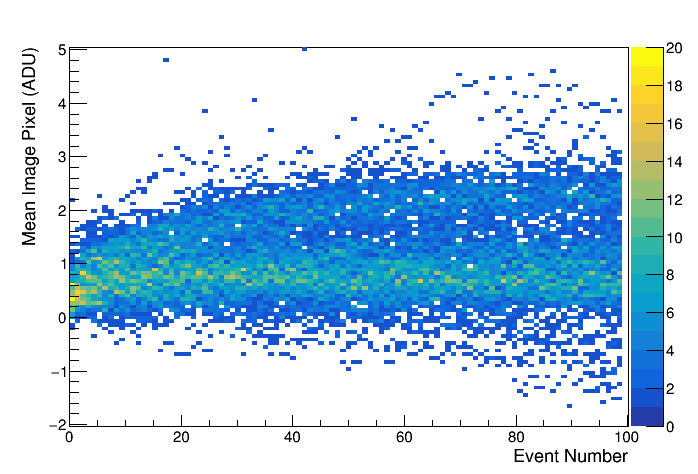}}
\subfloat[]{\label{hptpcPaper:sec:opticalgain:fig:Mean_vs_Event_RC}
\includegraphics[width=0.49\columnwidth, trim = 0 0 0 0, clip=true]{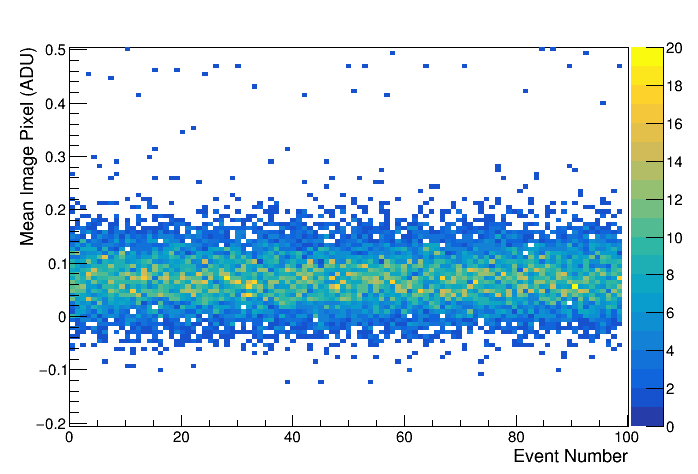}}
\caption{\label{hptpcPaper:sec:opticalgain:fig:Mean_vs_Event}Mean \si{ADU} value of exposure frames versus event number for 150 runs (of 100 events, \textit{i.e.} frames) taken over a number of days \protect\subref{hptpcPaper:sec:opticalgain:fig:Mean_vs_Event_Not_RC} before row correction (demonstrating pixel pedestal drift) \protect\subref{hptpcPaper:sec:opticalgain:fig:Mean_vs_Event_RC} after row correction. The latter demonstrates the correction of the pedestal drift by the row correction procedure.}
\end{figure}
Applying row correction to a super bias subtracted image also corrects for any time dependent drift of pixel intensities (\textit{e.g.} because of temperature). \Figref{hptpcPaper:sec:opticalgain:fig:Mean_vs_Event} show an example of the pedestal drift in 100 subsequent exposures (events) for 150 different runs before and after row correction. The bi-modal nature of \figref{hptpcPaper:sec:opticalgain:fig:Mean_vs_Event_Not_RC} is likely due to temperature differences on different days. It should be noted that the row correction can fail for regions on the CCD where a differential pedestal drift is present. The lower left corner of the CCDs experiences such a non differential shift and care has been taken to ensure that the source positions do not overlap with affected rows.\\
Overall, super bias frame subtraction significantly reduces the pixel intensity variance in an event which is normally introduced by the classical bias frame subtraction, because the mean pixel value error in a super bias frame is reduced by $1/\sqrt{N_{\text{frames}}}$.
When integrating a region of interest of $9\times9$ pixels, \textit{i.e.} $N_{\text{pixel}}=81$, this \SI{100}{\%} correlated uncertainty for a super bias frame constructed from 1000 (bias) frames ($N_{\text{bias}}=1000$) can be calculated using the following equation:
\begin{equation}
\sigma_{\text{bias}} = \sqrt{\frac{\sigma_{\text{pixel}}^{2} \cdot N_{\text{pixel}}}{N_{\text{bias}}}} \quad.
\label{eq:sigma_bias}
\end{equation}
For the standard deviation of a single pixel ($\sigma_{\text{pixel}}$) a typical value of \SI{40}{ADU} can be used to estimate $\sigma_{\text{bias}}$. The resulting $\sigma_{\text{bias}}=\SI{11.4}{ADU}$ is significantly smaller as for \textit{e.g.} the case where 5 bias frames are used.

\subsection{Light Yields for Different Gas Mixes}
\label{hptpcPaper:sec:hptpcPerfomrance:opticalRO}

An important question when operating gaseous detectors with optical readout is which gas mixtures will yield the most light from the interactions of interest. For this measurement light from a single $^{241}\textrm{Am}$ source in the overlap region of the top two cameras was used. Since the alpha particles from the decays travel only a few \si{\centi\meter} at the pressures considered (\textit{cf.} the beginning of the section), a small region around the source location was considered for the light yield measurement. One of the cameras (top left, \figrefbra{hptpcPaper:sec:hptpcPerfomrance:fig:calibSources:light}) had a lower conversion gain than the other three so only the top right camera was used for this analysis. The trialled gas mixes were: pure argon (3 and \SI{4}{bar} absolute), argon with carbon dioxide (\SI{4}{bar} absolute, \arcois{99/1} and (99.25/0.75)), argon with nitrogen (\SI{3}{bar} absolute, \arnis{98/2}) and argon with nitrogen and carbon dioxide (\SI{4}{bar} absolute \arconis{98.75/0.75/0.50} and \SI{4.9}{bar} absolute \arconis{96/2/2}).\\
\begin{figure}
\centering
\subfloat[]{\label{hptpcPaper:subsec:hptpcPerfomrance:lightGain:fig:lightGain:constChargeGain}
\includegraphics[width=0.8\columnwidth]{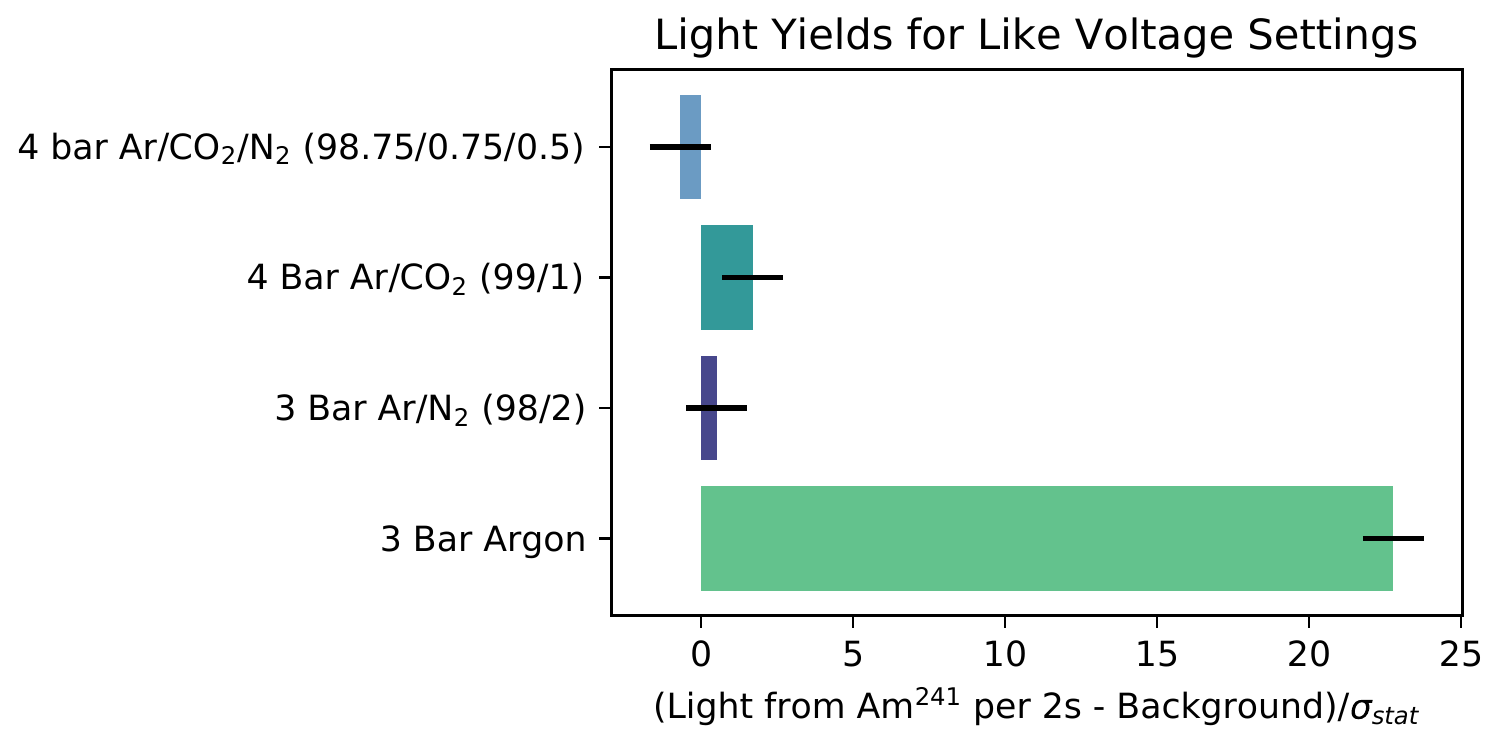}}\\
\subfloat[]{\label{hptpcPaper:subsec:hptpcPerfomrance:lightGain:fig:lightGain:maxLightGain}
\includegraphics[width=0.8\columnwidth]{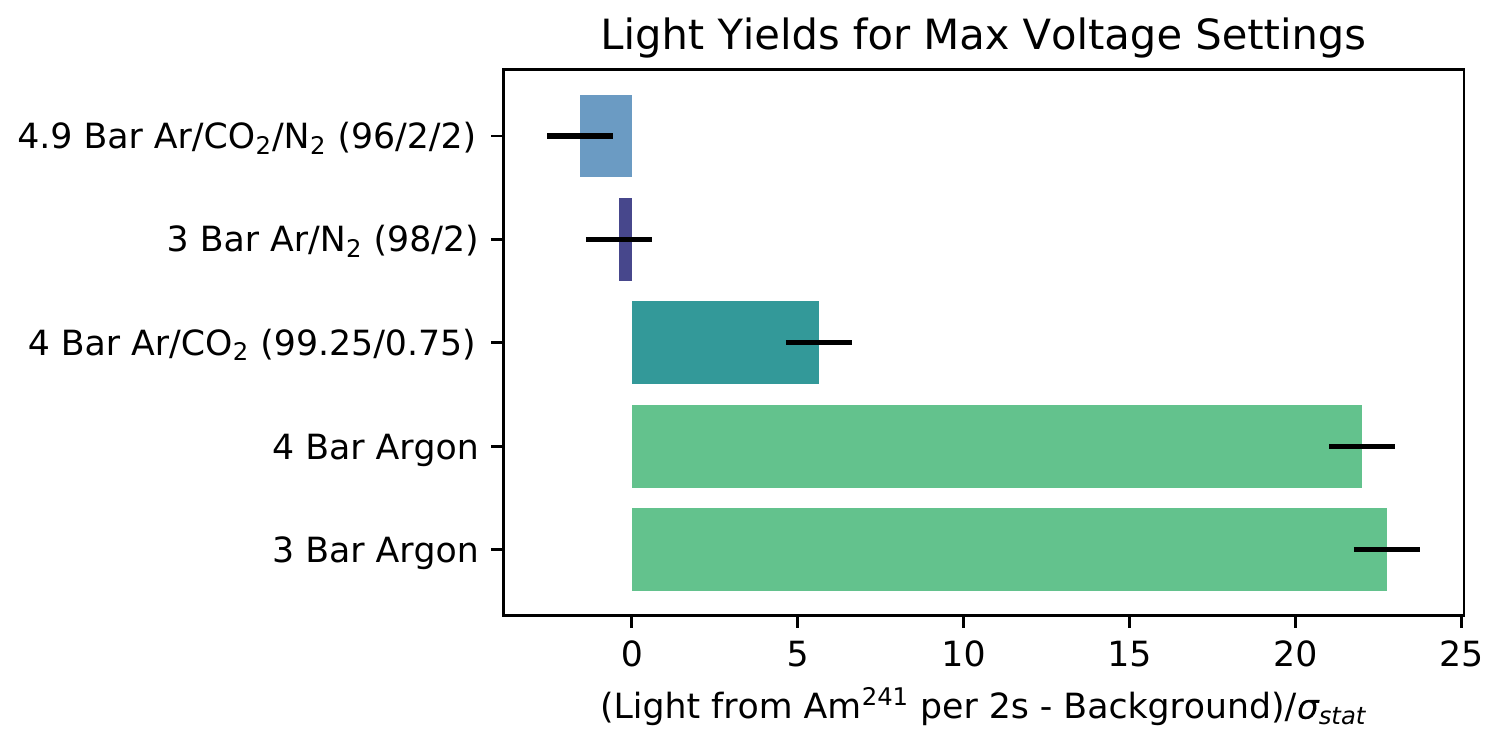}}
\caption{\label{hptpcPaper:subsec:hptpcPerformance:lightGain:fig:lightGain}Light yield measured for an $^{241}$Am source with different gas mixtures \protect\subref{hptpcPaper:subsec:hptpcPerfomrance:lightGain:fig:lightGain:constChargeGain} at near constant anode and cathode voltages and \protect\subref{hptpcPaper:subsec:hptpcPerfomrance:lightGain:fig:lightGain:maxLightGain} the maximal light yield achieved. The voltages used during these measurements are listed in \tabref{hptpcPaper:subsec:hptpcPerformance:lightGain:tab:lightGain}.}
\end{figure}
\begin{table}
\centering
\subfloat[]{\label{hptpcPaper:subsec:hptpcPerfomrance:lightGain:tab:lightGain:constChargeGain}
\begin{tabular}{l|c|c|c|c|c}
mixture or gas & $P$ & $V_{\text{a}1}$ $[\si{\volt}]$ & $V_{\text{a}2}$ $[\si{\volt}]$ & $V_{\text{a}3}$ $[\si{\volt}]$ & $V_{\text{c}}$ $[\si{\volt}]$ \\ \hline
\arconis{98.75/0.75/0.5} & \SI{4}{bar} & 1000 & 2000 & 4000 & -7000 \\ 
\arcois{99/1}       & \SI{4}{bar}   & 1200 & 2400 & 4000 & -7000 \\ 
\arnis{98/2}        & \SI{3}{bar}   & 1200 & 2800 & 4000 & -7000 \\ 
\ar{}               & \SI{3}{bar}   & 1500 & 2100 & 4500 & -5250 \\ 
\end{tabular}
}\\
\subfloat[]{\label{hptpcPaper:subsec:hptpcPerfomrance:lightGain:tab:lightGain:maxLightGain}
\begin{tabular}{l|c|c|c|c|c}
mixture or gas & $P$ & $V_{\text{a}1}$ $[\si{\volt}]$ & $V_{\text{a}2}$ $[\si{\volt}]$ & $V_{\text{a}3}$ $[\si{\volt}]$ & $V_{\text{c}}$ $[\si{\volt}]$ \\ \hline
\arconis{96/2/2}    & \SI{4.9}{bar} & 3000 & 5900 & 7600 & -8500 \\ 
\arnis{98/2}        & \SI{3}{bar}   & 1550 & 3300 & 5000 & -5000 \\ 
\arcois{99.25/0.75} & \SI{4}{bar}   & 1200 & 2500 & 4800 & -7000 \\ 
\ar{}               & \SI{4}{bar}   & 1000 & 1750 & 2800 & -5700 \\ 
\ar{}               & \SI{3}{bar}   & 1500 & 2100 & 4500 & -5250 \\ 
\end{tabular}
}
\caption{\label{hptpcPaper:subsec:hptpcPerformance:lightGain:tab:lightGain}Voltage settings for the result plot shown in \figref{hptpcPaper:subsec:hptpcPerformance:lightGain:fig:lightGain}: \protect\subref{hptpcPaper:subsec:hptpcPerfomrance:lightGain:tab:lightGain:constChargeGain} shows the voltages used for the settings shown in \figref{hptpcPaper:subsec:hptpcPerfomrance:lightGain:fig:lightGain:constChargeGain}, while \protect\subref{hptpcPaper:subsec:hptpcPerfomrance:lightGain:tab:lightGain:maxLightGain} the settings used for the data in \figref{hptpcPaper:subsec:hptpcPerfomrance:lightGain:fig:lightGain:maxLightGain}. The absolute pressure is quoted.}
\end{table}
To determine the light gain the calibration procedures in \secref{hptpcPaper:subsec:readoutAndDAQ:CCDSparkDetect} and \secref{hptpcPaper:subsec:readoutAndDAQ:CCDnoise} are applied to the relevant data runs. A $20\times20$ pixels ($\sim\!\!3.8\times\SI{3.8}{\centi\meter\squared}$) region of interest around the source position is examined in the bias subtracted and calibrated exposure frames. 
All light recorded in the region of interest is integrated. The results of this study can be seen in \figref{hptpcPaper:subsec:hptpcPerformance:lightGain:fig:lightGain}. For each result presented in the two plots of \figref{hptpcPaper:subsec:hptpcPerformance:lightGain:fig:lightGain} one data taking run has been used. All data is normalised to the same integrated exposure time.\\
Two different comparisons were made, one at fixed anode voltages (\figrefbra{hptpcPaper:subsec:hptpcPerfomrance:lightGain:fig:lightGain:constChargeGain}) and one at the maximum anode voltages reached during stable operation (\figrefbra{hptpcPaper:subsec:hptpcPerfomrance:lightGain:fig:lightGain:maxLightGain}). The voltage settings for both data sets are shown in \tabref{hptpcPaper:subsec:hptpcPerformance:lightGain:tab:lightGain}. In both cases the light yield from the $^{241}\textrm{Am}$ source was found to be highest in pure argon. The pure argon results shows also that a high relative light gain can be achieved with lower voltages as compared to the gas mixtures with a quencher. The high light gain of the \SI{4}{bar} pure argon measurement in \figref{hptpcPaper:subsec:hptpcPerfomrance:lightGain:fig:lightGain:maxLightGain} is surprising since intuitively a lower light gain than for the \SI{3}{bar} gas mixture would have been expected. Even more so as the voltages (normalised by pressure) applied during the \SI{4}{bar} measurement are lower than in the \SI{3}{bar} case. A saturation of the light gain at a given voltage setting can explain such findings. Furthermore the fields during the \SI{4}{bar} measurements could allow for the incoming and amplified electrons to create more excitations and thus photons on the cost of ionisations, as compared to the \SI{3}{bar} case.

\subsection{Light yield in argon at various voltage settings}
\label{hptpcPaper:subsec:hptpcPerfomrance:lightGain}

Having identified pure argon as the brightest gas among the mixtures set out in \secref{hptpcPaper:sec:gasChoice}, we now examine how the optical gain in this gas is affected by the operational settings of our amplification stage. Doing so we use pure argon at a pressure of \SI{3}{bar} absolute, following the findings in the previous section. Precisely how each of the multiple anodes contributes to the gain depends -- among other parameters as \textit{e.g.} the voltage settings -- on the anode's relative alignment, which for meshes is difficult to model analytically and numerically. The aim of the optical gain measurements in this section is to understand how the light gain of the TPC is affected by \textbf{a)} the absolute voltage of the three anode meshes $V_{a1}$, $V_{a2}$ and $V_{a3}$ when the potential difference between meshes is kept constant; \textbf{b)} the potential difference between anode meshes 2 and 3 ($\Delta V_{a23}$); and \textbf{c)} the potential difference between anode meshes 2 and 1 ($\Delta V_{a12}$). To do this we chose three voltage schemes which are as follows: 
\begin{itemize}
\item Scheme A - Constant $\Delta V_{a12}$ and constant $\Delta V_{a23}$; 
\item Scheme B - Constant $\Delta V_{a12}$ and varied $\Delta V_{a23}$;
\item Scheme C - Varied $\Delta V_{a12}$ and constant $\Delta V_{a23}$.
\end{itemize}
To conduct a light measurement, a single $^{241}\textrm{Am}$ source is used, positioned so that it can be imaged by camera 2, the bottom left camera. The source has an activity of \SI{10(1)}{\kilo\becquerel} as has been determined by an independent measurement, which was validated using $^{241}\textrm{Am}$ sources with known decay rates. 
An exposure time of 2 seconds per frame was chosen to balance reduction of readout noise with reduction in dead time due to lost frames from sparking.

\subsubsection{Optical Gain Analysis}
\label{hptpcPaper:subsec:hptpcPerfomrance:lightGain:subsubsec:results}

\begin{figure}
\centering
\includegraphics[width=0.6\columnwidth, trim = 0 0 0 0, clip=true]{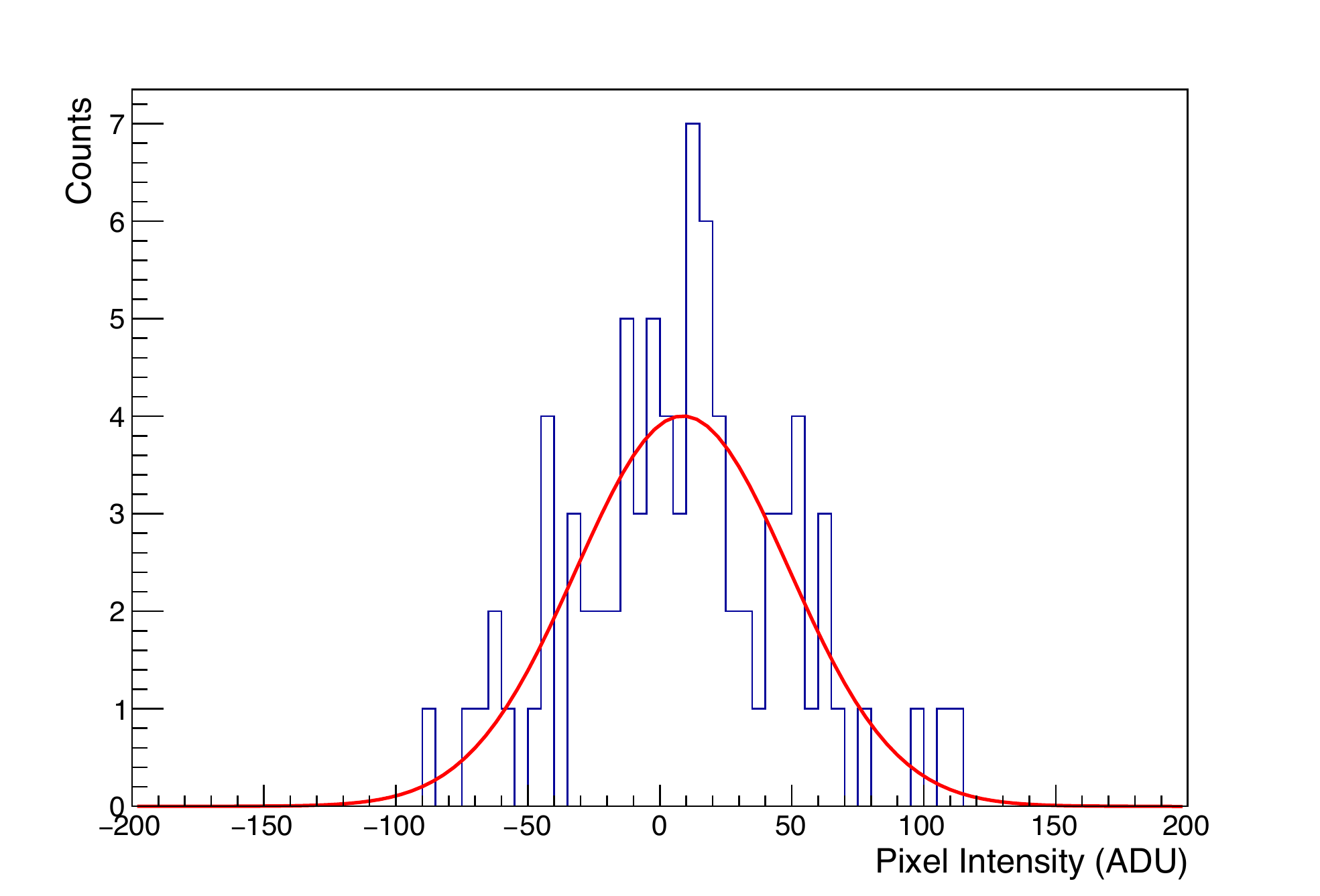}
\caption{\label{hptpcPaper:sec:opticalgain:fig:Box_Pix_ADU_V1}Intensity distribution of pixels within the source box for a single event.}
\end{figure}

Schemes A, B and C (as described in the beginning of \secref{hptpcPaper:subsec:hptpcPerfomrance:lightGain}) consist respectively of 5, 11 and 9 integrated ADU measurements taken at different anode voltage configurations with step-sizes of \SI{200}{\volt} or \SI{400}{\volt}. Each voltage configuration has between 1000 and 1500 events with one frame per camera each.
% (taken over 10 to 15 runs containing 100 events each). 
Four sets of 1000 TPC-off shutter open bias frames were also taken to produce four super bias frames. One taken before Scheme A and then one taken after each of the three schemes. First the calibrations and checks detailed in \secref{hptpcPaper:subsec:readoutAndDAQ:CCDSparkDetect} and \secref{hptpcPaper:subsec:hptpcPerfomrance:lightGain:subsubsect:lightgaincalibration} are applied. Doing so, all exposure frames recorded within one voltage scheme are independently subtracted with the super bias frame taken before and after the respective scheme. The more suitable super bias frame is selected for each scheme based on the Gaussian nature of the pixels' \si{ADU} distribution and on the flatness of the $x$ and $y$ projections of the \si{ADU} distribution of frames within a scheme. Two super bias frames were selected resulting in Schemes B and C sharing the same super bias frame.\\
Then a region of interest around the source is defined (referred to as \textit{source box}). 
The source box's size is optimised to contain as few pixels as possible whilst not rejecting any signal. The analysis found a nine by nine pixels ($16.56\times\SI{16.56}{\milli\meter\squared}$) source box to be optimal. 
After a loose pixel ADU cut, a Gaussian is fitted to the ADU values of the $N_{\text{pixel}}=81$ pixels in the box for a given frame (as shown in \figrefbra{hptpcPaper:sec:opticalgain:fig:Box_Pix_ADU_V1}). The integrated ADU per frame is then calculated by: $I_{\text{frame}} = \hat\mu_{\text{pixel}} \cdot N_{\text{pixel}}$, where $\hat\mu_{\text{pixel}}$ is the mean of the fitted Gaussian.
The integrated ADU measurement for a run ($I_{\text{run}}$) is calculated by fitting a Gaussian to the distribution of $I_{\text{frame}}$ values in that run. $I_{\text{run}}$ is given by the mean of the fit and its uncertainty ($\sigma_{I_{\text{run}}}$) by the standard deviation on that mean. The final step takes the $I_{\text{run}}$ values of the 10 to 15 runs in each configuration and calculates their weighted mean ($\bar{I}_{w}$) and weighted standard error ($\sigma_{w}$). The mean and standard error for each voltage configuration are weighted by $w_{i} = {1/\sigma_{i}^{2}}$ where $\sigma_{i}$ is the standard deviation per run ($\sigma_{I_{\text{run}}}$) of the $i^{th}$ run in the configuration. The weighted mean and weighted standard error are calculated as follows:
\begin{equation}
\bar{I}_{w} = \frac{\sum_{i = 1}^{N_{\text{run}}}  w_{i}I_{i}}{\sum_{i = 1}^{N_{\text{run}}} w_{i}}
\label{eq:intADU:average}
\end{equation}
\begin{equation}
\sigma_{w}^{\text{m}} = \frac{\sigma_{w}}{\sqrt{N_{\text{run}}}} = \sqrt{\frac{\sum_{i=1}^{N_{\text{run}}} (I_{i} -\bar{I}_{w})^{2}}{(N_{\text{run}}-1)\sum_{i=1}^{N_{\text{run}}}  w_{i}}}
\label{eq:intADU:sigma}
\end{equation}
where $I_{i}$ is the integrated ADU value per run ($I_{\text{run}}$) of the $i^{th}$ run in the configuration, $N_{\text{run}}$ is the total number of runs in the configuration and $\sigma_{w}$ is the weighted standard deviation. $\bar{I}_{w}$ and $\sigma_{w}^{\text{m}}$ give the final integrated ADU value of the voltage configuration ($I_{\text{config}}$) and its uncertainty ($\sigma_{I_{config}}$). In theory one could calculate $\sigma_{I_{config}}$ without the intermediate step of calculating $I_{\text{run}}$. However, examining $I_{\text{run}}$ ensures that run to run instabilities are accounted for in the the uncertainty of the final measurement.\\

\subsubsection{Light gain as function of voltage}
\label{hptpcPaper:subsec:hptpcPerfomrance:lightGain:subsubsec:results:result}

\begin{figure}
\centering
\subfloat[]{\label{hptpcPaper:sec:opticalgain:fig:schemeA_source}
\includegraphics[width=0.49\columnwidth, trim = 0 0 0 0, clip=true]{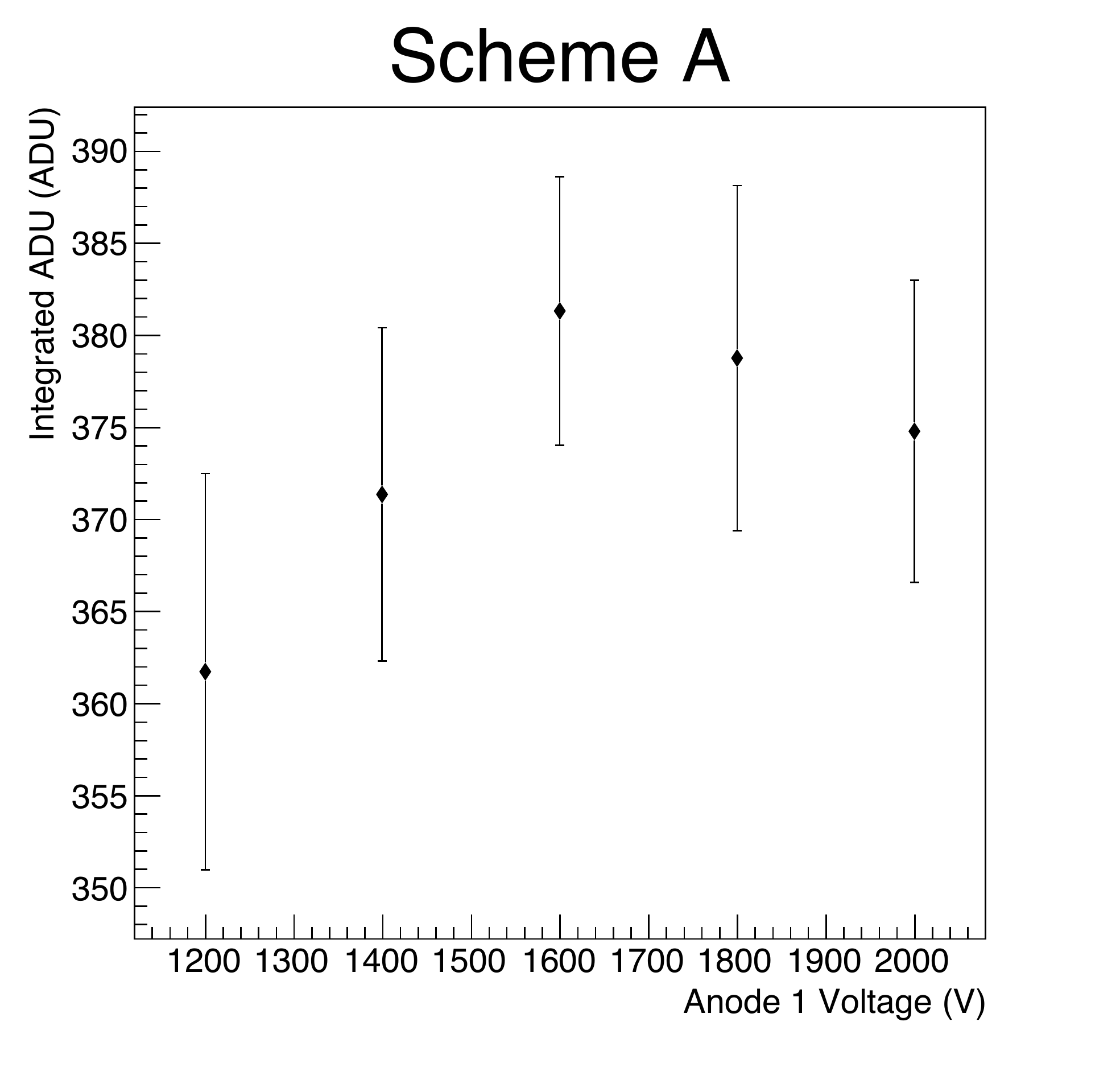}}
\subfloat[]{\label{hptpcPaper:sec:opticalgain:fig:schemeB_source}
\includegraphics[width=0.49\columnwidth, trim = 0 0 0 0, clip=true]{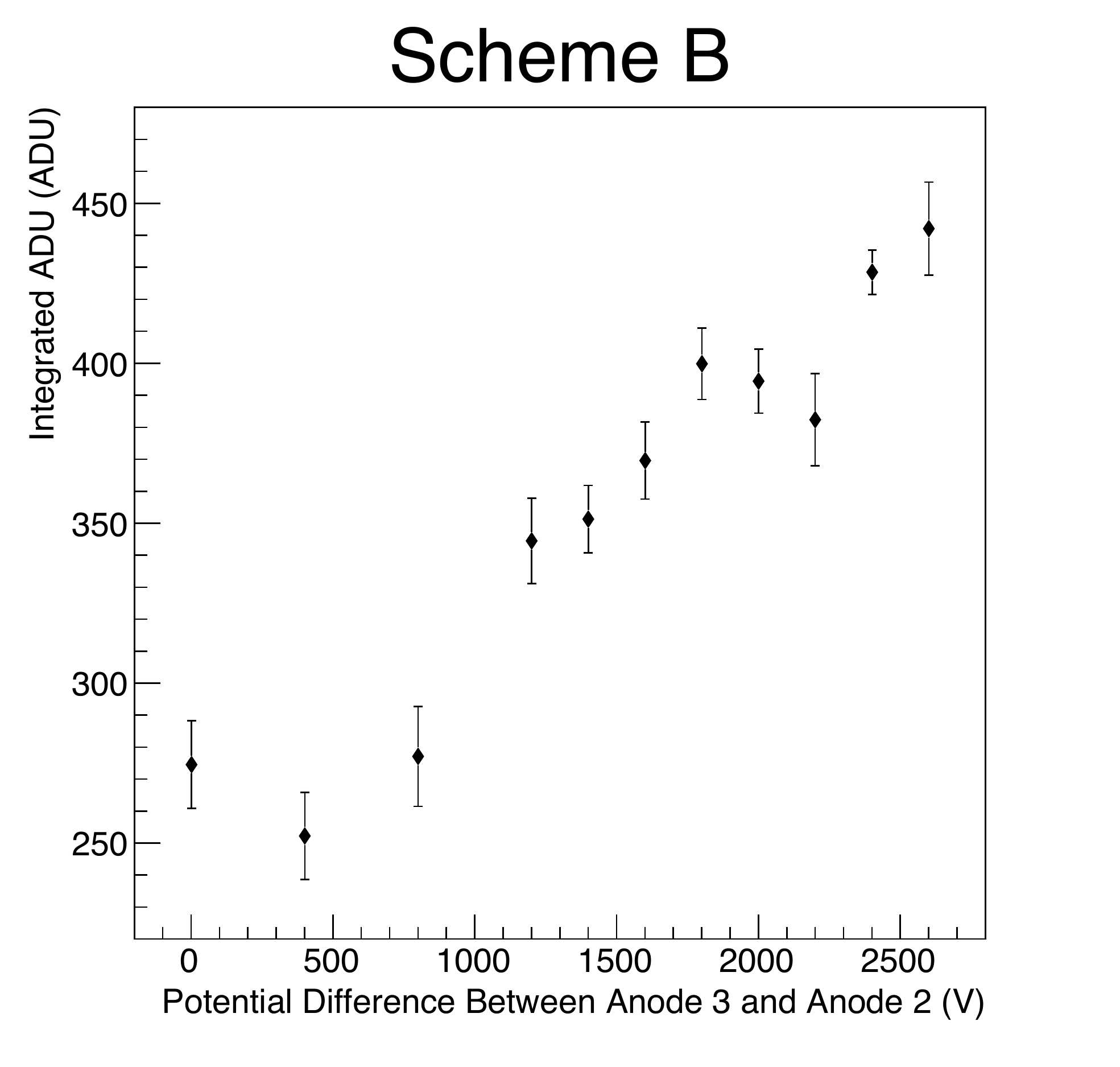}}\\ 
\subfloat[]{\label{hptpcPaper:sec:opticalgain:fig:schemeC_source}
\includegraphics[width=0.49\columnwidth, trim = 0 0 0 0, clip=true]{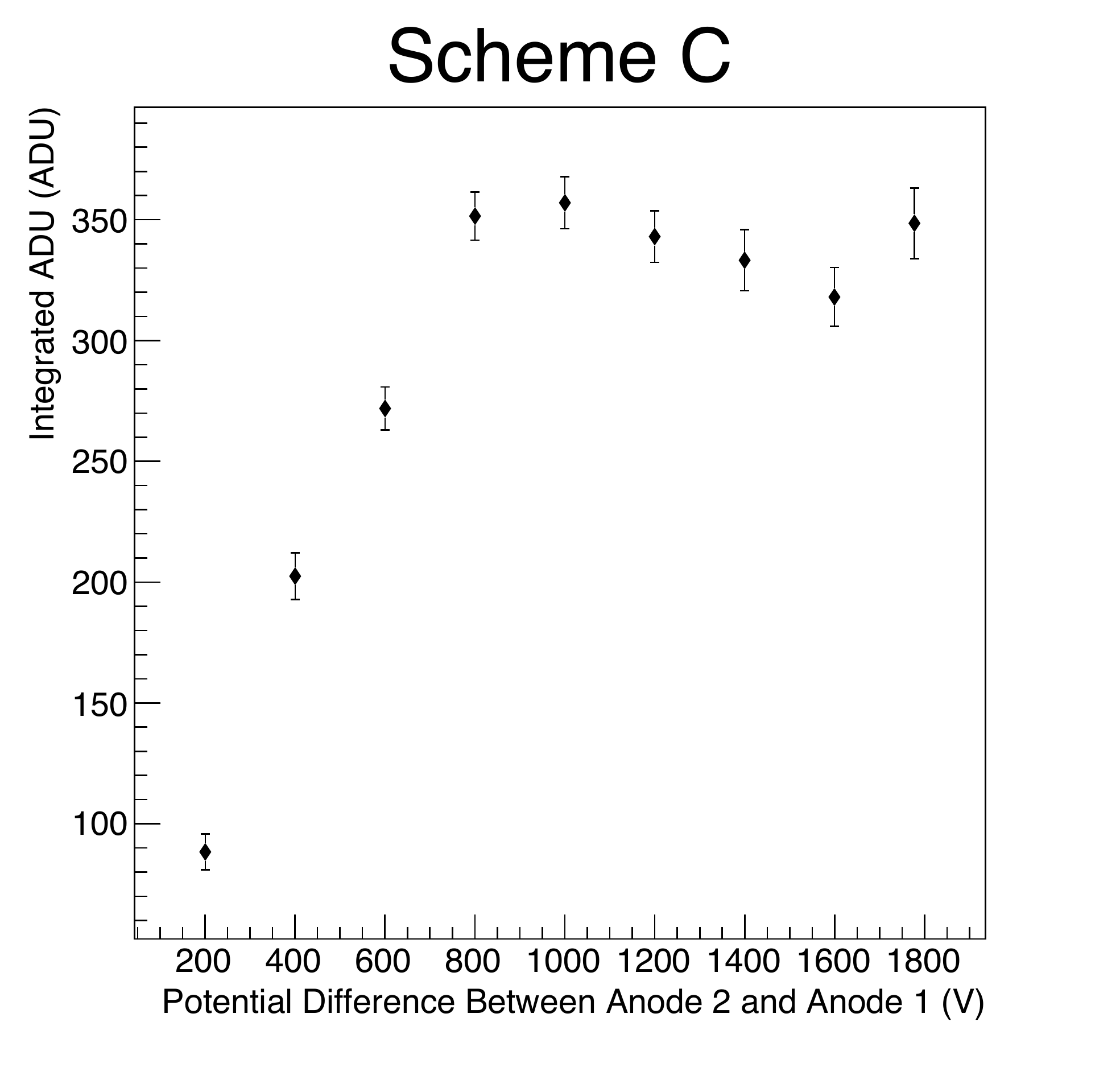}}
\caption{\label{hptpcPaper:sec:opticalgain:fig:schemeABC_source}Light gain measurements of integrated ADU from $^{241}\textrm{Am}$ source  \protect\subref{hptpcPaper:sec:opticalgain:fig:schemeA_source} vs anode 1 voltage where the voltage difference between anode 1 and 2 (anode 2 and 3) is kept constant at $\Delta V_{a12} = \Delta V_{a23} = \SI{1200}{\volt}$, \protect\subref{hptpcPaper:sec:opticalgain:fig:schemeB_source} vs voltage potential difference between anode 2 and anode 3 whilst the voltage difference between anodes 1 and 2 is maintained at \SI{1200}{\volt}, and \protect\subref{hptpcPaper:sec:opticalgain:fig:schemeC_source} vs potential difference between anodes 1 and 2 whilst the potential difference between anode 2 and 3 is maintained at \SI{1200}{\volt}. The figure is shown with a logistic function fit which is chosen on an empirical basis. All measurements have been performed in the same fill of pure argon at \SI{3}{bar} absolute pressure.}
\end{figure}
The final results of the light gain measurements can be seen in \figref{hptpcPaper:sec:opticalgain:fig:schemeABC_source}. In Scheme A, shown in \figref{hptpcPaper:sec:opticalgain:fig:schemeA_source}, the potential differences between the three anode meshes are held constant at\SI{1200}{\volt} while the voltages of all three are varied in \SI{200}{\volt} steps. Scheme A is consistent with the light gain having no dependence on the absolute voltage of the three anodes while $\Delta V_{a12}$ and $\Delta V_{a23}$ are fixed at \SI{1200}{\volt}, suggesting the amplification is driven by the voltage differences between the anode meshes. Across all Scheme B voltage configurations, shown in \figref{hptpcPaper:sec:opticalgain:fig:schemeB_source}, $V_{a1}=$ \SI{1200}{\volt} and $V_{a2}=$ \SI{2400}{\volt} respectively, while $V_{a3}$ and thus $\Delta V_{a23}$, is varied. Scheme B shows a clear linear dependence of light gain on $\Delta V_{a23}$ over the range \SI{0}{\volt} to \SI{2500}{\volt} with a gradient of \SI{0.074(5)}{ADU\per\volt} (\SI{1.50(1)e-4}{ADU\per{(\volt\per m})}). Across all Scheme C (\figrefbra{hptpcPaper:sec:opticalgain:fig:schemeC_source}) voltage configurations $V_{a1}$ and $\Delta V_{a23}$ are equal to \SI{1200}{\volt}, whilst $\Delta V_{a12}$ and thus $V_{a2}$ and $V_{a3}$ are varied. The results of Scheme C suggest the light gain has a positive linear dependence on $\Delta V_{a12}$ up to $\sim\!\!\SI{800}{\volt}$ where the light gain plateaus to a value of \SI{343.0(47)}{ADU}. The gradient of this linear region is \SI{0.45(4)}{ADU\per\volt} (\SI{3.75(4)e-4}{ADU\per{(\volt\per m})}) when fitting a first order polynomial to the first four points of the scheme.\\
We speculate on the origin of the plateau after $\sim\!\!\SI{700}{\volt}$ observed in Scheme C.
One hypothesis is that the plateau occurs when the electric field between anodes 1 and 2 ($E_{a12}$) equals that between anodes 2 and 3 ($E_{a23}$). When $E_{a12}>E_{a23}$ fewer electrons will be able to move from the gap between anode 1 and anode 2 into the gap between anode 2 and 3 and thus there are fewer electrons available for amplification and/or excitation.
The analysis of the circuit response and the inferred capacitances (\secrefbra{hptpcPaper:sec:dataAnalysis:subsec:chargeCalib}) suggest that the distance between anode 1 and 2 is \SI{1.20(5)}{\milli\meter} and the distance between anode 2 and anode 3 is \SI{2.0(2)}{\milli\meter}.
Using these distances we obtain $E_{a12}=\SI{5.83(87)}{\kilo\volt\per\centi\meter}$ and $E_{a23}=\SI{6.0(6)}{\kilo\volt\per\centi\meter}$. As both value agree with each other, we find $E_{a12}=E_{a23}$ where the plateau occurs.
The fact that the rise in light gain stops when $E_{a12}=E_{a23}$, could thus be related to a change in electron transparency of anode 2.
Observing a plateau and not a simple drop in the light gain's gradient with increasing voltage is however surprising, because for a plateau to arise the hypothesised electron loss needs to be exactly compensated by an increased light yield from the electrons in the anode 1 and 2 gap. During the Scheme B measurements $E_{a12}$ was held at a value of \SI{10.00(416)}{\kilo\volt\per\centi\meter}, using the distances discussed before. $E_{a23}$ was scanned from 0 to \SI{13.0(13)}{\kilo\volt\per\centi\meter}. $E_{a23}>E_{a12}$ is fulfilled from a $\Delta V_{a23}$ of \SI{2000(200)}{\volt} onwards and a plateau should be visible as in the case of Scheme C.
The data in \figref{hptpcPaper:sec:opticalgain:fig:schemeB_source} is not sufficient to conclude that the trend reaches a plateau at said value nor the opposite as the plateau's expected position is too close to the end of the $\Delta V_{a23}$ voltage scan. With the maximal $E_{a12}$ in its error-bars, a $\Delta V_{a23}$ of \SI{2800}{\volt} would be required to reach the cross over between the rising and the plateau region.\\
The conclusion drawn form this study is that the light gain in the amplification region depends most strongly on the potential differences between the meshes, rather than the absolute voltage on the mesh wires.

\subsubsection{Number of Photons in Amplification Region Per Primary Electron}
\label{hptpcPaper:subsec:hptpcPerfomrance:lightGain:subsubsec:results:electroncount}

In order to calculate how many photons are produced in the amplification region per primary electron in the drift volume it is necessary to make use of additional measurements and some assumptions. In this work we do not attempt to calculate the relationship between primary electrons in the drift volume and the number of electrons in the amplification region directly (by considering diffusion, mesh transit and charge gain) as we cannot externally constrain all the variables. Instead we calculate the number of photons per second in the amplification region ($N_{\gamma}$) from the $^{241}\text{Am}$ source using the observed ADU in the CCD as follows:
\[
N_{\gamma} = \frac{\textrm{ADU}_{\textrm{obs}}}{(\textrm{Conversion Gain})\times {QE^{\ast}}(\varepsilon)\times\Omega\times\left[T_{\textrm{window}} \times T_{\textrm{lens}}\times T_{\textrm{cathode}} \times T^{2}_{\textrm{anode}}\right]} \quad,
\]
where $\textrm{ADU}_{\textrm{obs}}$ is the observed ADU from the CCD per second in the region around the source. In our measurements the maximum value reached was \SI{225(10)}{ADU\per\second}, \textit{cf}. \figref{hptpcPaper:sec:opticalgain:fig:schemeB_source} divided by the exposure time of \SI{2}{\second}. The conversion gain provided by the manufacturer is $1.5$ ADU per electron. The quantum efficiency ${QE}^{\ast}$ is a function of the incident photon energy, for light in the near infrared the manufacturer specifies $\SI{60(10)}{\%}$ of photons converted into electrons. In principle we could be seeing light in the visible, infrared and ultraviolet from the argon scintillation. However, since both the quantum efficiency of the CCD and the transmission probability through the quartz windows has a rapid drop off below \SI{350}{\nano\meter} we assume in this calculation that we are not sensitive to the UV light. Some photons will not make it from the amplification region to the CCD. The geometric acceptance of the system, $\Omega$ was calculated to be $(1.1 \pm 0.11) \times 10^{-4}$. $T_{\textrm{lens}} = \SI{80(10)}{\%}$ and $T_{\textrm{window}} = 97^{+3}_{-4} \%$ are the transmission probabilities through the lens and quartz window respectively. In addition, all of the photons imaged from the amplification must pass through at least one cathode and one anode mesh, with the majority of them passing through two anode meshes. The transmission probabilities through the cathode and anode meshes are $T_{\text{cathode}} = 0.97$ and $T_{\textrm{anode}} = 0.89$ respectively. This results in $N_{\gamma} = (3.8 \pm 0.7) \times 10^{6}$ photons per second in the amplification region.\\
We then make a calculation of the expected primary electrons in the drift volume per second, $N_{e}$, based on the measured activity of our $^{241}\text{Am}$ source as follows:
\[
N_{e} = N_{\alpha} \times \frac{\langle \varepsilon_{\alpha}\rangle}{W}
\]
where $N_{\alpha} = 10 \pm 1 \textrm{kBq}$ is the activity of our alpha source and $\langle \varepsilon_{\alpha}\rangle = \SI{4.56}{\mega\electronvolt}$ is the expected energy deposited by the alpha particles after exiting the source and the energy required for ionisation in argon is $W = \SI{26.4(3)}{\electronvolt\per electron}$. This results is $N_{e} = (1.7 \pm 0.2) \times 10^{9}$ electrons per second in the imaged part of the drift volume. Combining these two results we expect there to be a total of $(2.2 \pm 0.5) \times 10^{-3}$ photons in the NIR in the amplification region per primary electron in the drift volume.

%% file: sec_6_chargeAnalysis.tex
\section{Charge readout analysis and performance}
\label{hptpcPaper:sec:chargeAnalysis}

In this section, we discuss the raw data obtained from the charge readout and the analysis which turns this raw data into physical quantities. We explain the calibration of the TPC charge readout with radioactive sources and cosmic radiation, and report the charge gain obtained with different high voltage settings.

\subsection{Anatomy of a waveform}
\label{hptpcPaper:sec:chargeAnalysis:subsec:raw}

\begin{figure}
\centering
\subfloat[]{
  \begin{tikzpicture}
  \node[anchor=south west,inner sep=0] (image) at (-0.3,0) {
  \includegraphics[width=0.49\columnwidth, trim = 0 0 0 30, clip=true]{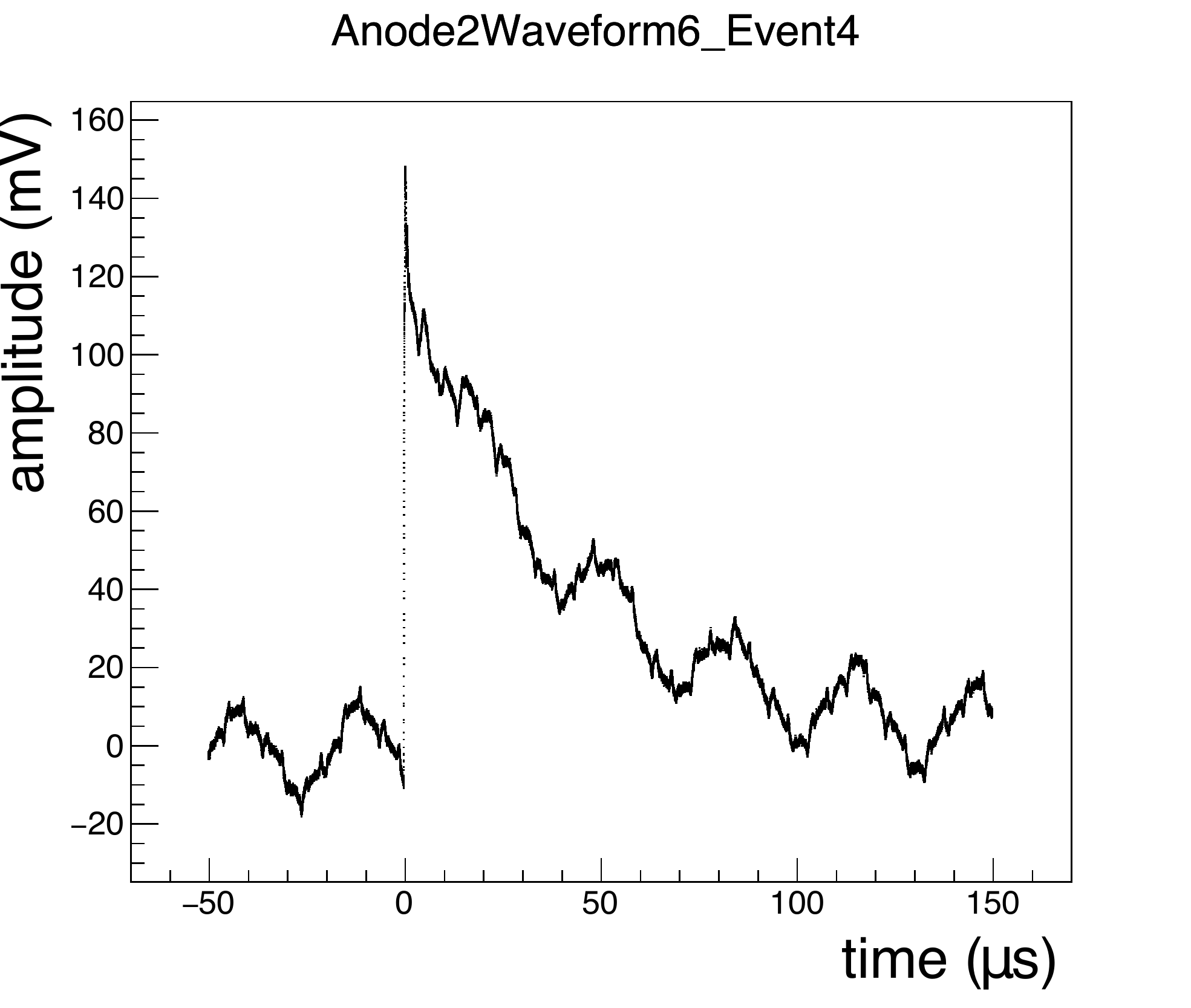}
  };
  \draw [ultra thick, blue, -]    (0.3, 1.650) -- ++(6.5, 0) node [align=center, pos=0.5] {};
  \node[blue] (baseline) at (3.2, 1.650) {\raisebox{-1.5\baselineskip}{Baseline}};
  \draw [ultra thick, red, -]     (0.3, 2.025) -- ++(6.5, 0) node [align=center, pos=0.5] {};
  \node[red] (ninty) at (3.2, 2.025) {\raisebox{\baselineskip}{$\SI{10}{\%}$}};
  \draw [ultra thick, red, -]     (0.3, 5.025) -- ++(6.5, 0) node [align=center, pos=0.5] {\raisebox{-1.5\baselineskip}{$\SI{90}{\%}$}};
  \draw [ultra thick, blue, -]    (0.3, 5.400) -- ++(6.5, 0) node [align=center, pos=0.5] {\raisebox{\baselineskip}{Maximal Value}};
  \draw [ultra thick, magenta, |-|]  (1.0, 3.200) -- ++(1.3, 0) node [align=center, pos=0.5] {};
  \node[magenta] (hello) at (1.4, 3.150) {\raisebox{-2.0\baselineskip}{Pre-trigger}};
  \node[magenta] (hullu) at (1.4, 2.750) {\raisebox{-2.0\baselineskip}{region}};
  \draw [thick, magenta, -]       (2.3, 0.500) -- ++(0, 5.5) node [align=center, pos=1.05] {};
  \draw [thick, magenta, -]       (4.0, 0.500) -- ++(0, 5.5) node [align=center, pos=-0.05] {\rotatebox{90}{$t_{\text{f},10}$}};
  \end{tikzpicture}
  \label{hptpcPaper:sec:dataAnalysis:subsec:CR:fig:genericWaveform:full}
}   
\subfloat[]{
  \begin{tikzpicture}
  \node[anchor=south west,inner sep=0] (image) at (-0.3,0) {
  \includegraphics[width=0.49\columnwidth, trim = 0 0 0 30, clip=true]{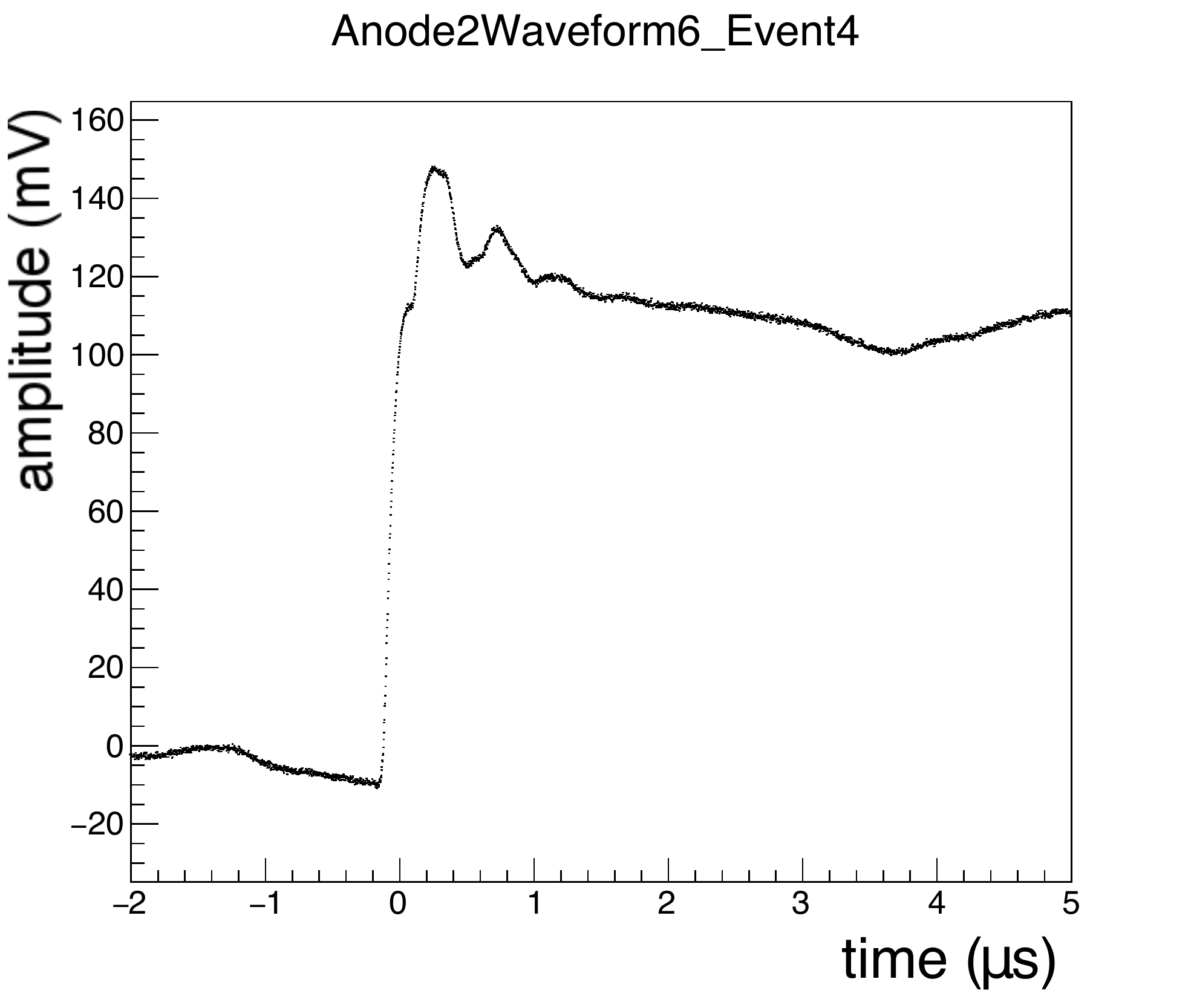}
  };
  \draw [ultra thick, blue, -]  (0.30, 1.650) -- ++(6.5, 0) node [align=center, pos=0.5] {\raisebox{-1.5\baselineskip}{Baseline}};
  \draw [ultra thick, red, -]   (0.30, 2.025) -- ++(6.5, 0) node [align=center, pos=0.5] {\raisebox{\baselineskip}{$\SI{10}{\%}$}};
  \draw [ultra thick, red, -]   (0.30, 5.025) -- ++(6.5, 0) node [align=center, pos=0.5] {};
  \node[red] (salut) at (5.0, 5.025) {\raisebox{-1.5\baselineskip}{$\SI{90}{\%}$}};
  \draw [ultra thick, blue, -]  (0.30, 5.400) -- ++(6.5, 0) node [align=center, pos=0.5] {};
  \node[blue] (hallo) at (5.0, 5.40) {\raisebox{\baselineskip}{Maximal Value}};
  \draw [thick, magenta, -]     (2.20, 0.500) -- ++(0, 5.5) node [align=center, pos=-0.05] {\rotatebox{90}{$t_{\text{r},10}$}};
  \draw [thick, magenta, -]     (2.40, 0.500) -- ++(0, 5.5) node [align=center, pos=1.050] {\rotatebox{90}{$t_{\text{r},90}$}};
  \draw [thick, magenta, -]     (2.60, 0.500) -- ++(0, 5.5) node [align=center, pos=-0.05] {\rotatebox{90}{$t_{\text{f},90}$}};
  \end{tikzpicture}
  \label{hptpcPaper:sec:dataAnalysis:subsec:CR:fig:genericWaveform:zoom}}
\caption{\label{hptpcPaper:sec:dataAnalysis:subsec:CR:fig:genericWaveform}Example for a charge signal, a waveform -- \protect\subref{hptpcPaper:sec:dataAnalysis:subsec:CR:fig:genericWaveform:full} and \protect\subref{hptpcPaper:sec:dataAnalysis:subsec:CR:fig:genericWaveform:zoom} zoom -- with some of its defining features indicated. See the text for more explanations. The first vertical line in Figure \protect\subref{hptpcPaper:sec:dataAnalysis:subsec:CR:fig:genericWaveform:full} shows the approximate position of all the vertical lines in the zoomed plot in \protect\subref{hptpcPaper:sec:dataAnalysis:subsec:CR:fig:genericWaveform:zoom}.}
\end{figure}
Signals from the three anode meshes are decoupled from their respective HV line as described in \secref{hptpcPaper:subsubsec:hptpc:amplifiers}, fed into a pre-amplifier, and digitised. \Figref{hptpcPaper:sec:dataAnalysis:subsec:CR:fig:genericWaveform} shows an example of a digitised waveform, as a trace of voltage versus time. We define the quantities $V_i$ and $t_i$ to be the digitised voltage and time, respectively, at the $i^{\text{th}}$ time sample.\\
A waveform is comprised of three characteristic regions in time, shown on the sample waveform: the period before the digitiser has triggered (\textit{pre-trigger}), the time at which the digitiser triggered and the period after (\textit{post-trigger}). The pre-trigger region -- that is, sample 1 to sample $N_{\text{pre-trig}}$ corresponding to $t=0$ -- is used to calculate a mean baseline (\textit{Baseline} in \figrefbra{hptpcPaper:sec:dataAnalysis:subsec:CR:fig:genericWaveform}) and baseline RMS for a waveform. 

The pre-amplifiers are charge-integrating; thus the maximum voltage of the charge waveform is proportional to the total charge collected on an anode. The amplitude (negative amplitude) of a waveform is taken to be the largest (smallest) $V_i$ value of the waveform, $\text{max}\left(V_{i=0\dots N}\right)$ ($\text{min}\left(V_{i=0\dots N}\right)$), subtracted by the mean baseline. We distinguish properties of negative polarity pulses from positive ones by adding a \textit{negative} where appropriate to the respective property's name.\\
The start-time ($t_{r,10}$ in \figrefbra{hptpcPaper:sec:dataAnalysis:subsec:CR:fig:genericWaveform:zoom}) of a pulse is found by looking backwards in time (examining samples with decreasing sample number $i$) from the sample with the maximum (minimum) $V_i$ value to the point in time where the waveform reaches \SI{10}{\%} of its amplitude value. The point at which the waveform reaches \SI{90}{\%} ($t_{r,90}$ in \figrefbra{hptpcPaper:sec:dataAnalysis:subsec:CR:fig:genericWaveform:zoom}) of its amplitude is identified in the same manner. From the maximum voltage, the waveform decays exponentially with a time-constant depending on the pre-amplifier chip used. Likewise, the $t_{f,10}$ and $t_{f,90}$ points on the tail of the waveform are found by finding the point after the maximum (minimum) where the amplitude first falls below \SI{90}{\%} and \SI{10}{\%} of the peak value. A pulse's rise time (fall-time) is calculated as the difference between $t_{r,10}$ and $t_{r,90}$ ($t_{f,90}$ and $t_{f,10}$).\\
In addition to the above the RMS of a charge waveform is calculated as:
\begin{equation}
\label{hptpcPaper:sec:chargeAnalysis:subsec:raw:eq:WaveformRMS}
  \textit{WaveformRMS} = \sqrt{\frac{1}{N} \sum_{i=1}^{j} V_{i}^{2}} 
\end{equation}
where $j$ is the number of the last sample. The \textit{BaselineRMS} is calculated in a similar manner, but only taking $V_i$ in the pre-trigger region into account.

\subsection{Waveform cleaning}
\label{hptpcPaper:sec:dataAnalysis:subsubsec:cleaning}

\begin{figure}
\centering
\subfloat[]{\label{hptpcPaper:sec:dataAnalysis:subsubsec:cleaning:fig:BaselineSpectrumUnclean}
\includegraphics[width=0.49\columnwidth, trim = 0 0 0 30, clip=true]{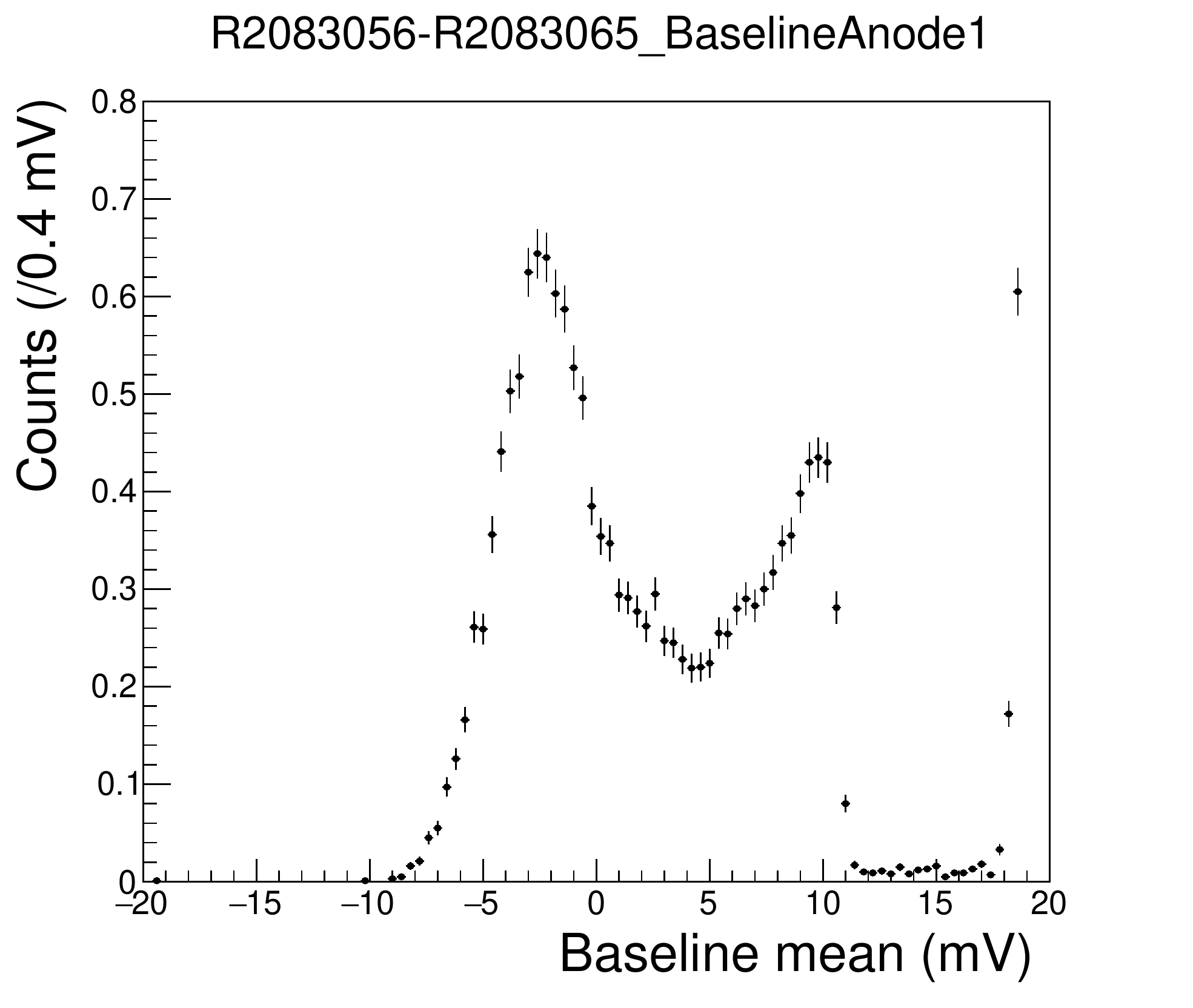}}
\subfloat[]{\label{hptpcPaper:sec:dataAnalysis:subsubsec:cleaning:fig:BaselineSpectrumClean}
\includegraphics[width=0.49\columnwidth, trim = 0 0 0 30, clip=true]{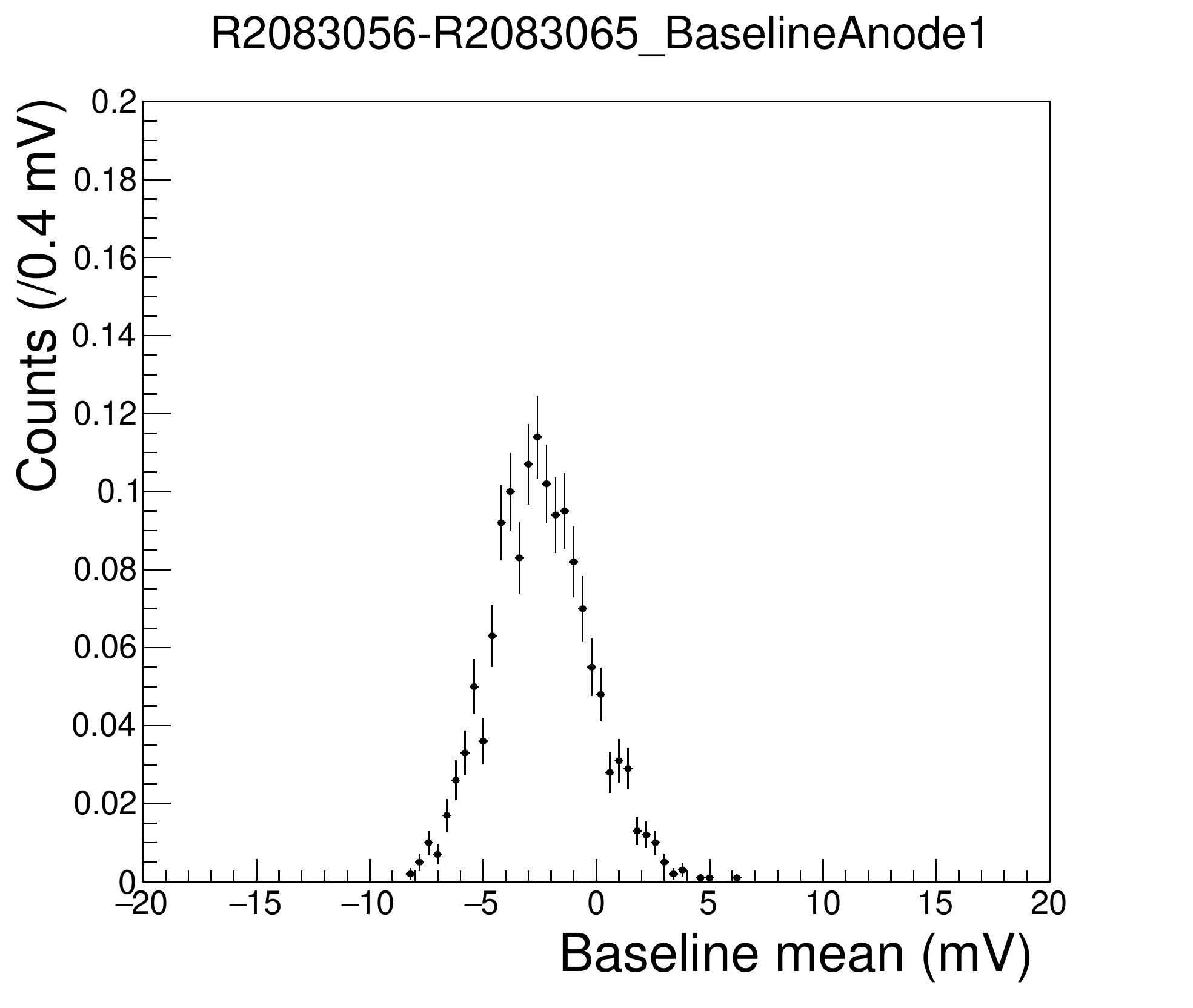}}
\caption{\label{hptpcPaper:sec:dataAnalysis:subsubsec:cleaning:fig:BaselineSpectra}Anode 1 \textit{Baseline} spectrum \protect\subref{hptpcPaper:sec:dataAnalysis:subsubsec:cleaning:fig:BaselineSpectrumUnclean} before cleaning, and \protect\subref{hptpcPaper:sec:dataAnalysis:subsubsec:cleaning:fig:BaselineSpectrumClean} after cleaning. Waveforms with large \textit{Baseline} values are cut, which removes spark events.}
\end{figure}

In order to get the most accurate values of the parameters described above, we apply a series of cleaning steps to the waveforms before calculating the parameters. 
Cuts are made to select waveforms based on their \textit{Baseline} and \textit{BaselineRMS}. The mean of the \textit{Baseline} values of all waveforms in a run is calculated. If the baseline mean of a waveform is not within a \SI{5}{RMS} interval of the mean of all \textit{Baseline} values, the waveform is rejected. Similarly, if the \textit{BaselineRMS} of waveform is not within a \SI{5}{RMS} interval of the mean of all \textit{BaselineRMS} values, the waveform is cut. This cut allows to remove all waveforms with anomalous fluctuations of the baseline, as occur \textit{e.g.} during sparks. \Figref{hptpcPaper:sec:dataAnalysis:subsubsec:cleaning:fig:BaselineSpectra} shows a spectrum of the anode 1 \textit{Baseline} values before and after applying these cuts.\\
Waveforms with a \textit{Baseline} above the trigger threshold are cut. Furthermore, a set of simultaneously recorded waveforms is rejected when the maximum $V_i$ value of the anode 3 waveform is below the trigger threshold. 
This is because we trigger the simultaneous readout of all three anodes with the anode 3 signal. In cases where only the anode 3 waveform's maximal $V_i$ is above its trigger threshold, the corresponding anode 1 and 2 waveforms can still be used in an amplitude measurement.\\
Checks are made to identify events containing sparks and such events with a damaged pre-amplifier. An ``event'' contains all waveforms recorded during the exposure time of the simultaneously taken CCD frames. In case of sparks the pre-amplifiers' baselines moves substantially 
and it takes time for the pre-amplifier to return to the pre-spark status. Thus an event is flagged as spark event when it contains more than 5 waveforms with a \textit{Baseline} above the trigger threshold. When a pre-amplifier gets damaged the result is flat waveform. So we flag events where the maximum value is very close to the baseline $(\text{max}(V_i) < Baseline \times 1.02)$ and $(\text{max}(V_i) > Baseline \times 0.98)$ as having been taken with a damaged preamp.\\
Finally, waveforms are accepted or rejected based on their rise time and \textit{Peak Time}. The \textit{Peak Time} is the time value $t_j$ in a waveform for which $V_j = \text{max}(V_i)$, $i\in0\dots N$. We calculate the rise time as the time difference between: $t_{\text{r},90}$ and $t_{\text{r},10}$. For anodes 2 and 3, waveforms with a rise time above \SI{5}{\micro\second} or a \textit{Peak Time} which is not within a \SI{5}{\micro\second} interval around $t=0$ are cut. For anode 1, the peaks are not always visible above the noise. Waveforms with a long rise time or a \textit{Peak Time} outside of $t<\pm\SI{5}{\micro\second}$ are rejected for the anode 1 amplitude measurement, but the corresponding waveforms in anode 2 and 3 are not cut. These time values are conservative cuts, chosen far above the pre-amplifiers specified rise time of \SI{3}{\nano\second}, which help to remove waveforms which have been triggered by noise.
\Tabref{hptpcPaper:sec:dataAnalysis:subsubsec:cleaning:tab:cleaningCutsNoSparks} and \tabref{hptpcPaper:sec:dataAnalysis:subsubsec:cleaning:tab:cleaningCutsSparks} show the fraction of analysed waveforms rejected by each data cleaning cut.
\begin{table}
\centering
\begin{tabular}{c|c|c}
Cut           & \multicolumn{2}{c}{surviving signals}         \\ 
                                                    & single cut        & cuts applied subsequently \\ \hline
No Cuts                                             & 100  $\%$ &   100  $\%$                       \\
$Baseline < \text{Trigger-threshold}$               & 99.97 $\%$ &  99.97  $\%$                     \\
$Baseline \text{ within mean interval}$             & 99.97 $\%$ &  99.97  $\%$                     \\
$BaselineRMS \text{ within interval}$               & 99.99 $\%$ &  99.97  $\%$                     \\
$\text{max}(V_i) > \text{Trigger}\ \text{threshold}$& 11.25 $\%$ &  11.23  $\%$                     \\
$\text{rise\ time} < \SI{5}{\micro\second}$         & 61.23 $\%$ &  9.59  $\%$                      \\
${Peak\ Time} < \SI{5}{\micro\second}$              & 20.56 $\%$ &  9.59  $\%$                      \\
\end{tabular}
\caption{\label{hptpcPaper:sec:dataAnalysis:subsubsec:cleaning:tab:cleaningCutsNoSparks}{Fraction of analysed waveforms rejected for each data cleaning cut for a run where no sparking was observed.}}
\end{table}
\begin{table}
\centering
\begin{tabular}{c|c|c}
Cut           & \multicolumn{2}{c}{surviving signals}         \\ 
                                                  & single cut     & cuts applied subsequently \\ \hline
No Cuts                                           & 100  $\%$   &  100  $\%$                   \\
$Baseline < \text{Trigger-threshold}$             & 53.26  $\%$ &  53.26  $\%$                 \\
$Baseline \text{ within mean interval}$           & 26.85  $\%$ &  26.85  $\%$                 \\
$BaselineRMS \text{ within interval}$             & 68.29  $\%$ &  26.85  $\%$                 \\
$\text{max}(V_i) > \text{Trigger threshold}$      & 5.64  $\%$  &  5.22  $\%$                  \\
$\text{rise\ time} < \SI{5}{\micro\second}$       & 51.53  $\%$ &  4.35  $\%$                  \\
${Peak\ Time} < \SI{5}{\micro\second}$            & 14.92  $\%$ &  4.25  $\%$                  \\
\end{tabular}
\caption{\label{hptpcPaper:sec:dataAnalysis:subsubsec:cleaning:tab:cleaningCutsSparks}{Fraction of analysed waveforms rejected for each data cleaning cut for a run containing spark events.}}
\end{table}

\subsection{Gas gain measurement}

In this section the charge gain of the three anode amplification stage is calculated from the amplitude spectra discussed above. Features in the spectra have to be related to a known energy deposition inside the HPTPC. A known energy deposit can be realised using a radioactive source \textit{e.g.} $^{241}\textrm{Am}$ (\textit{cf}. \secrefbra{hptpcPaper:sec:ccdAnalysis:subsec:energyDeposit}). Primary ionisation electrons ($Q_{\textrm{e}}$) from converted $\gamma$-rays or $\alpha$ particles drift towards the anode meshes and are amplified there. The amplification factor, the charge gain of the amplification region $G_{\text{amp}}$, depends on the meshes' configuration such as inter mesh distance and HV settings. After charge signals are decoupled from the HV line, they are amplified by the pre-amplifiers ($G_{\textrm{preamp}}$). The amplitude $A$ of a waveform thus relates to $Q_{\textrm{e}}$ as
\begin{align}
A \left[\si{\milli\volt}\right] = f \cdot G_{\textrm{preamp}} \left[\si{\milli\volt\per\pico\coulomb}\right] \cdot  G_{\text{amp}} \cdot Q_{\textrm{e}} \left[\si{\pico\coulomb}\right] \quad .
\label{hptpcPaper:sec:dataAnalysis:eq:gain}
\end{align}
The factor $f$ is another dimensionless factor which we introduce in order to describe (attenuating) effects of the readout circuit on the signal height. $G_{\textrm{preamp}}$ and $f$ are determined with dedicated measurements to calibrate the readout circuit.

\subsubsection{Pre-amplifier and circuit calibration}
\label{hptpcPaper:sec:dataAnalysis:subsec:chargeCalib}

\begin{figure}
\centering
\subfloat[]{\label{fig:Gain:vsInAmp}
\includegraphics[width=0.65\columnwidth]{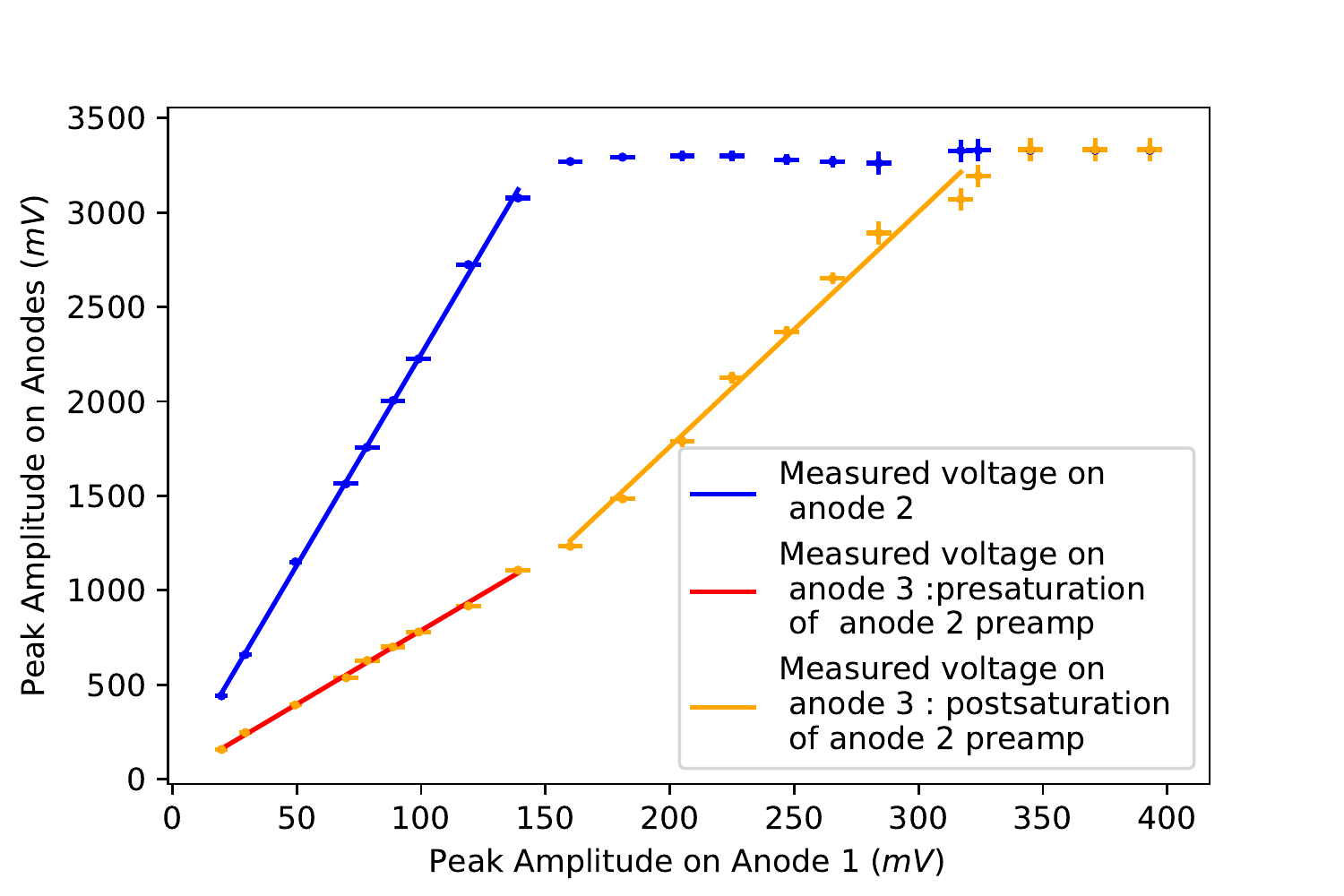}}\\
\subfloat[]{\label{fig:Gain:vsCharge}
\includegraphics[width=0.65\columnwidth,trim= 0 0 0 20, clip = true]{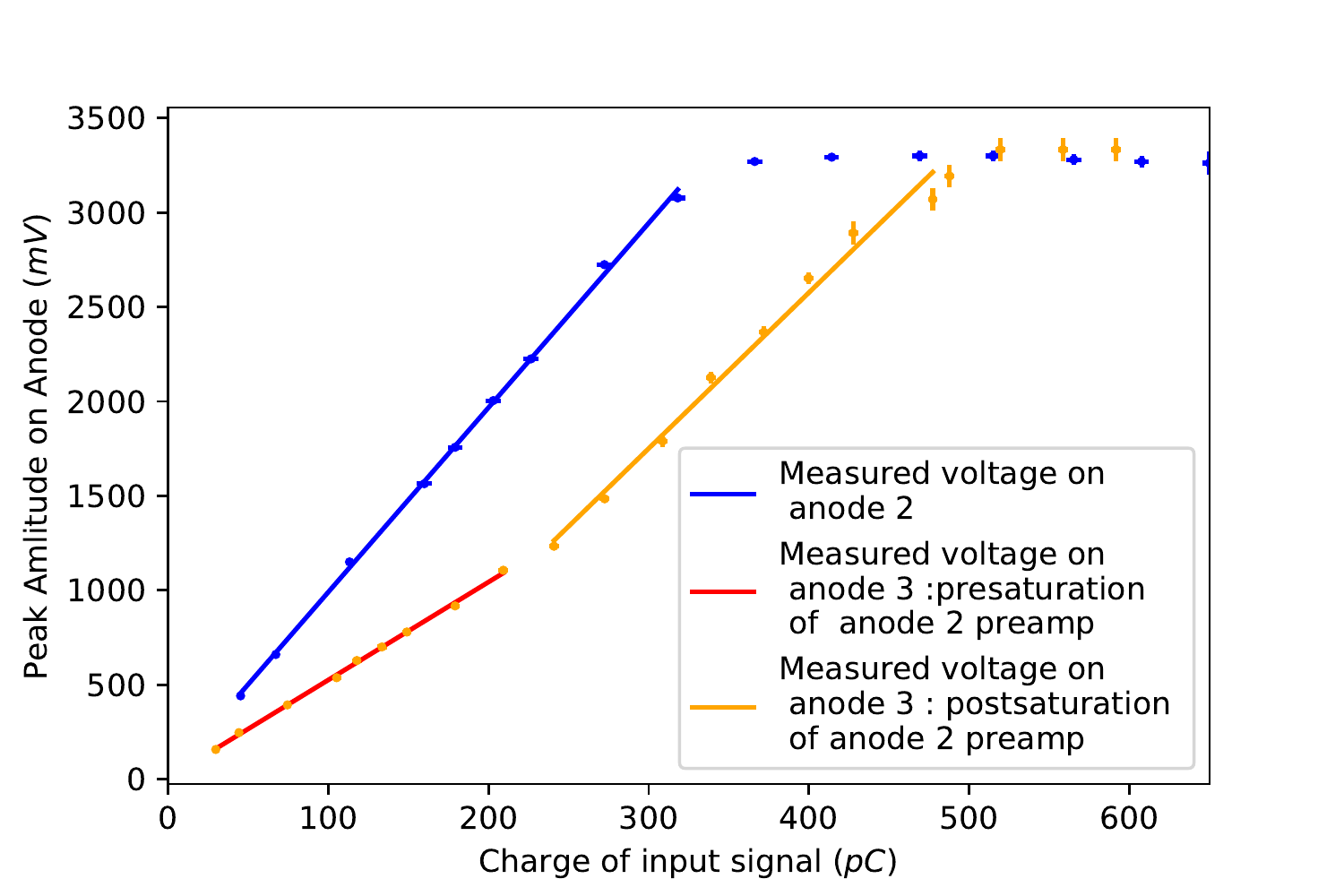}}
\caption{\label{fig:Gain}Peak height ($V_{\textrm{output}}^{\text{amp}}$) measured by the anode 2 and anode 3 readout channel (with pre-amplifier) for test pulses injected into the amplification region via the anode 1 mesh. Both plots show the same data with different units on the horizontal axis: \protect\subref{fig:Gain:vsInAmp} $V_{\textrm{output}}^{\text{amp}}$ as function of input test pules signal height ($V_{\textrm{input}}$) and \protect\subref{fig:Gain:vsCharge} as a function of the charge seen at the pre-amplifier input. One polynomial of order ($P1$) one is fitted to the anode 2 (blue) measurement and two separate $P1$s are fitted to the different regions on anode 3. One in the pre-saturation region of the anode 2 pre-amplifier (red) and one in the post-saturation region of anode 2 pre-amplifier (orange).}
\end{figure}

The pre-amplifier chips employed are Cremat CR-112 and CR-113 charge sensitive pre-amplifiers, hosted on a CR-150-R5 evaluation board. The gain of the pre-amplifier chips is calibrated by injecting pulses into the evaluation board test input (\SI{1}{\pico\farad} input capacitance). We chose rectangular pulses with a pulse height $V_{\text{input}}$, and a low frequency and long width as compared to the pre-amplifiers decay time of a few \SI{100}{\micro\second}. These pulses are recorded with the HPTPC's data acquisition system and analysed with the analysis chain descried above, but without applying cleaning since no noise signals are present when not applying HV to the detector. For a given test pulse height, the resulting amplitude spectrum features one peak. The ratio of the peak's mean amplitude to the input pulse height gives $G_{\textrm{preamp}}$ when taking the pre-amplifiers input capacitance into account. Testing several chips, the average gain of the CR-112 chips is measured to be $G_{\textrm{preamp}}^{\textrm{CR}-112}=\SI{11.7(6)}{\milli\volt\per\pico\coulomb}$, and the average gain of the CR-113 chips was measured to be $G_{\textrm{preamp}}^{\textrm{CR}-113}=\SI{1.24(6)}{\milli\volt\per\pico\coulomb}$. These values are consistent with the values provided by the supplier of \SI{13}{\milli\volt\per\pico\coulomb} and \SI{1.3}{\milli\volt\per\pico\coulomb}, respectively.\\
\label{hptpcPaper:sec:dataAnalysis:subsubsec:circuitResponse}
A detailed description of the HPTPC’s circuit response to test pulses can be found in \cite{maAdamTarrant2020}. To perform these tests one of the three pre-amplifiers is disconnected from its bias box ($S_{\text{anode}i}$ in \figrefbra{hptpcPaper:subsubsec:hptpc:HV:fig:HV}) and test pulses are injected where it is usually connected. Doing so induces signals on the other two anode meshes, which are read out. \Tabref{hptpcPaper:sec:dataAnalysis:subsubsec:cleaning:tab:capmeas} shows the inter-mesh capacitances measured with a digital multimeter as well as the results from a fit to the data obtained during the test-pulse campaign.
\begin{table}
  \centering
  \begin{tabular}{l|l|l|l|l|l}
    Measurement taken                & \begin{tabular}[c]{@{}l@{}}Capacitance's \\ between \\ Anode 1/2 [\si{\nano\farad}]\end{tabular} & \begin{tabular}[c]{@{}l@{}}\(\chi^2/N_{\text{dof}}\) of \\ Anode 1/2\end{tabular} & \begin{tabular}[c]{@{}l@{}}Capacitances \\ between \\ Anode 2/3 [\si{\nano\farad}]\end{tabular} & \begin{tabular}[c]{@{}l@{}}\(\chi^2/N_{\text{dof}}\) of \\ Anode 2/3\end{tabular} & \begin{tabular}[c]{@{}l@{}}Capacitances \\ between \\ Anode 1/3 [\si{\nano\farad}]\end{tabular} \\ \hline
    fit                        & 7.3\(\pm\)0.3   & 0.76 & 4.4\(\pm\)0.4   & 0.35 & -               \\ 
    Multimeter reading         & 6.06\(\pm\)0.05 & -    & 3.72\(\pm\)0.05 & -    & 2.16\(\pm\)0.05 \\
  \end{tabular}
  \caption{\label{hptpcPaper:sec:dataAnalysis:subsubsec:cleaning:tab:capmeas}Mesh capacitances determined by a fit \cite{maAdamTarrant2020} and by a direct measurement with a multimeter.}
\end{table}
The capacitance determined by measuring pulse amplitudes and by multimeter measurement differ by \SI{17}{\%}. 
This difference is likely due to the fact that the multimeter measurement is performed close to the detector, \textit{i.e.} no long cables and other parasitic capacitances are present. The distance between the mesh planes can be determined knowing the capacitances:
\begin{equation}
  C=\epsilon_o\frac{A}{d},
\end{equation}
where \(C\) is the capacitance, \(\epsilon_0\) the vacuum permittivity, $A$ the area of the mesh planes and $d$ the distance between two mesh planes. This assumes that the meshes can be approximated as a parallel plate capacitor. Inserting our mesh geometry into the calculations in \cite{deloach1971study} shows that such an approximation overestimates (underestimates) the actual capacitance (mesh distance) by less than \SI{10}{\%}. Furthermore we use $\epsilon_{\text{Ar}}=1$ which is accurate to a level better than \SI{1}{\permil} \cite{Bergstr_m_1976}, hence $\epsilon_{\text{Ar}}\epsilon_{0} = \epsilon_{0}$. We can calculate that anode 1 and 2 are \SI{1.20(5)}{\milli\meter} apart and the distance between anode 2 and anode 3 is \SI{2.0(2)}{\milli\meter}. These values are likely too small, since the used approximation underestimates the distance as mentioned before. During construction we aimed for a spacing of \SI{0.5}{\milli\meter} (\SI{1}{\milli\meter}) between anode 1 and 2 (anode 2 and 3) (\textit{cf.} \secrefbra{hptpcPaper:subsec:hptpc:AmplificationStage}). The values determined here have the right order of magnitude and are close to the design values. The difference can be due to the fact that the exact thickness of the glue layers in the amplification region is not known, therefore the design values are most likely a lower limit.\\
\begin{figure}
\centering
\includegraphics[width=0.54\columnwidth, trim = 0 0 0 30, clip=true]{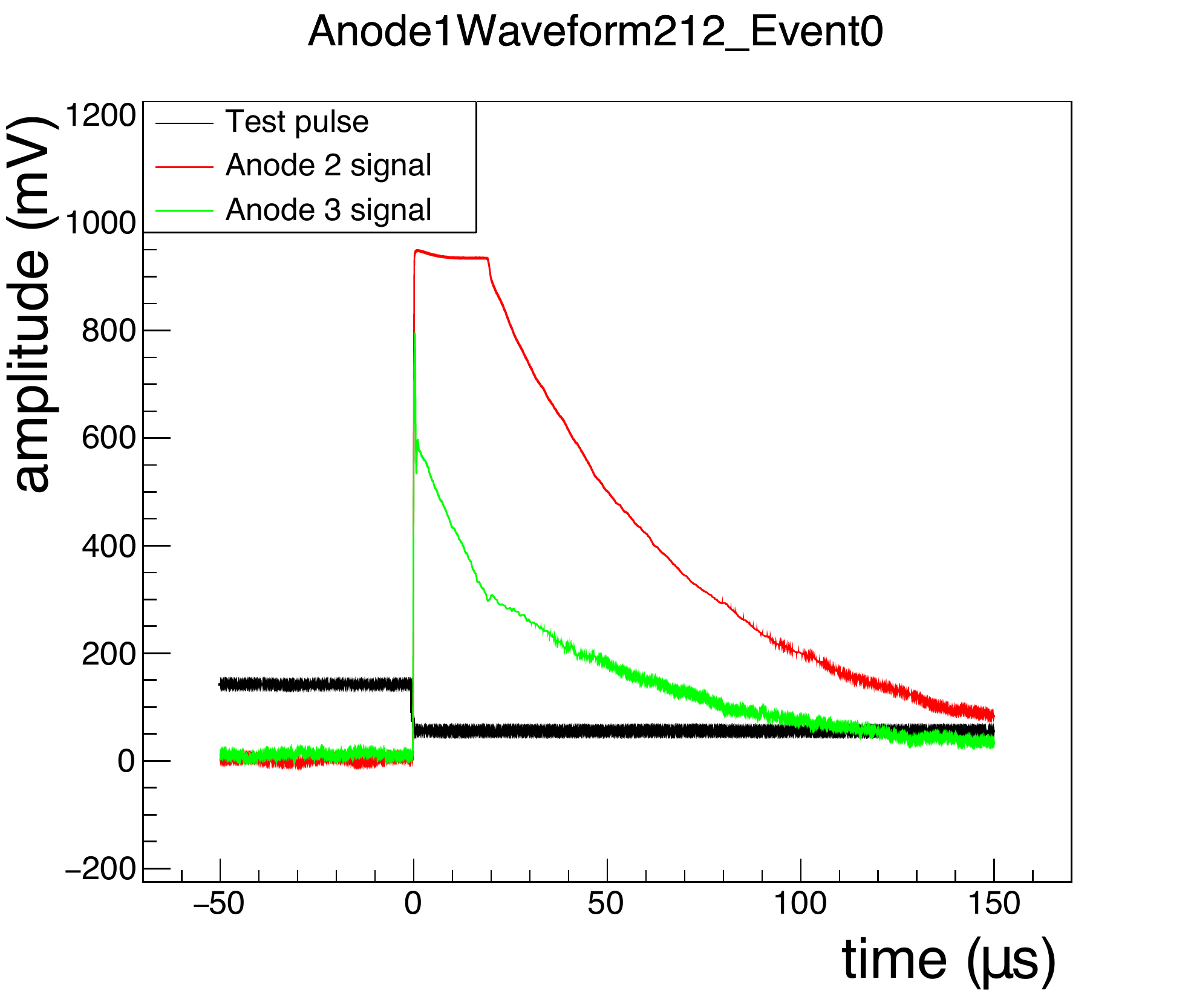}
\caption{\label{hptpcPaper:sec:dataAnalysis:subsubsec:preampAndCircuitResponse:fig:testPulseAndOutput}Waveform of a test pulse, coupled into the anode 1 mesh and the resulting amplified pulses (CR-112) as digitised by the HPTPC's data acquisition system.}
\end{figure}
\Figref{hptpcPaper:sec:dataAnalysis:subsubsec:preampAndCircuitResponse:fig:testPulseAndOutput} shows an example where a test pulse is coupled into $S_{\text{anode}1}$ at the anode 1 bias box while the anode 2 and anode 3 signals are amplified and digitised. In the figure the anode 2 signal is saturated, whilst the decay of the anode 3 signal shows a change of decay constant around $\sim\!\!\SI{20}{\micro\second}$. A systematic study of the amplified signals' peak height ($V_{\textrm{output}}^{\text{amp}}$) revealed that as soon as one pre-amplifier is saturated, the signal on the other pre-amplifier shows a modified decay similar to what is visible in \figref{hptpcPaper:sec:dataAnalysis:subsubsec:preampAndCircuitResponse:fig:testPulseAndOutput} \cite{maAdamTarrant2020}. This behaviour affects the measured amplitude as shown in \figref{fig:Gain}: The points for anode 2 feature two distinct regions: An initial region of linear increase up until an output voltage of \SI{3330(20)}{\milli\volt} where the pre-amplifier saturates and the region after that.
The saturation value is in line with the manufacturer's technical specification for output swing of \(\pm\)\SI{3}{\volt}. \Figref{fig:Gain:vsCharge} shows the same data as in \figref{fig:Gain:vsInAmp}, but as function of the charge which arrives at the input of the respective pre-amplifier. The charge is calculated using $V_{\text{input}}$ and the circuit elements shown in \figref{hptpcPaper:subsubsec:hptpc:HV:fig:HV}. The slope of the anode 2 data before saturation in the plot gives the pre-amplifier gain ($G_{\text{preamp}}^{\text{CR}-112}=\SI{11.7(6)}{\milli\volt\per\pico\coulomb}$) multiplied by $f^{a2}$, which describes signal attenuation and losses in the circuit (\textit{cf}. Eq. \eqref{hptpcPaper:sec:dataAnalysis:eq:gain}). Fitting of a polynomial of order one to the data points, corresponding to the anode 2 line in \figref{fig:Gain:vsInAmp}, yields $f^{a2}\cdot G_{\text{preamp}}^{\text{CR}-112}=\SI{9.8(1)}{\milli\volt\per\pico\coulomb}$, the corresponding value for $f^{a2}$ is shown in \tabref{tab:Attenuation}.\\
For anode 3, however, three regions can be identified in \figref{fig:Gain:vsInAmp}. There are two regions of distinct linear increase but with different gradients. The first region -- up to a $V_{\textrm{input}}$ of \SI{150}{\milli\volt} -- ends at the point when the anode 2 pre-amplifier saturates. From this point onwards two decay constants are observed in anode 3 waveforms similar to what is shown in \figref{hptpcPaper:sec:dataAnalysis:subsubsec:preampAndCircuitResponse:fig:testPulseAndOutput}. In the second region the rise is still linear, but with a different slope than in the first region and the third region covers the saturation of the anode 3 pre-amplifier. When the anode 2 pre-amplifier saturates, the AC signal current can no longer simply flow through its input and feedback capacitor and the signal sharing is modified. This feedback is then seen in the detector as more charge being measured by the anode 3 pre-amplifier than expected.
It has been confirmed that this behaviour is indeed due to the anode 2 pre-amplifier saturating. Removing this pre-amplifier from the circuit results in anode 3 signals with only one decay constant and no change in gradient -- similar to what is shown for anode 2 in \figref{fig:Gain:vsInAmp}. A fit of a polynomial of order one yields \(f^{a3}\cdot G_{\textrm{preamp}}^{\text{CR}-112}=\SI{5.18(7)}{\milli\volt\per\pico\coulomb}\) for anode 3 before the anode 2 saturation and \(f^{a3}_{\text{post}}\cdot G_{\textrm{preamp}}^{\text{CR}-112}=\SI{8.3(4)}{\milli\volt\per\pico\coulomb}\) after the saturation. \Tabref{tab:Attenuation} shows the circuit response $f$ obtained by comparing the measurements of $f\cdot G_{\textrm{preamp}}$ to the bare $G_{\textrm{preamp}}^{\text{CR}-112}$ measurements at the beginning of this section.\\
\begin{table}
  \centering
  \begin{tabular}{c|c|c}
    Anode           & \begin{tabular}[c]{@{}c@{}} $f\cdot G_{\textrm{preamp}}$ \\ $[\si{\milli\volt}/\si{\pico\coulomb}]$\end{tabular} & \begin{tabular}[c]{@{}c@{}}Modification factor\\ \(f\)\end{tabular} \\ \hline
    anode 2                 & 9.8 \(\pm\)0.1  & 0.754\(\pm\)0.007 \\
    anode 3 Pre-saturation  & 5.18\(\pm\)0.07 & 0.398\(\pm\)0.005 \\
    anode 3 Post-saturation & 8.3 \(\pm\)0.4  & 0.64\(\pm\)0.03 \\
  \end{tabular}
  \caption{\label{tab:Attenuation}Using the measured pre-amplifier without the circuit response ($G_{\textrm{preamp}}^{\textrm{CR}-112}=\SI{11.7(6)}{\milli\volt\per\pico\coulomb}$) and the measurements of the pre-amplifiers connected to the detector $f\cdot G_{\textrm{preamp}}$, the circuit response modification-factor $f$ is determined \cite{maAdamTarrant2020}.}
\end{table}
Finding $f=1$ would imply that there are no signal losses or attenuation effects in the circuit. The $f$ values measured here show a substantial attenuation which can be corrected for since $f$ is constant as a function of amplitude. These losses may occur through the resistive elements shown in \figref{hptpcPaper:subsubsec:hptpc:HV:fig:HV}. The change in $f$ on anode 3 when the anode 2 pre-amplifier is saturated makes this correction slightly more complicated. The capacitances of the amplification region and the available pulse generator did not allow to drive the anode 3 pre-amplifier into saturation to examine whether a similar feed-back occurs on anode 2. In general, events with either pre-amplifier being saturated occur only rarely, as do events where the signal amplitude on anode 2 is higher than on anode 3 due to the way the amplification region is biased.

\subsubsection{Charge gain of the amplification region}
\label{hptpcPaper:subsec:hptpcPerfomrance:chargeGain}

\begin{figure}
\centering
\subfloat[]{\label{hptpcPaper:subsec:hptpcPerformance:chargeGain:fig:FittedAmplitudeSpectraAnode1}
\includegraphics[width=0.30\columnwidth, trim = 0 0 0 30, clip=true]{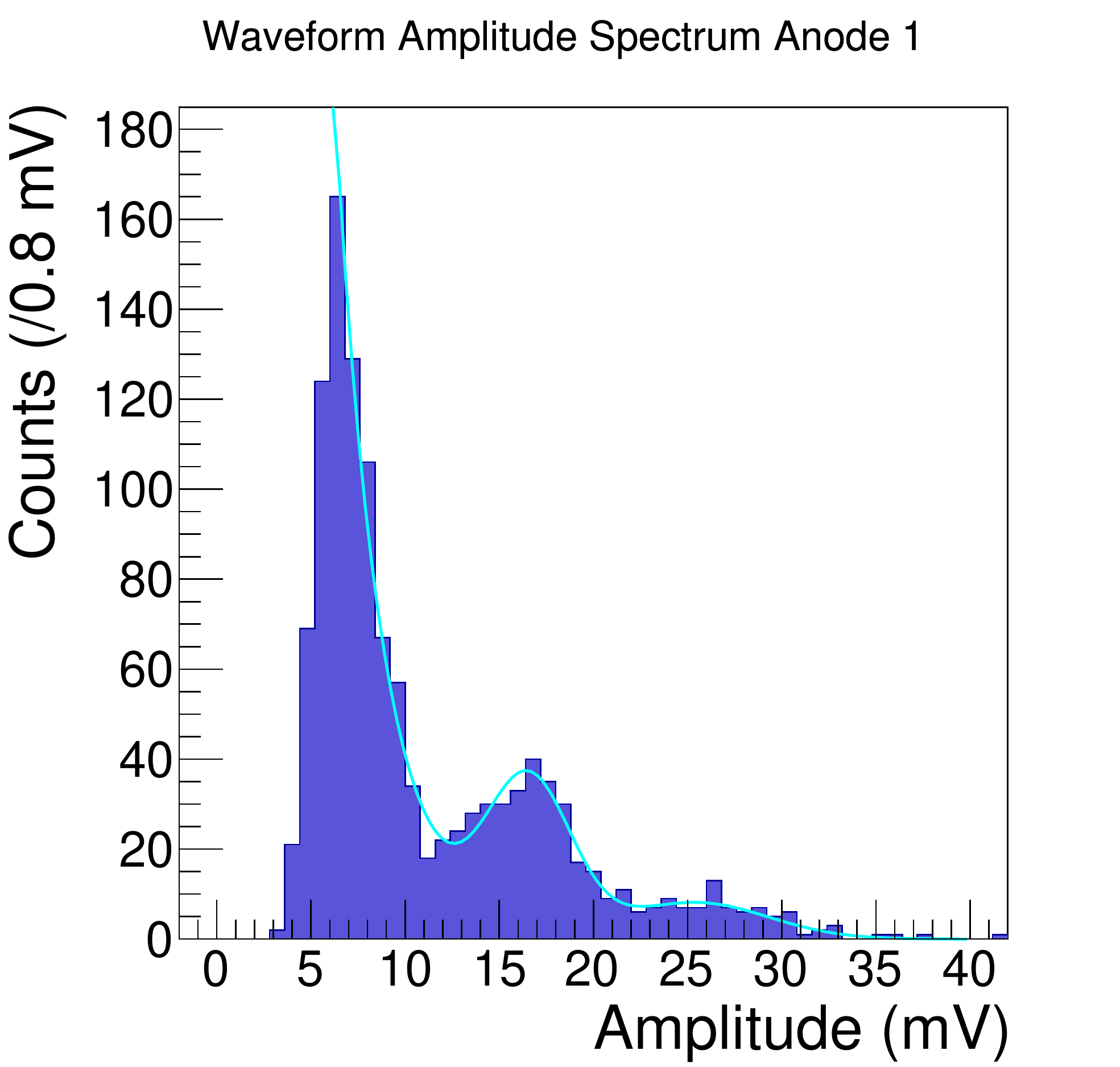}}
\subfloat[]{\label{hptpcPaper:subsec:hptpcPerformance:chargeGain:fig:FittedAmplitudeSpectraAnode2}
\includegraphics[width=0.30\columnwidth, trim = 0 0 0 30, clip=true]{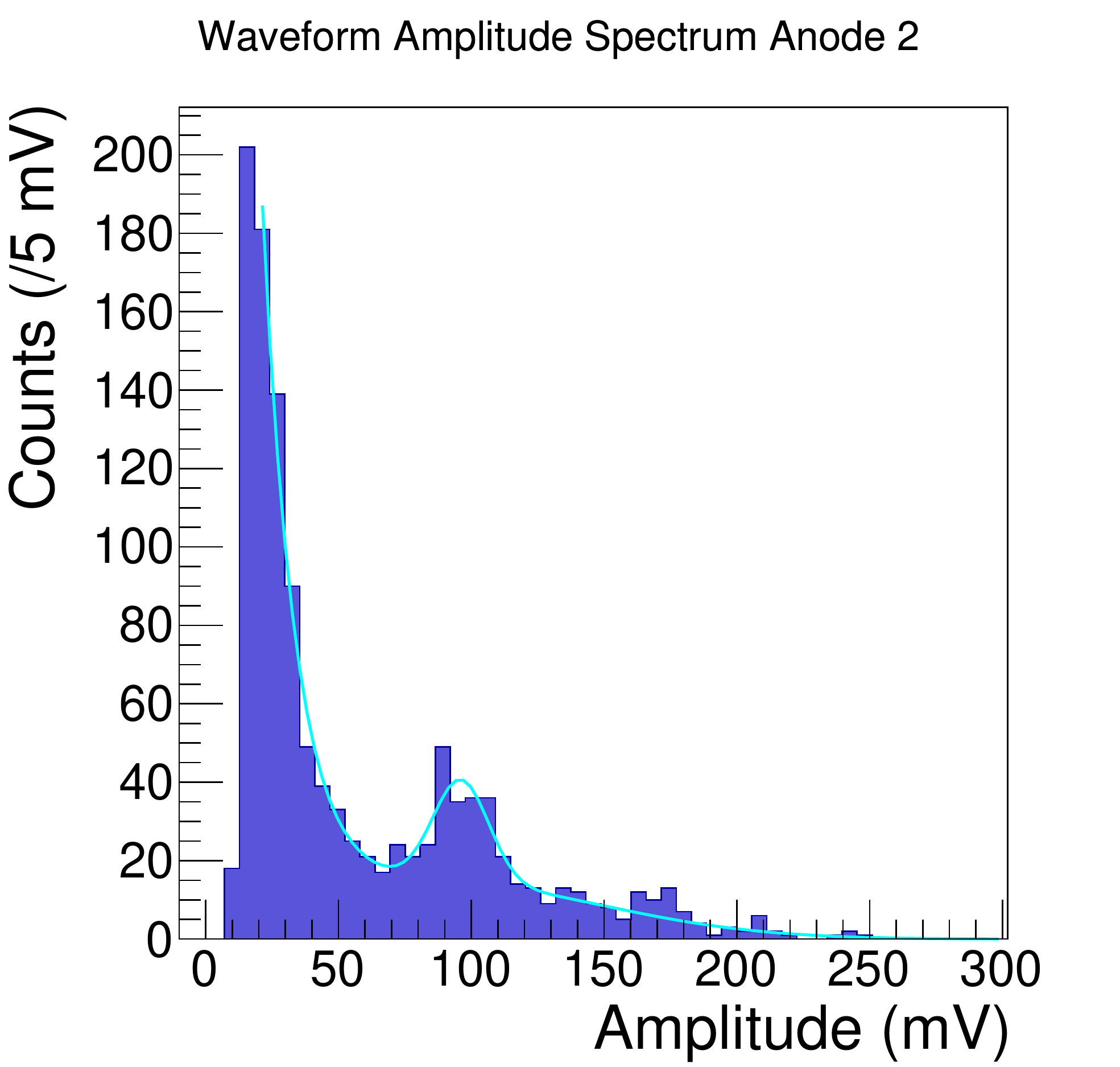}}
\subfloat[]{\label{hptpcPaper:subsec:hptpcPerformance:chargeGain:fig:FittedAmplitudeSpectraAnode3}
\includegraphics[width=0.30\columnwidth, trim = 0 0 0 30, clip=true]{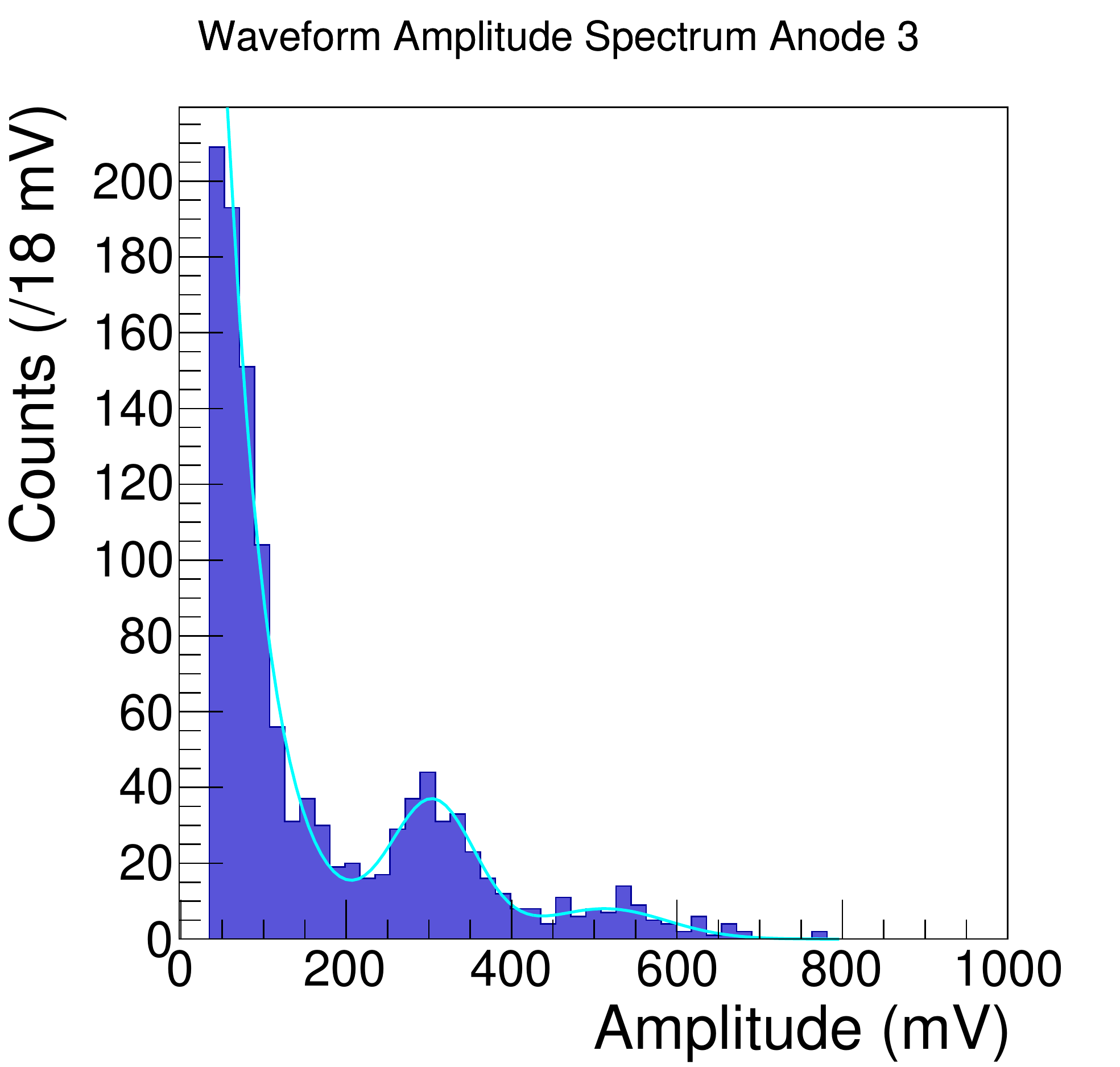}}
\caption{\label{hptpcPaper:subsec:hptpcPerformance:chargeGain:fig:FittedAmplitudeSpectra}Waveform amplitude spectra for anodes \protect\subref{hptpcPaper:subsec:hptpcPerformance:chargeGain:fig:FittedAmplitudeSpectraAnode1} 1, \protect\subref{hptpcPaper:subsec:hptpcPerformance:chargeGain:fig:FittedAmplitudeSpectraAnode2} 2, and \protect\subref{hptpcPaper:subsec:hptpcPerformance:chargeGain:fig:FittedAmplitudeSpectraAnode3} 3. On the vertical axis counts are shown, normalised to the time of one CCD exposure, \textit{i.e.} \SI{2}{\second}. The spectra are fitted with an exponential plus two Gaussian functions. Amplitude spectra shown are summed data over 15 consecutive runs taken at the same voltage settings, $V_{a1}=\SI{1200}{\volt}$, $V_{a2}=\SI{2400}{\volt}$, $V_{a3}=\SI{3600}{\volt}$, $V_{c}=-\SI{6000}{\volt}$.}
\end{figure}
This section discusses the analysis of charge waveform data taken simultaneously as the CCD frames used for the light analysis described in \secref{hptpcPaper:subsec:hptpcPerfomrance:lightGain}. All data was taken in the same TPC fill of pure argon at \SI{3}{bar} absolute pressure. The three voltage schemes -- A, B, and C -- are described in detail in the previous section.

\paragraph{\textbf{Determining the Am-241 alpha decay peak amplitude}}

After data cleaning (\secref{hptpcPaper:sec:dataAnalysis:subsubsec:cleaning}) and taking into account the calibration discussed in \secref{hptpcPaper:sec:dataAnalysis:subsec:chargeCalib} we create amplitude spectra for each anode per voltage setting as shown in \figref{hptpcPaper:subsec:hptpcPerformance:chargeGain:fig:FittedAmplitudeSpectra}. The number of entries in the amplitude spectra for each voltage configuration varies between 100 and 2500, with an average of 910 entries per configuration. This variation is caused by the fact that at higher voltage settings a greater number of waveforms are recorded and survive the cleaning cuts.
From the results presented in \secref{hptpcPaper:subsec:hptpcPerfomrance:lightGain} (and \secref{hptpcPaper:sec:hptpcPerfomrance:opticalRO}), we are confident that we should see the $\alpha$ particles from the $^{241}\textrm{Am}$ decay in the waveform amplitude spectra. Whilst the qualitative example spectrum in \figref{hptpcPaper:sec:hptpcPerfomrance:fig:energyDepostiInDet} does neither account for the gas gain and the electronic noise, the measured amplitude spectra should show some resemblance to this simulation. The measured amplitude spectra (\figrefbra{hptpcPaper:subsec:hptpcPerformance:chargeGain:fig:FittedAmplitudeSpectra}) appear as an exponentially falling background, with a clear peak. This peak corresponds to the deposit of the $\sim\!\!\SI{4.5}{\mega\electronvolt}$ $\alpha$ particles from the $^{241}\text{Am}$ decay. The exponential background is a mix of the expected cosmic radiation background, of the $^{241}\text{Am}$ x-ray signals and noise triggers. The amplitude spectra are fitted with the function
\begin{equation}
\label{hptpcPaper:subsec:hptpcPerformance:chargeGain:eqn:fitting_function}
\begin{split}
  s(\text{amplitude}) = \exp{\left\{p_0 + p_1 \cdot \text{amplitude}\right\}} + p_2 \cdot \exp{\left\{-0.5 \cdot {\left(\frac{\text{amplitude} - p_3}{p_4}\right)}^2\right\}}\\
   + p_5 \cdot \exp{\left\{-0.5 \cdot {\left(\frac{\text{amplitude} - p_6}{p_7}\right)}^2\right\}} \quad,
\end{split}
\end{equation}
\begin{figure}
\centering
\subfloat[Anode 1, Scheme A]{\label{hptpcPaper:subsec:hptpcPerformance:chargeGain:fig:PeakPostionVsVoltageSchemeAAnode1}
\includegraphics[width=0.325\columnwidth, trim = 0 0 0 36, clip=true]{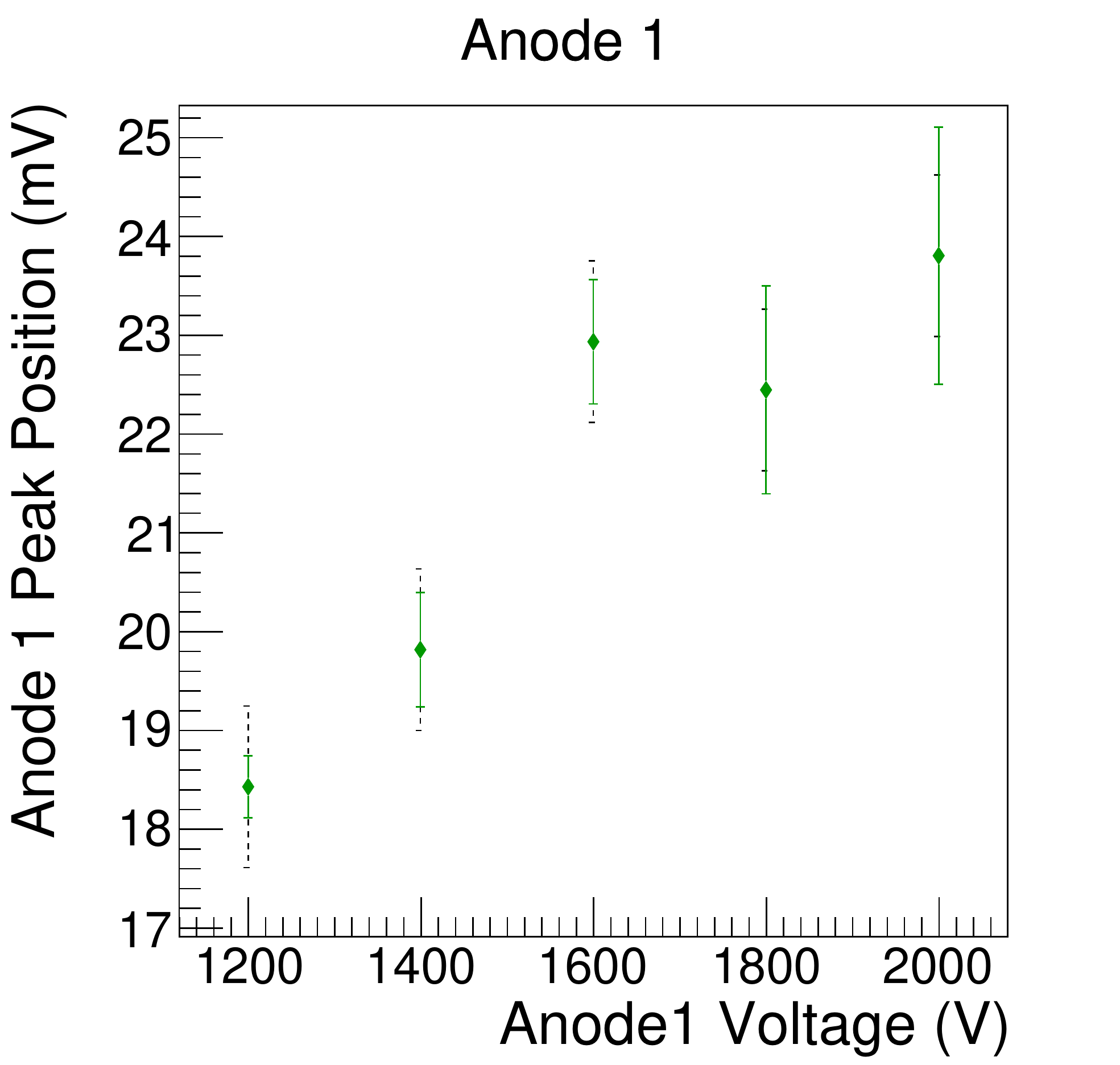}}
\subfloat[Anode 2, Scheme A]{\label{hptpcPaper:subsec:hptpcPerformance:chargeGain:fig:PeakPostionVsVoltageSchemeAAnode2}
\includegraphics[width=0.325\columnwidth, trim = 0 0 0 36, clip=true]{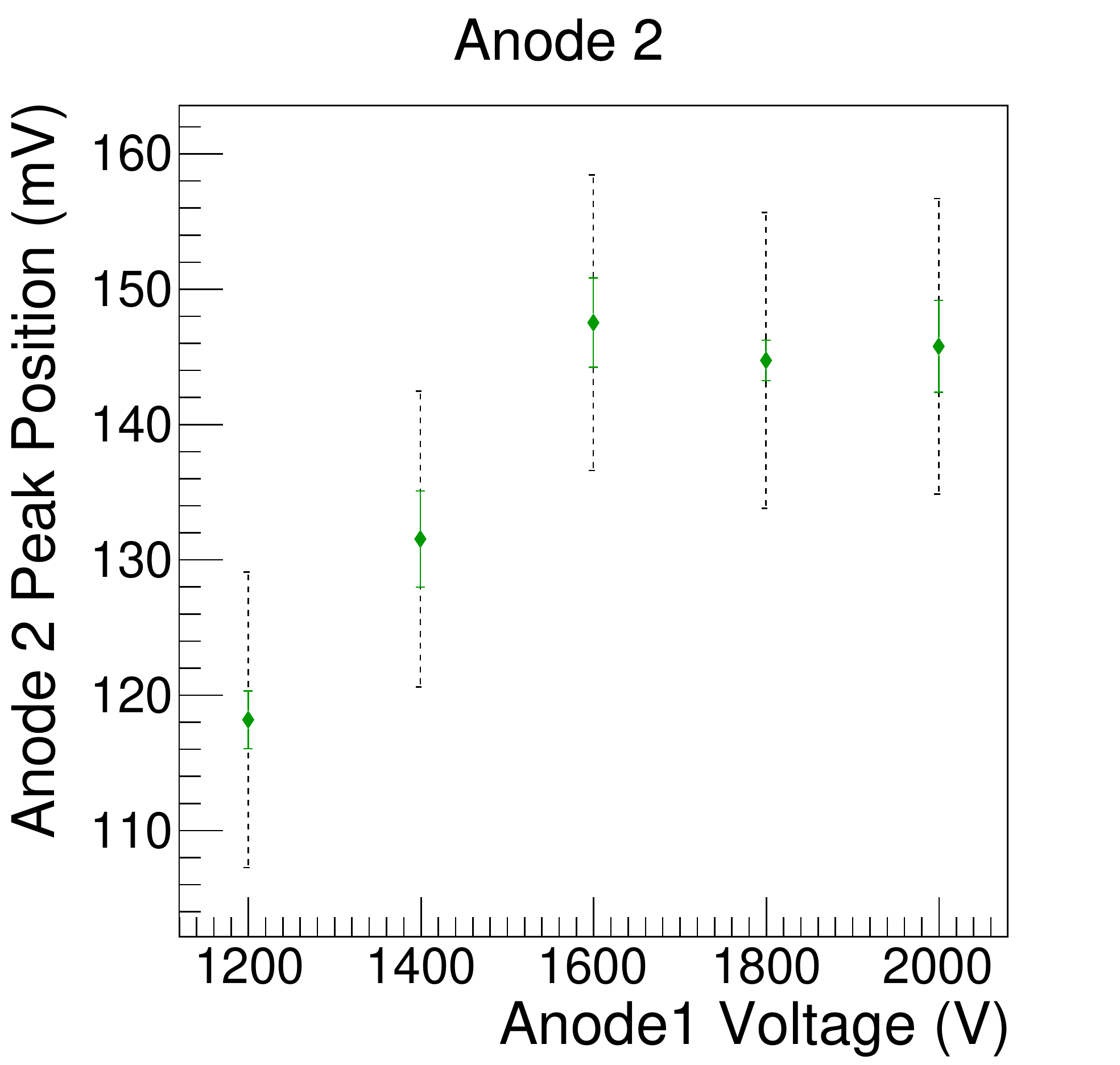}}
\subfloat[Anode 3, Scheme A]{\label{hptpcPaper:subsec:hptpcPerformance:chargeGain:fig:PeakPostionVsVoltageSchemeAAnode3}
\includegraphics[width=0.325\columnwidth, trim = 0 0 0 36, clip=true]{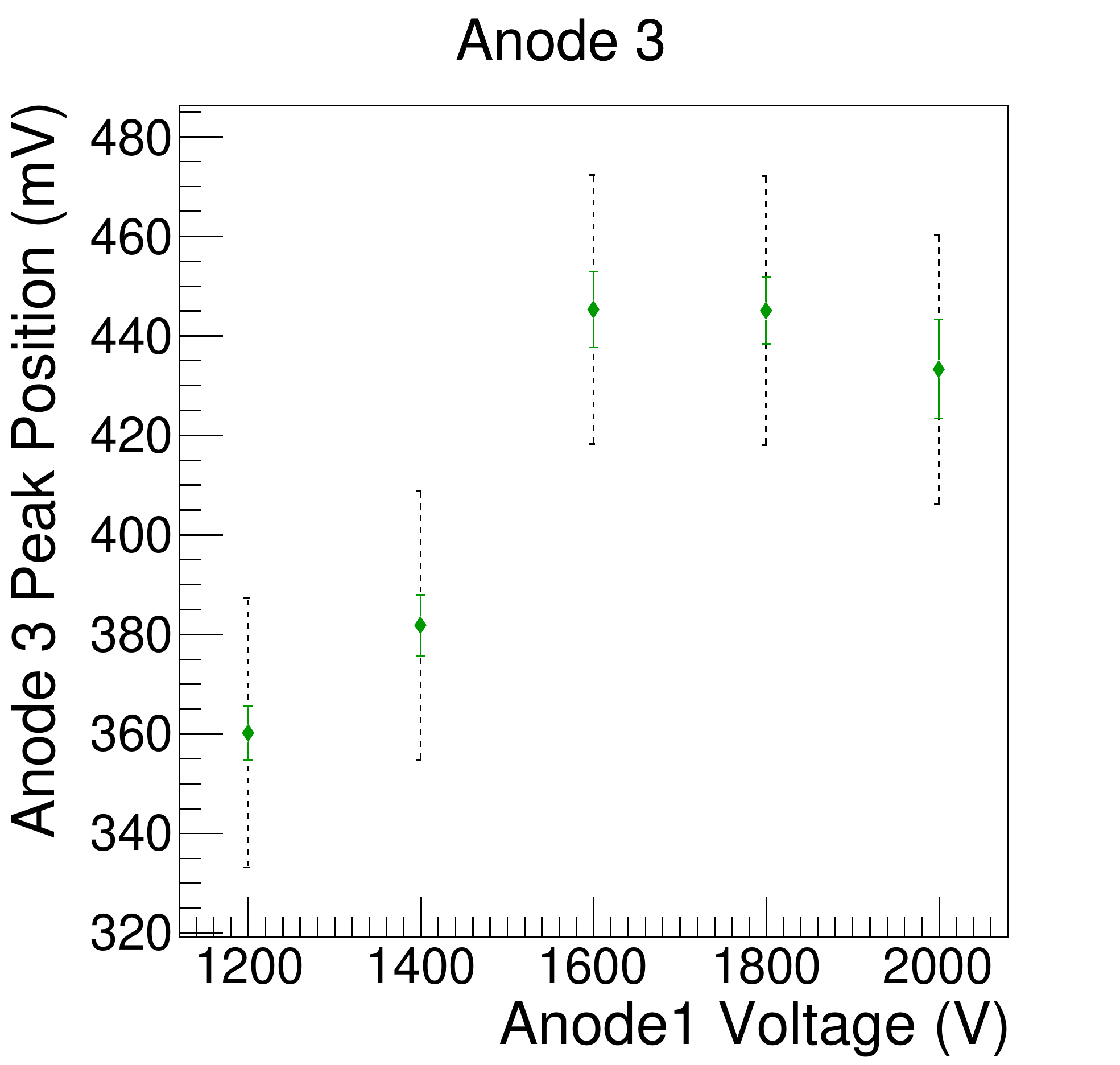}}\\
\subfloat[Anode 1, Scheme B]{\label{hptpcPaper:subsec:hptpcPerformance:chargeGain:fig:PeakPostionVsVoltageSchemeBAnode1}
\includegraphics[width=0.325\columnwidth, trim = 0 0 0 36, clip=true]{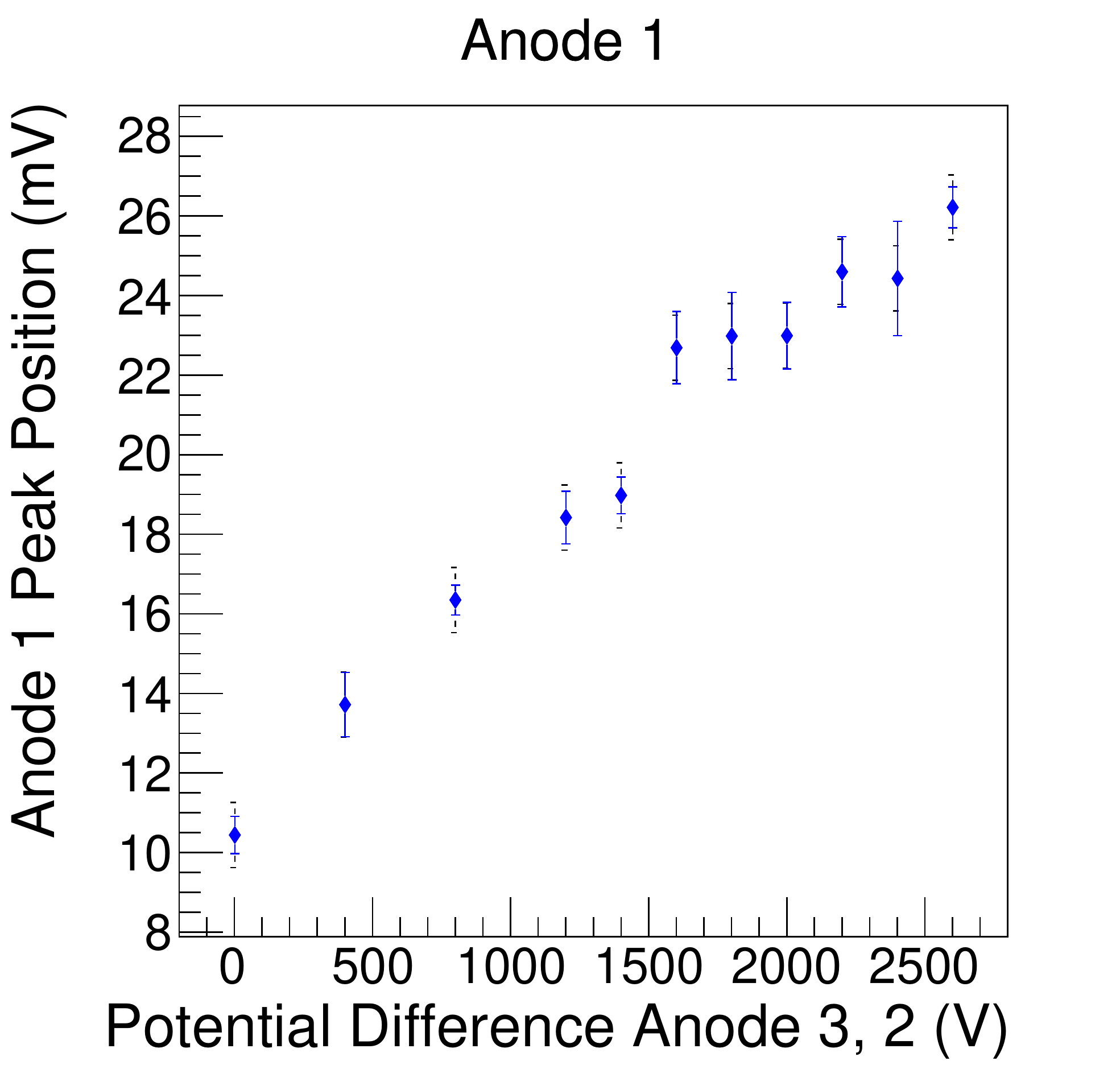}}
\subfloat[Anode 2, Scheme B]{\label{hptpcPaper:subsec:hptpcPerformance:chargeGain:fig:PeakPostionVsVoltageSchemeBAnode2}
\includegraphics[width=0.325\columnwidth, trim = 0 0 0 36, clip=true]{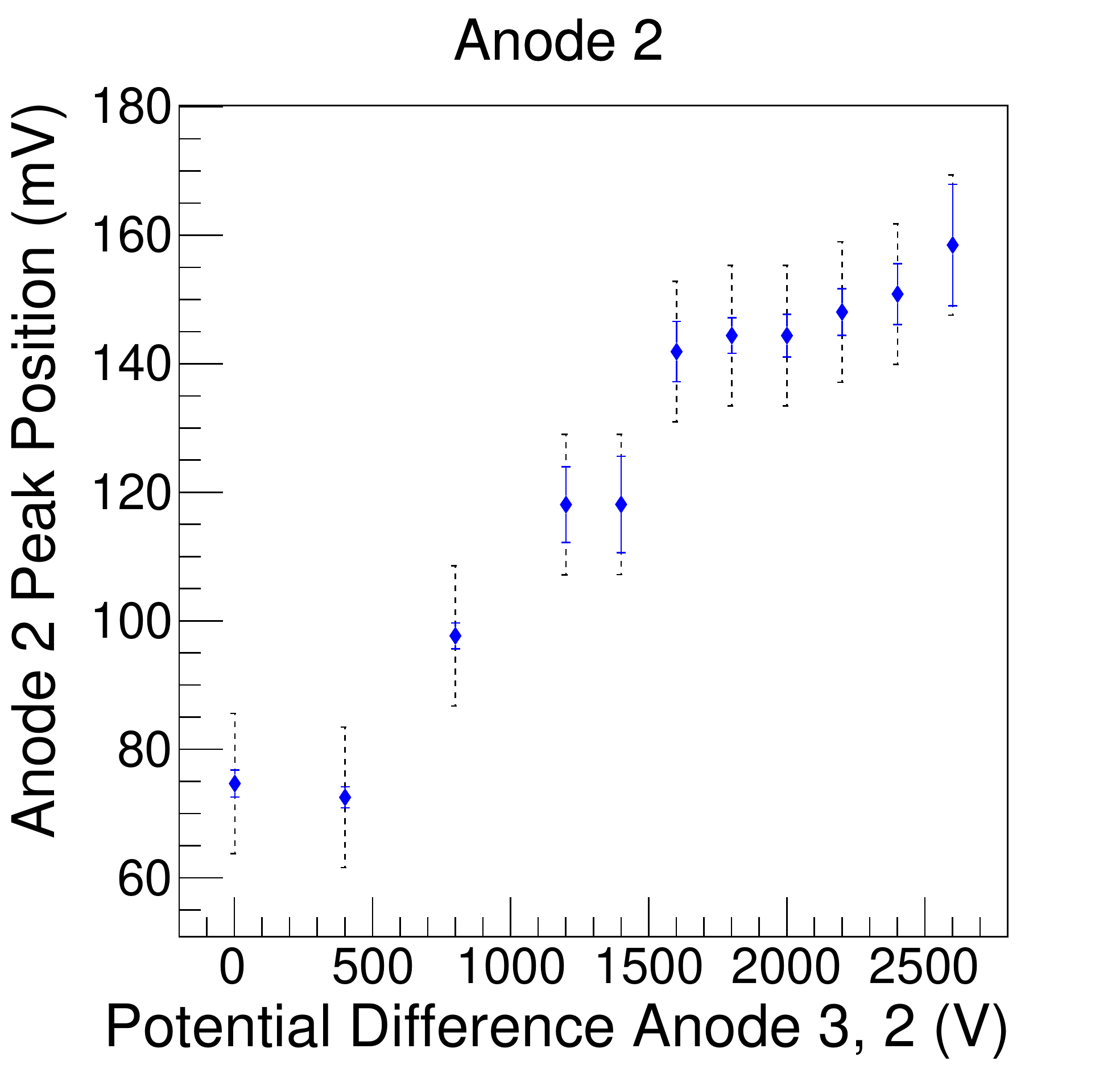}}
\subfloat[Anode 3, Scheme B]{\label{hptpcPaper:subsec:hptpcPerformance:chargeGain:fig:PeakPostionVsVoltageSchemeBAnode3}
\includegraphics[width=0.325\columnwidth, trim = 0 0 0 36, clip=true]{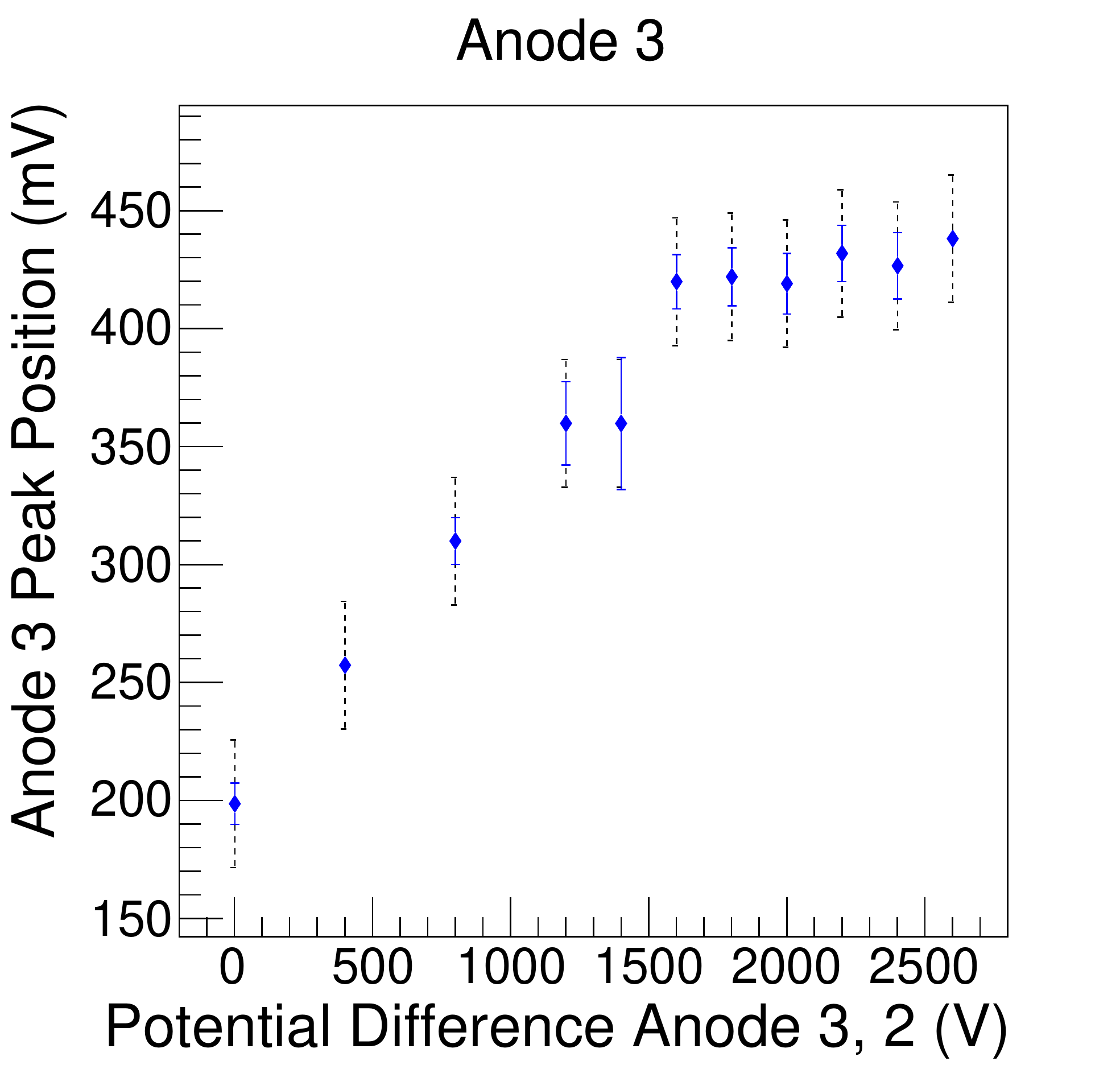}}\\
\subfloat[Anode 1, Scheme C]{\label{hptpcPaper:subsec:hptpcPerformance:chargeGain:fig:PeakPostionVsVoltageSchemeCAnode1}
\includegraphics[width=0.325\columnwidth, trim = 0 0 0 36, clip=true]{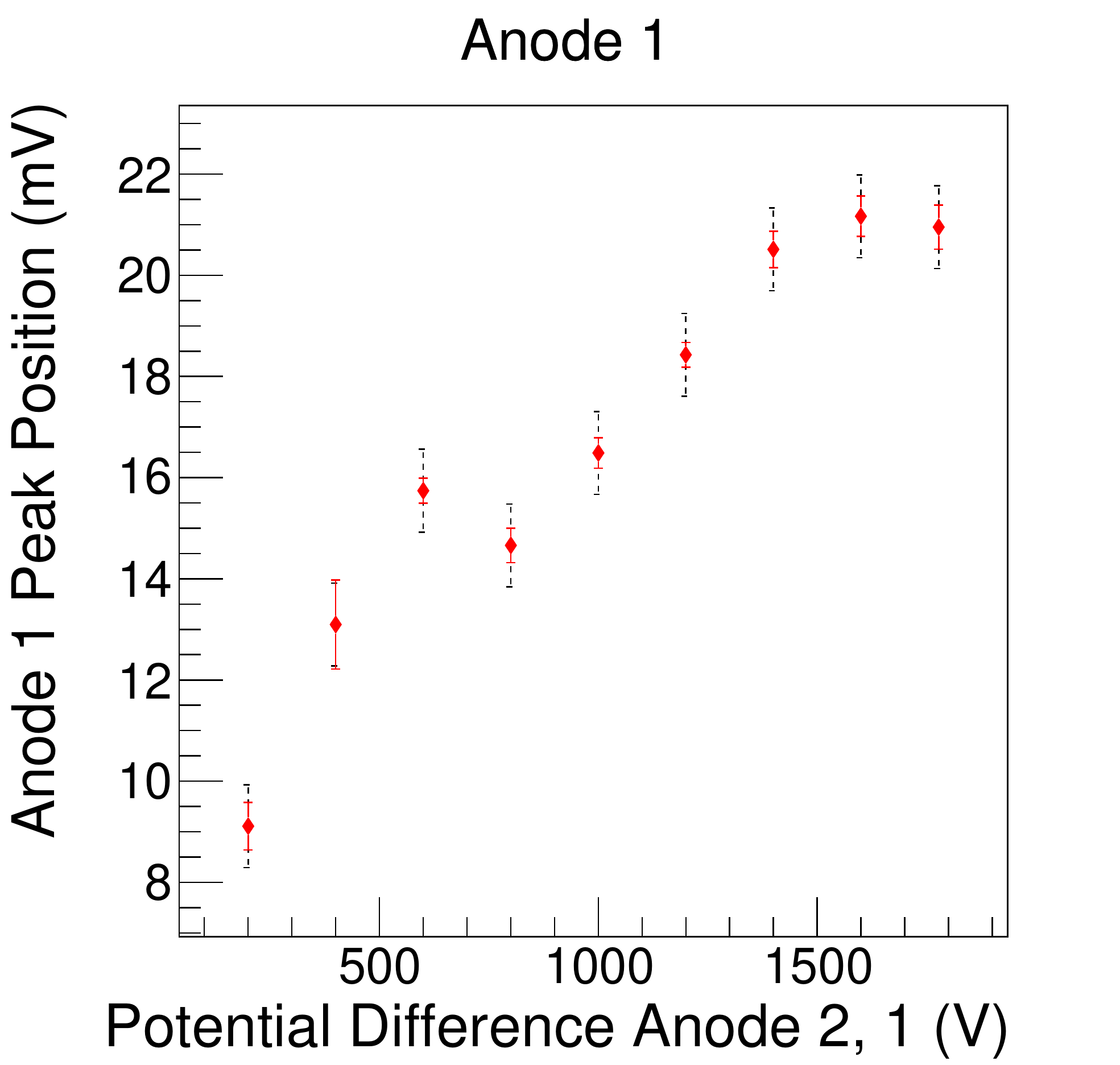}}
\subfloat[Anode 2, Scheme C]{\label{hptpcPaper:subsec:hptpcPerformance:chargeGain:fig:PeakPostionVsVoltageSchemeCAnode2}
\includegraphics[width=0.325\columnwidth, trim = 0 0 0 36, clip=true]{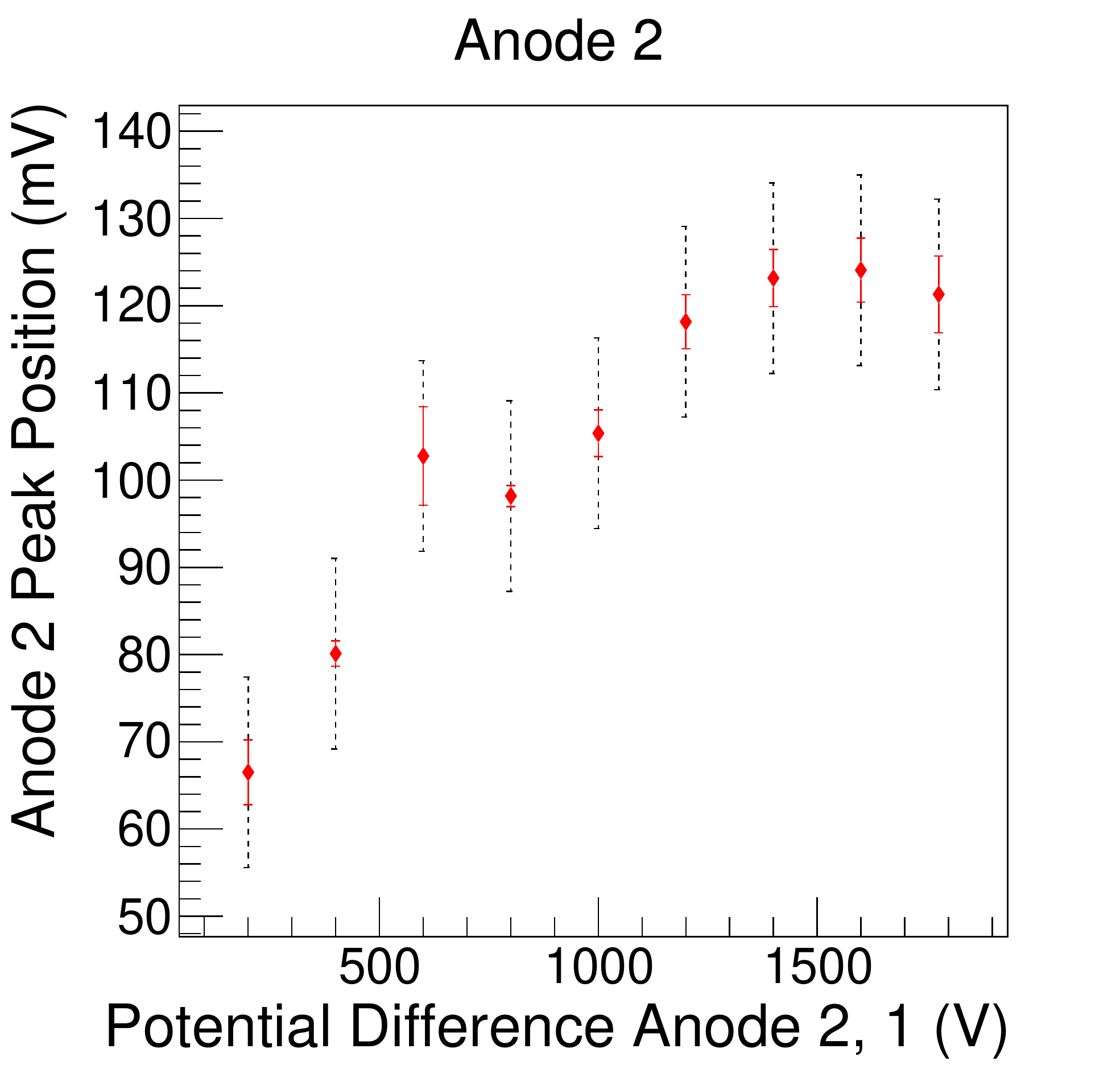}}
\subfloat[Anode 3, Scheme C]{\label{hptpcPaper:subsec:hptpcPerformance:chargeGain:fig:PeakPostionVsVoltageSchemeCAnode3}
\includegraphics[width=0.325\columnwidth, trim = 0 0 0 36, clip=true]{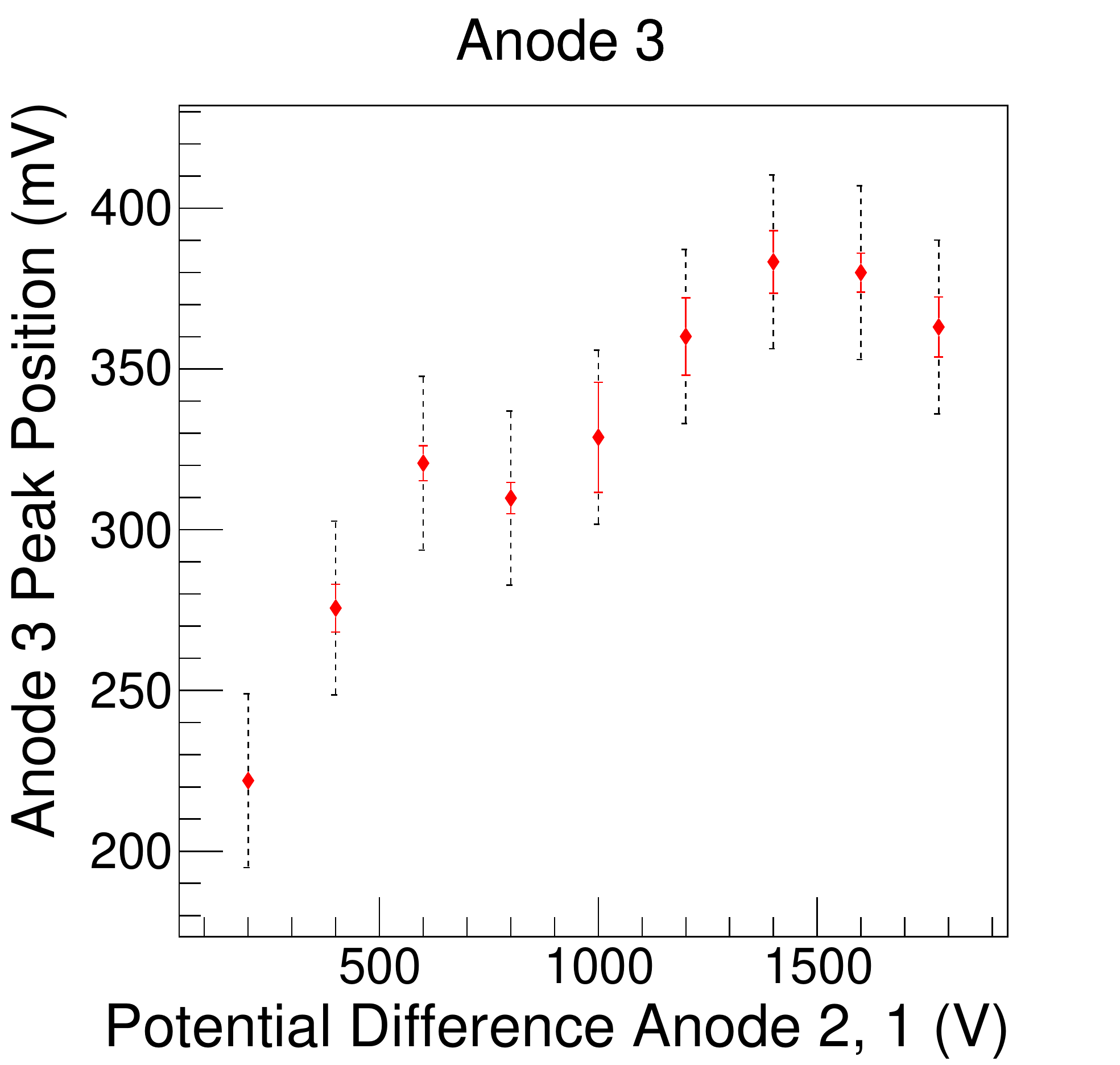}}
\caption{
\label{hptpcPaper:subsec:hptpcPerformance:chargeGain:fig:PeakPostionVsVoltageSchemes}Plots of the position of the $\alpha$-peak in the respective amplitude spectra. In the first row (\protect\subref{hptpcPaper:subsec:hptpcPerformance:chargeGain:fig:PeakPostionVsVoltageSchemeAAnode1}, \protect\subref{hptpcPaper:subsec:hptpcPerformance:chargeGain:fig:PeakPostionVsVoltageSchemeAAnode2}, and \protect\subref{hptpcPaper:subsec:hptpcPerformance:chargeGain:fig:PeakPostionVsVoltageSchemeAAnode3}) the peak position is plotted vs anode 1 voltage (Scheme A). During Scheme A, the voltages of all three anodes are increased in steps of \SI{200}{\volt} while the potential difference between anodes is kept constant. In the second row (Scheme B: \protect\subref{hptpcPaper:subsec:hptpcPerformance:chargeGain:fig:PeakPostionVsVoltageSchemeBAnode1}, \protect\subref{hptpcPaper:subsec:hptpcPerformance:chargeGain:fig:PeakPostionVsVoltageSchemeBAnode2}, and \protect\subref{hptpcPaper:subsec:hptpcPerformance:chargeGain:fig:PeakPostionVsVoltageSchemeBAnode3}) the peak position is plotted vs the potential difference between anodes 2 and 3 ($\Delta V_{a23}$). During the measurement $V_{a1}$, $V_{a2}$ and $\Delta V_{a12}$ are kept constant. Third row (Scheme C: \protect\subref{hptpcPaper:subsec:hptpcPerformance:chargeGain:fig:PeakPostionVsVoltageSchemeCAnode1}, \protect\subref{hptpcPaper:subsec:hptpcPerformance:chargeGain:fig:PeakPostionVsVoltageSchemeCAnode2}, and \protect\subref{hptpcPaper:subsec:hptpcPerformance:chargeGain:fig:PeakPostionVsVoltageSchemeCAnode3}): Peak position vs the potential difference between anodes 1 and 2 ($V_{a12}$) while $V_{a1}$ and $\Delta V_{a23}$ are kept constant. All measurements have been made in the same gas fill of \SI{3}{bar} absolute of pure argon.}
\end{figure}
where the first term is an exponential function to fit the noise, and x-ray and $\gamma$-ray background, and the second term is a Gaussian function to fit the $\alpha$-peak. The third term is a second Gaussian function which fits the higher amplitude entries of the spectra, where the spectra are shaped by cosmic muons. Examples of these fits are shown in \figref{hptpcPaper:subsec:hptpcPerformance:chargeGain:fig:FittedAmplitudeSpectra}.\\
The mean of the Gaussian fitting the $\alpha$-peak from the $^{241}\textrm{Am}$ decay, $p_3$, is extracted and taken as a measure for the mean energy deposit of the $\alpha$ particles. In \figref{hptpcPaper:subsec:hptpcPerformance:chargeGain:fig:PeakPostionVsVoltageSchemes} the $\alpha$-peak position is plotted against the varied voltage in the respective voltage scheme. The peak position uncertainty shown in the plots are the fit uncertainties on the mean of the Gaussian, scaled by the $\chi^{2}/N_{\text{dof}}$ of the fit, for fits where $\chi^{2}/N_{\text{dof}} > 1$.

For Scheme A the peak position is plotted against the voltage of anode 1 (\figrefbra{hptpcPaper:subsec:hptpcPerformance:chargeGain:fig:PeakPostionVsVoltageSchemes}, first row), for Scheme B the peak position is plotted against the potential difference between anodes 2 and 3 (\figrefbra{hptpcPaper:subsec:hptpcPerformance:chargeGain:fig:PeakPostionVsVoltageSchemes}, second row), and for Scheme C the peak position is plotted against the potential difference between anodes 1 and 2 (\figrefbra{hptpcPaper:subsec:hptpcPerformance:chargeGain:fig:PeakPostionVsVoltageSchemes}, third row).

\paragraph{\textbf{Gas gain against voltage for the three voltage schemes}}

\begin{figure}
\centering
\subfloat[Anode 1, Scheme A]{\label{hptpcPaper:subsec:hptpcPerformance:chargeGain:fig:GasGainVsVoltageSchemeAAnode1}
\includegraphics[width=0.325\columnwidth, trim = 0 0 0 36, clip=true]{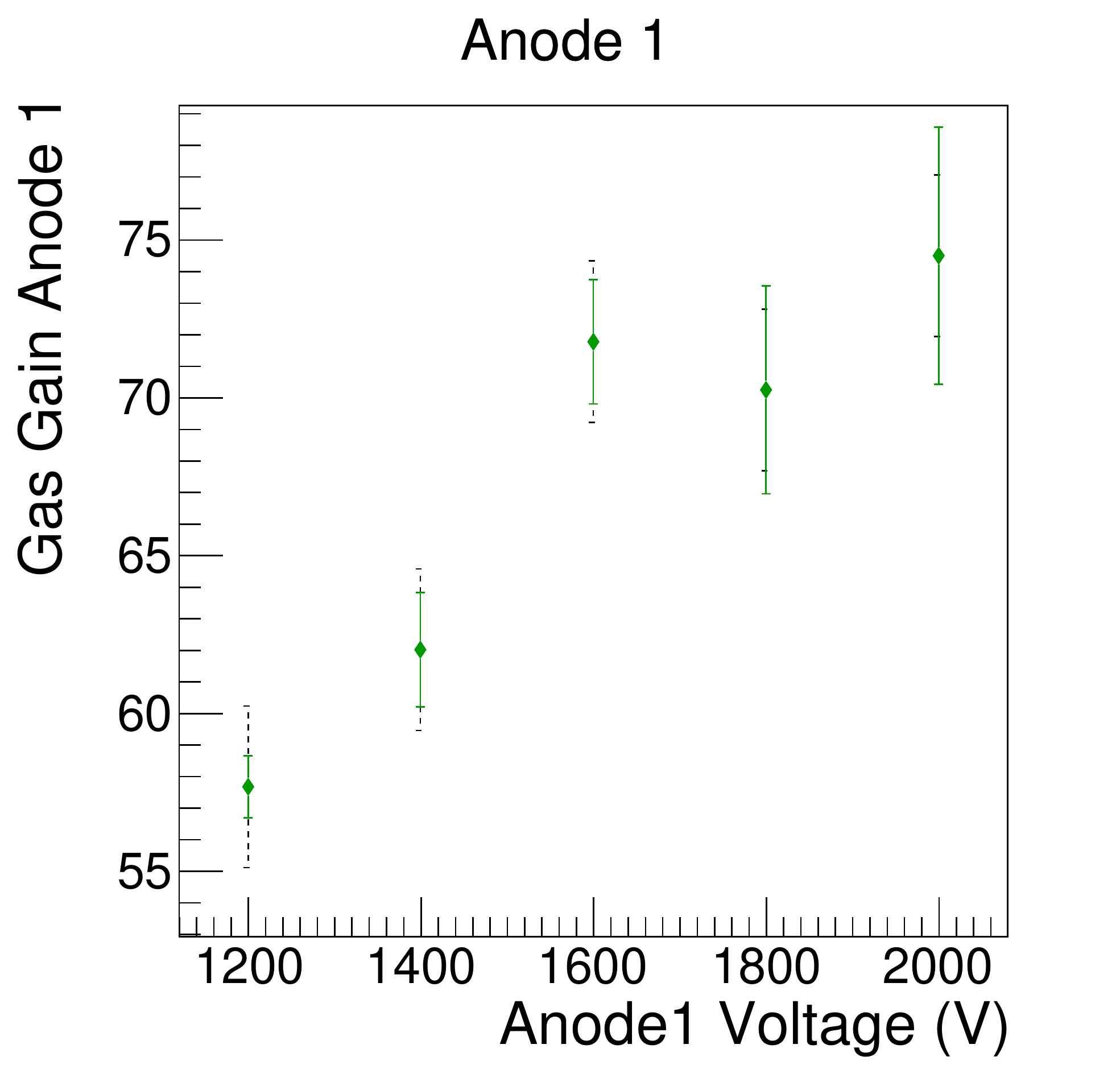}}
\subfloat[Anode 2, Scheme A]{\label{hptpcPaper:subsec:hptpcPerformance:chargeGain:fig:GasGainVsVoltageSchemeAAnode2}
\includegraphics[width=0.325\columnwidth, trim = 0 0 0 36, clip=true]{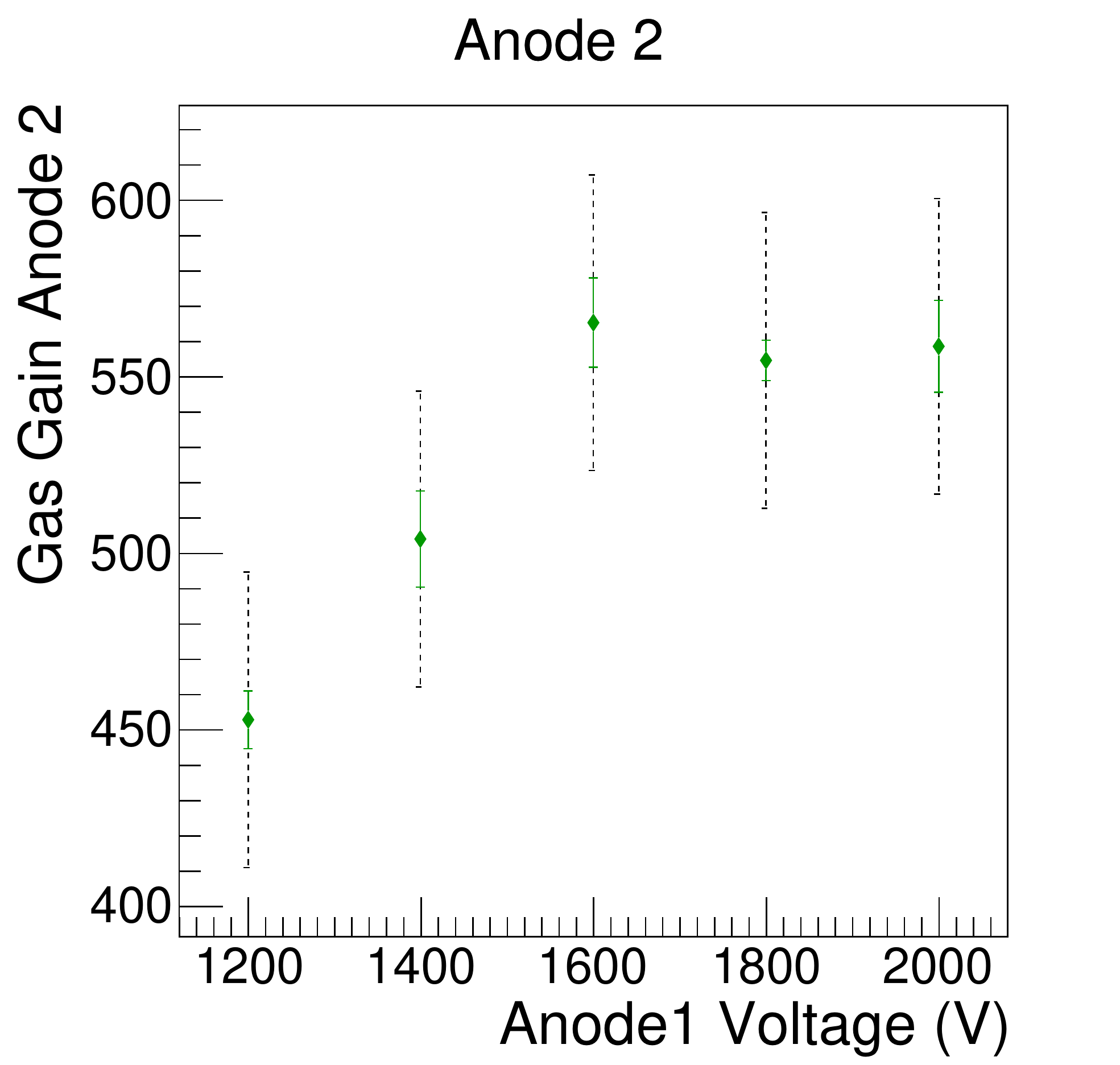}}
\subfloat[Anode 3, Scheme A]{\label{hptpcPaper:subsec:hptpcPerformance:chargeGain:fig:GasGainVsVoltageSchemeAAnode3}
\includegraphics[width=0.325\columnwidth, trim = 0 0 0 36, clip=true]{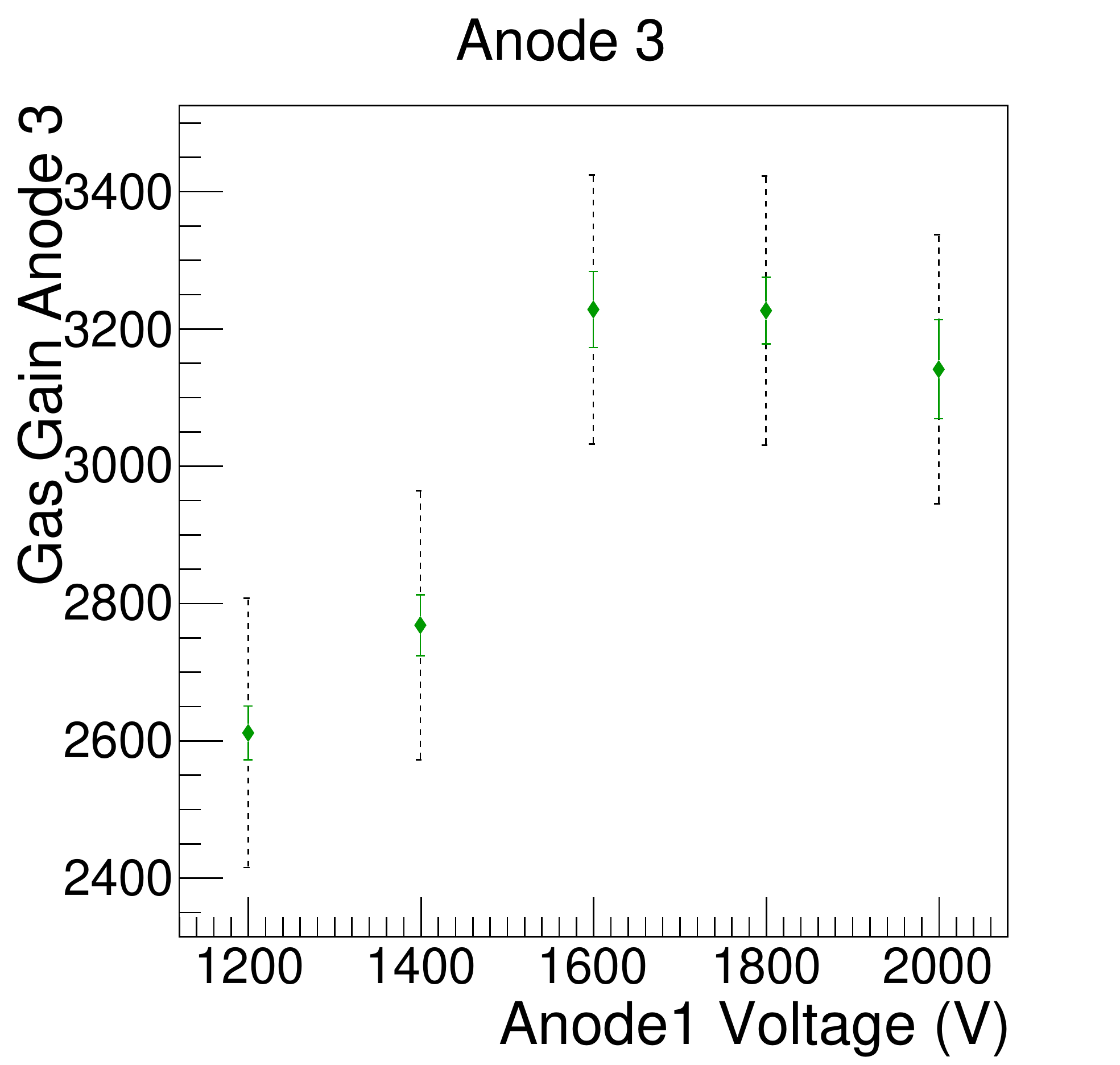}}\\
\subfloat[Anode 1, Scheme B]{\label{hptpcPaper:subsec:hptpcPerformance:chargeGain:fig:GasGainVsVoltageSchemeBAnode1}
\includegraphics[width=0.325\columnwidth, trim = 0 0 0 36, clip=true]{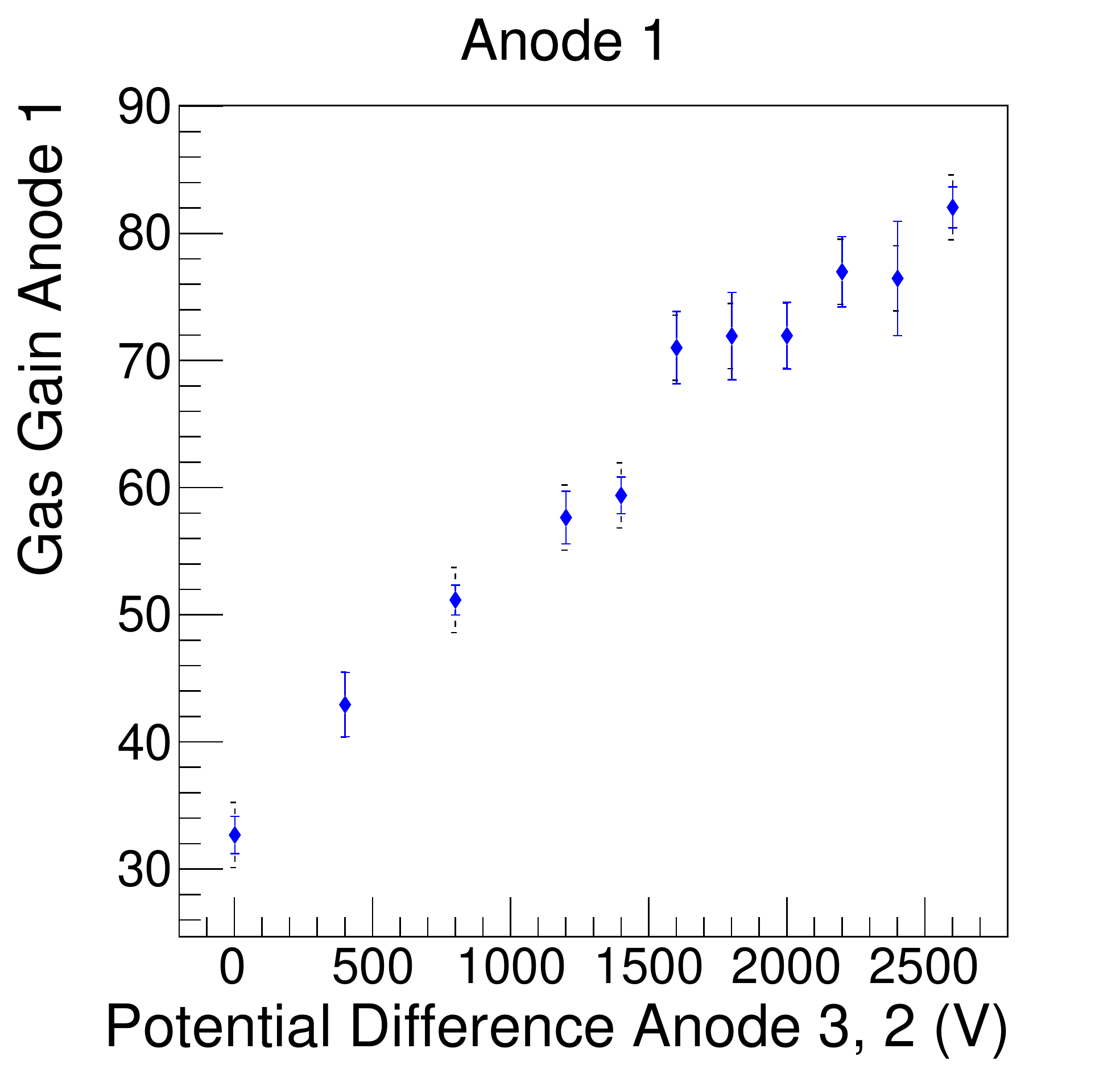}}
\subfloat[Anode 2, Scheme B]{\label{hptpcPaper:subsec:hptpcPerformance:chargeGain:fig:GasGainVsVoltageSchemeBAnode2}
\includegraphics[width=0.325\columnwidth, trim = 0 0 0 36, clip=true]{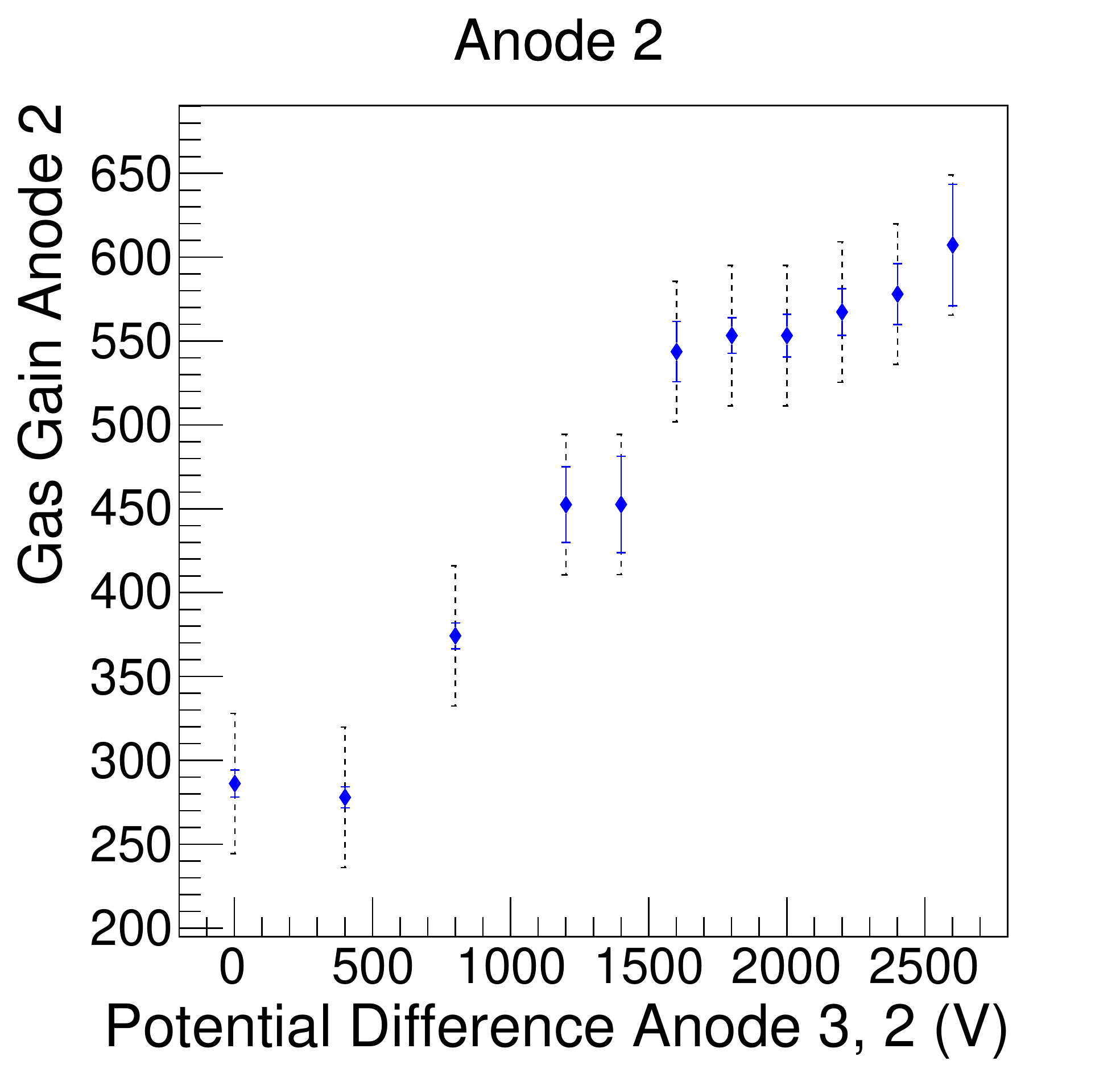}}
\subfloat[Anode 3, Scheme B]{\label{hptpcPaper:subsec:hptpcPerformance:chargeGain:fig:GasGainVsVoltageSchemeBAnode3}
\includegraphics[width=0.325\columnwidth, trim = 0 0 0 36, clip=true]{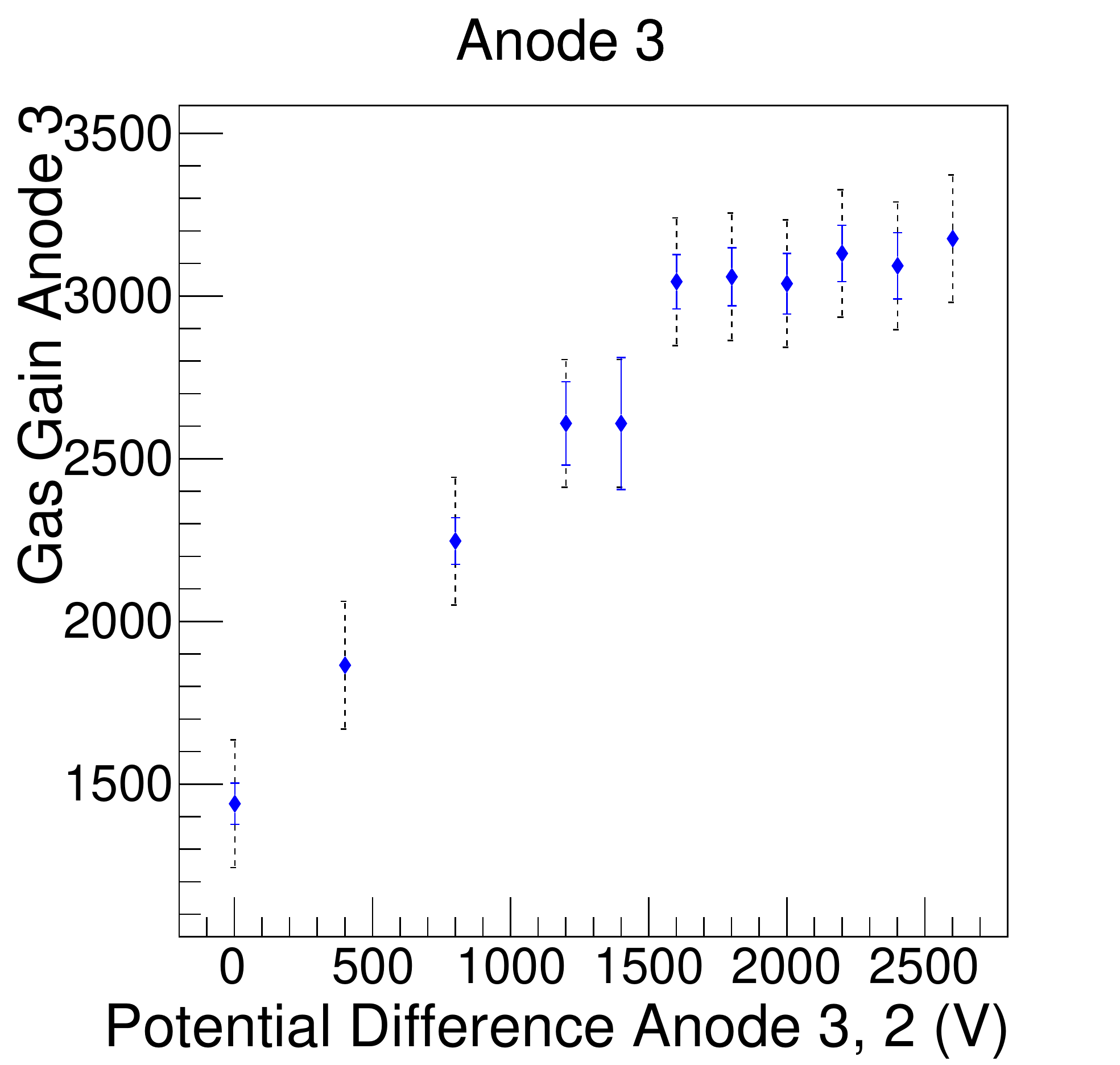}}\\
\subfloat[Anode 1, Scheme C]{\label{hptpcPaper:subsec:hptpcPerformance:chargeGain:fig:GasGainVsVoltageSchemeCAnode1}
\includegraphics[width=0.325\columnwidth, trim = 0 0 0 36, clip=true]{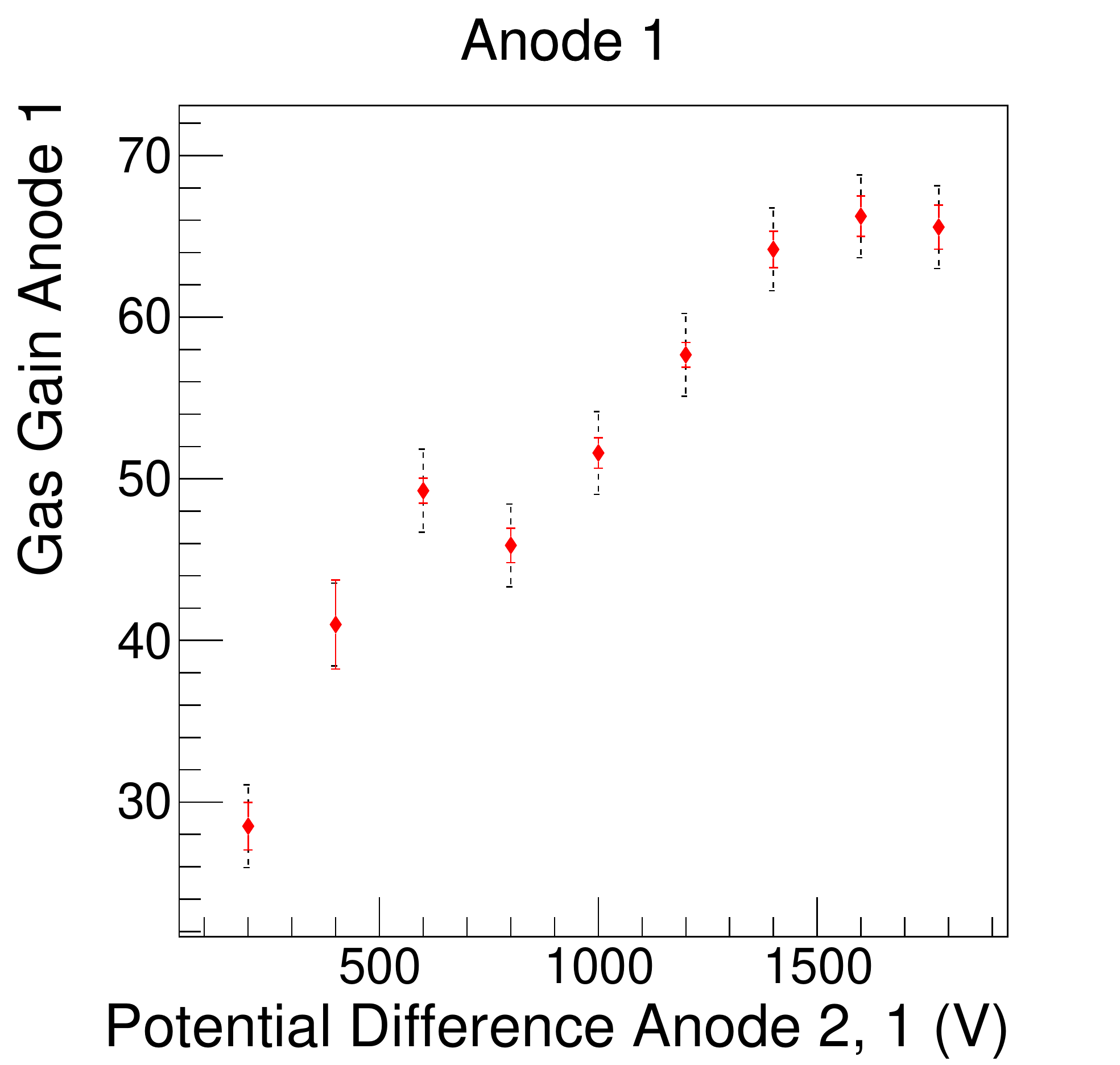}}
\subfloat[Anode 2, Scheme C]{\label{hptpcPaper:subsec:hptpcPerformance:chargeGain:fig:GasGainVsVoltageSchemeCAnode2}
\includegraphics[width=0.325\columnwidth, trim = 0 0 0 36, clip=true]{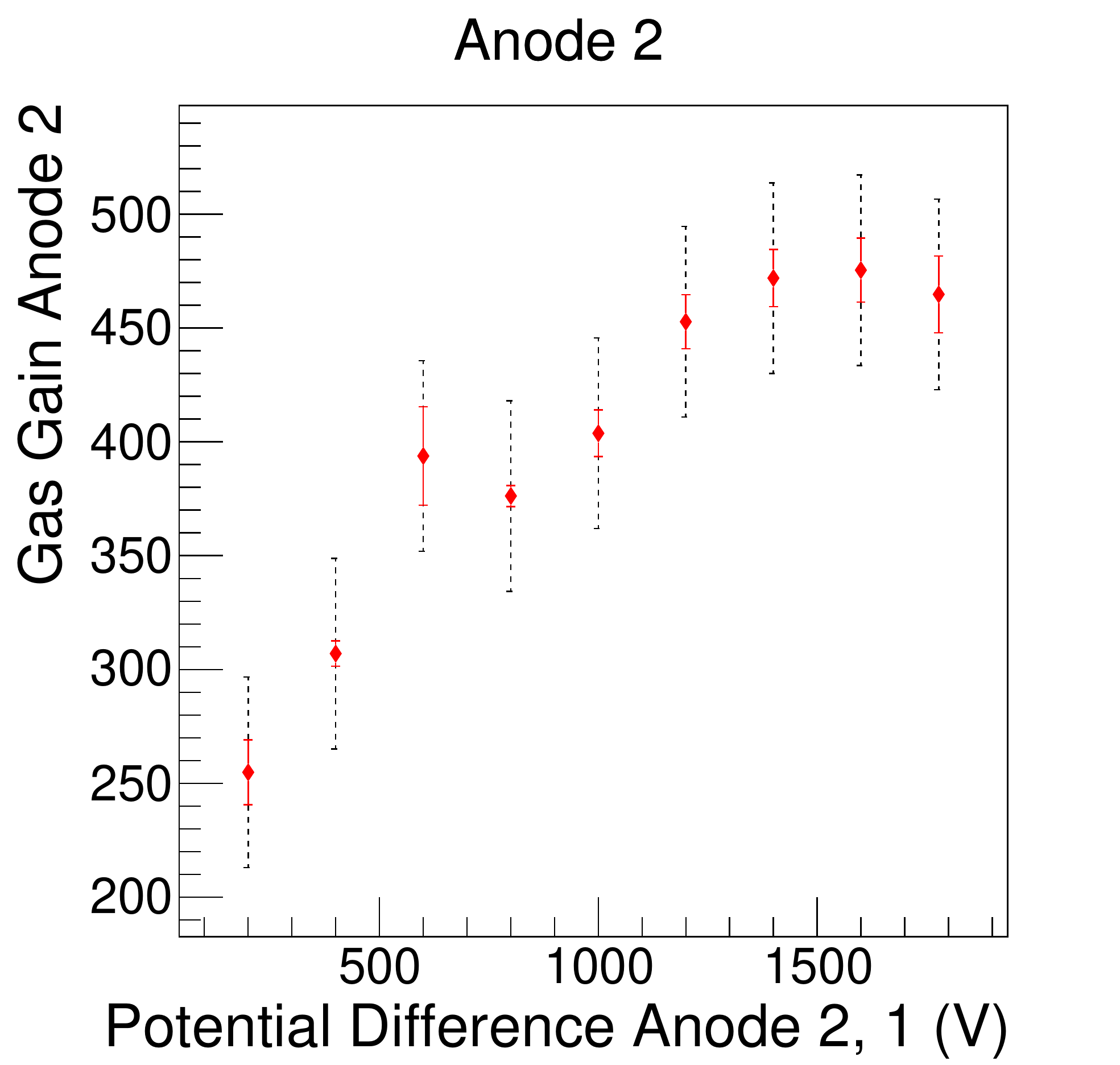}}
\subfloat[Anode 3, Scheme C]{\label{hptpcPaper:subsec:hptpcPerformance:chargeGain:fig:GasGainVsVoltageSchemeCAnode3}
\includegraphics[width=0.325\columnwidth, trim = 0 0 0 36, clip=true]{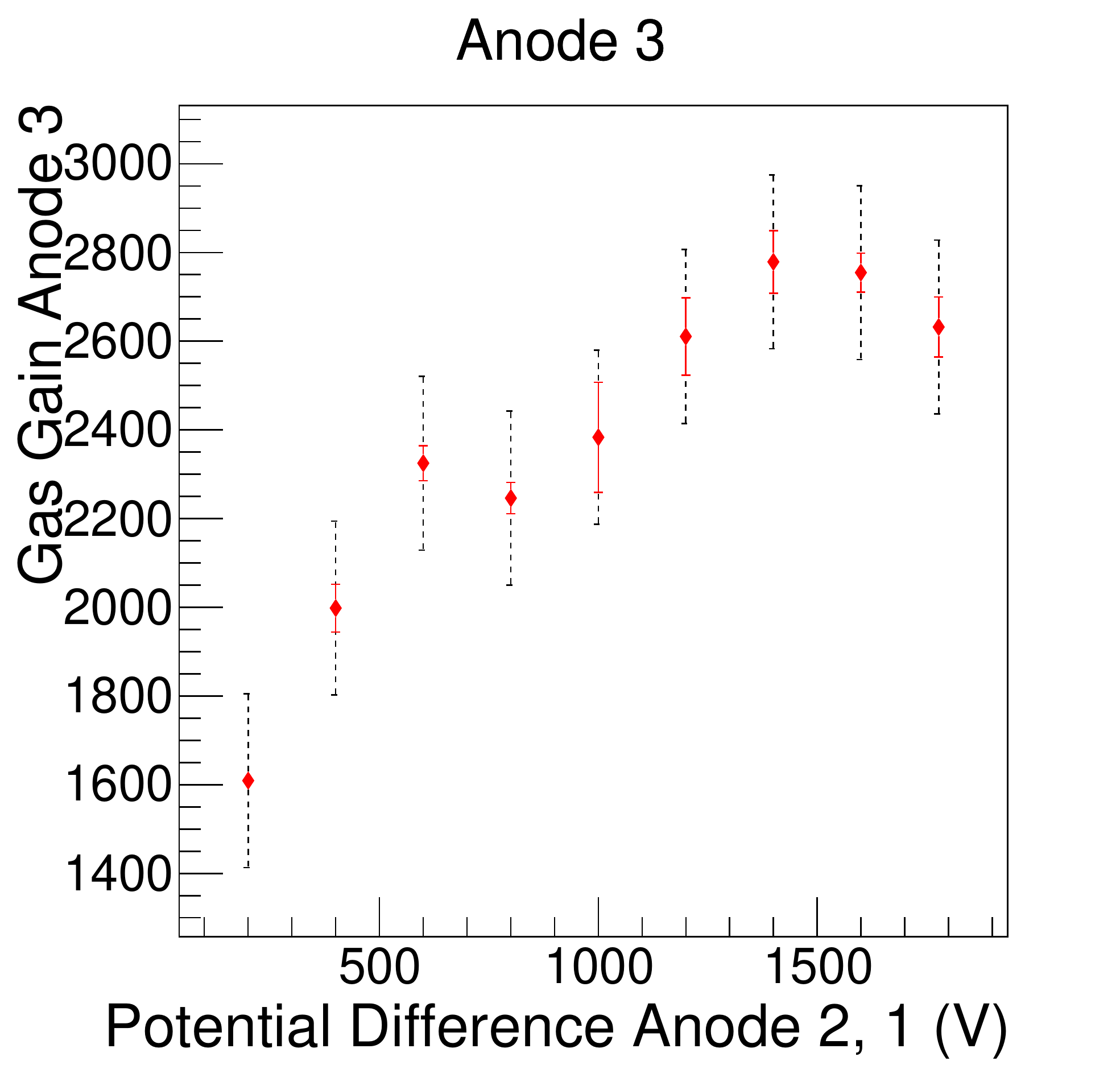}}
\caption{\label{hptpcPaper:subsec:hptpcPerformance:chargeGain:fig:GasGainVsVoltageSchemes}Plots of the calculated gas gain vs either anode voltage or inter-anode voltage difference. The gain is calculated from the data shown in the respective plot in \figref{hptpcPaper:subsec:hptpcPerformance:chargeGain:fig:PeakPostionVsVoltageSchemes}. First row (\protect\subref{hptpcPaper:subsec:hptpcPerformance:chargeGain:fig:PeakPostionVsVoltageSchemeAAnode1}, \protect\subref{hptpcPaper:subsec:hptpcPerformance:chargeGain:fig:PeakPostionVsVoltageSchemeAAnode2}, and \protect\subref{hptpcPaper:subsec:hptpcPerformance:chargeGain:fig:PeakPostionVsVoltageSchemeAAnode3}): Scheme A, gain vs anode 1 voltage ($V_{a1}$), $V_{a1}$, $V_{a2}$ and $V_{a3}$ are increased by the same amount whilst $\Delta V_{a12} = \Delta V_{a23} = \SI{1200}{\volt}$. Second row (\protect\subref{hptpcPaper:subsec:hptpcPerformance:chargeGain:fig:PeakPostionVsVoltageSchemeBAnode1}, \protect\subref{hptpcPaper:subsec:hptpcPerformance:chargeGain:fig:PeakPostionVsVoltageSchemeBAnode2}, and\protect\subref{hptpcPaper:subsec:hptpcPerformance:chargeGain:fig:PeakPostionVsVoltageSchemeBAnode3}): Scheme B, gain vs the voltage difference between anode 2 and 3 ($\Delta V_{a23}$), $V_{a3}$ and $\Delta V_{a23}$ are increased whilst $V_{a1}=\SI{1200}{\volt}$ and $V_{a2}=\SI{2400}{\volt}$. Third row (\protect\subref{hptpcPaper:subsec:hptpcPerformance:chargeGain:fig:PeakPostionVsVoltageSchemeCAnode1}, \protect\subref{hptpcPaper:subsec:hptpcPerformance:chargeGain:fig:PeakPostionVsVoltageSchemeCAnode2}, and \protect\subref{hptpcPaper:subsec:hptpcPerformance:chargeGain:fig:PeakPostionVsVoltageSchemeCAnode3}): Scheme C, gain vs the anode 1 to anode 2 voltage differences ($\Delta V_{a12}$), $V_{a2}$, $\Delta V_{a12}$ and $V_{a3}$ are increased whilst keeping $V_{a1}$ and $\Delta V_{a23}$ constant. All data has been taken in the same gas fill of \SI{3}{bar} absolute of pure argon.}
\end{figure}

Before calculating the gas gain for the three voltage schemes A, B and C from the values in the amplitude spectra, gas quality degradation needs to be considered. Degrading gas quality can have a significant effect on the gain measurements, and so we took data at identical gas, pressure and bias voltage settings every 24 hours to obtain calibration correction as the data used in this analysis was taken over three days.\\
We reconstruct the peak position in the amplitude spectra of these calibration runs. After the gas quality calibration is fit to these data points vs. the measurement time. The correction function is $ y(\text{time}) = m \cdot \text{time} + b$, where the values of ($m,b$) are (\SI{0.8(2)}{\milli\volt\per day}, -\SI{1(5)}{\milli\volt}), (\SI{11(2)}{\milli\volt\per day}, -\SI{135(42)}{\milli\volt}) and (\SI{27(7)}{\milli\volt\per day}, -\SI{277(141)}{\milli\volt}) for anode 1, anode 2, and anode 3 spectra respectively, and the calibration is normalized such that the non-calibrated data and the calibrated data have the same value at the beginning of Scheme C. We observe a drift in the peak position as can be seen from the $(m,b)$ pairs, however, the drift is such that no change could been observed when examining the amplitude spectra for each run in a voltage setting individually.\\
A systematic uncertainty contribution is assessed to account for this effect, represented by the dotted error bars in \figref{hptpcPaper:subsec:hptpcPerformance:chargeGain:fig:PeakPostionVsVoltageSchemes}. This contribution takes the expected peak position shift over the measurement time in each voltage scheme into account and is calculated as the standard deviation of the measured peak positions with respect to the peak position after correction.\\
Equation \eqref{hptpcPaper:sec:dataAnalysis:eq:gain} allows now to calculate the gas amplification factor of the amplification region, $G_{\text{amp}}$, using
\begin{align}
G_{\text{amp}} &= \frac{A}{f \cdot G_{\text{preamp}} \cdot Q_{\text{e}}} \label{hptpcPaper:subsec:hptpcPerformance:chargeGain:eqn:gas_gain} \quad \text{where} \\
Q_{\text{e}} &= \frac{\langle \varepsilon_{\alpha}\rangle}{W} \cdot \SI{1.6022E-19}{\coulomb} \quad .
\label{hptpcPaper:subsec:hptpcPerformance:chargeGain:eqn:q_meas}
\end{align}
We calculate $G_{\text{amp}}$ for the amplitude spectra measured at each mesh.
In the calculation we use the best-fit peak position of the $\alpha$-peak in the amplitude spectra to be $A$, corrected by the calibration procedure described above. \Figref{hptpcPaper:subsec:hptpcPerformance:chargeGain:fig:GasGainVsVoltageSchemes} shows the gas gain at each anode plotted against the respective voltage in the three voltage schemes.
\begin{table}
\centering
\begin{tabular}{c|c|c|c|c}
Scheme          & \multicolumn{2}{c}{Voltage setting (A1 / A2 / A3) [V]}   \vline  &  \multicolumn{2}{c}{Gas gain at anode 3 at voltage setting} \\ 
                & Lowest                & Highest               & at lowest setting                      & at highest setting       \\ \hline
A               & 1200 / 2400 / 3600    & 2000 / 3200 / 4400    & (2.61 $\pm$ 0.20) $\times$ $\num{e3}$  &  (3.14 $\pm$ 0.20) $\times$ $\num{e3}$ \\
B               & 1200 / 2400 / 2400    & 1200 / 3400 / 5000    & (1.44 $\pm$ 0.20) $\times$ $\num{e3}$  &  (3.18 $\pm$ 0.20) $\times$ $\num{e3}$ \\
C               & 1200 / 1400 / 2600    & 1200 / 3000 / 4200    & (1.61 $\pm$ 0.20) $\times$ $\num{e3}$  &  (2.63 $\pm$ 0.20) $\times$ $\num{e3}$ \\
\end{tabular}
\caption{\label{hptpcPaper:subsec:hptpcPerformance:chargeGain:tab:GainsOfEachScheme}{The charge gain measured at the highest at lowest voltage settings of each voltage scheme.}}
\end{table}
The goal of this analysis is to determine the dependence of the gain on the absolute voltages of the anodes ($V_{a1}$, $V_{a2}$, and $V_{a3}$) and on the potential differences between the anodes ($\Delta V_{a12}$ and $\Delta V_{a23}$). The results of the charge gain measurement for schemes A, B and C are shown in \figref{hptpcPaper:subsec:hptpcPerformance:chargeGain:fig:GasGainVsVoltageSchemes} in the first, second and third row, respectively, and the gas gains measured at the highest and lowest voltage settings for each scheme are presented in \tabref{hptpcPaper:subsec:hptpcPerformance:chargeGain:tab:GainsOfEachScheme}. The voltage range covered during the three schemes has been optimised for the light analysis, to the end that i) all voltage settings of the three schemes could be taken in one gas fill without the degradation of the gas fill, ii) to avoid to reach a voltage regime where sparking occurs, and iii) to have sufficient overlap between the three voltage schemes. As a result of this our study of the charge gain of the amplification region covers only a small gain range (\tabrefbra{hptpcPaper:subsec:hptpcPerformance:chargeGain:tab:GainsOfEachScheme} and \figrefbra{hptpcPaper:subsec:hptpcPerformance:chargeGain:fig:GasGainVsVoltageSchemes}).\\
For all three voltage schemes the measured gas gain increases from anode 1, to anode 2, to anode 3, as is expected from a cascade of amplification stages. The gas amplification factor in Scheme C is overall the lowest. Examining the multiplication factor between different meshes we find $G_{\text{amp}}^{\text{mesh}2}\sim 8 \cdot G_{\text{amp}}^{\text{mesh}1}$ ($G_{\text{amp}}^{\text{mesh}2}\sim 6.5 \cdot G_{\text{amp}}^{\text{mesh}1}$) and $G_{\text{amp}}^{\text{mesh}3}\sim 5.5 \cdot G_{\text{amp}}^{\text{mesh}2}$ ($G_{\text{amp}}^{\text{mesh}3}\sim 6 \cdot G_{\text{amp}}^{\text{mesh}2}$) in scheme A and Scheme B (Scheme C). The highest contribution to the combined gas gain $G_{\text{amp}}^{\text{mesh}1} \cdot G_{\text{amp}}^{\text{mesh}2} \cdot G_{\text{amp}}^{\text{mesh}3}$ is thus the contribution of the anode 1 mesh. The dependence of the gain on the various voltages shows a similar functional shape as the light gain reported in \secref{hptpcPaper:subsec:hptpcPerfomrance:lightGain:subsubsec:results:result}, \figref{hptpcPaper:sec:opticalgain:fig:schemeABC_source}.
Due to relatively large uncertainties, the results in Scheme A are consistent with either a slight dependence or no dependence of the gain on the absolute voltages of the anodes while $\Delta V_{a12}$ and $\Delta V_{a23}$ are fixed at \SI{1200}{\volt} and are therefore consistent with the conclusions of the light gain analysis. The results of schemes B and C are consistent with a positive correlation of gain on $\Delta V_{a23}$ and $\Delta V_{a12}$, this is again consistent with the conclusions drawn from the light gain analysis. The results of the charge gain analysis supports the conclusions of the light gain analysis, that the amplification is primarily driven by the electric field between the anodes.

%% file: sec_7_combinedCCDandChargeAnalysis.tex
\section{Combined Optical and Charge Readout Analysis}
\label{hptpcPaper:sec:ChargeLight}

\begin{figure}
\centering
\subfloat[]{\label{hptpcPaper:sec:ChargeLight:fig:schemeA_ChargeLight}
\includegraphics[width=0.49\columnwidth, trim = 0 0 0 0, clip=true]{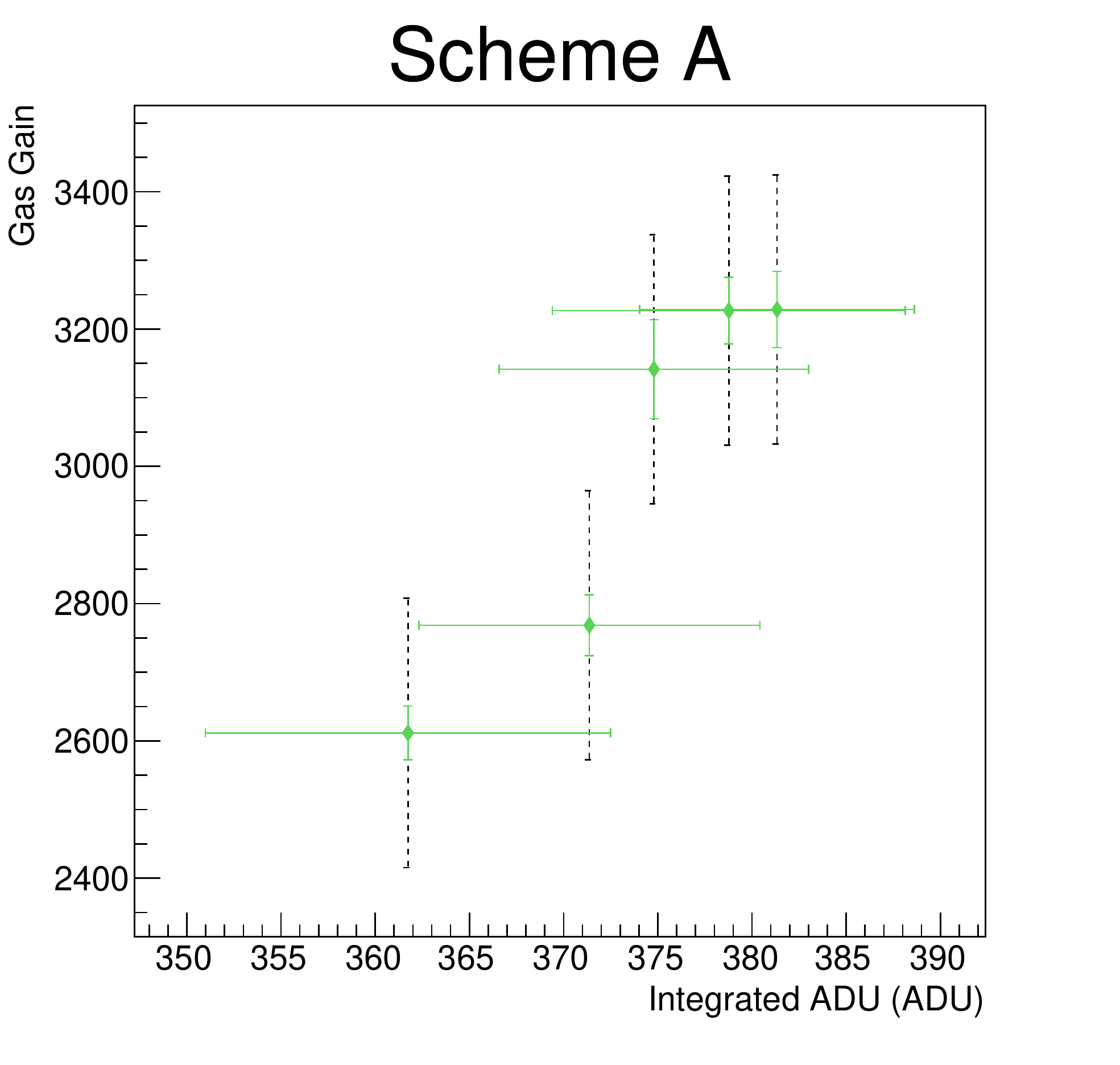}}
\subfloat[]{\label{hptpcPaper:sec:ChargeLight:fig:schemeB_ChargeLight}
\includegraphics[width=0.49\columnwidth, trim = 0 0 0 0, clip=true]{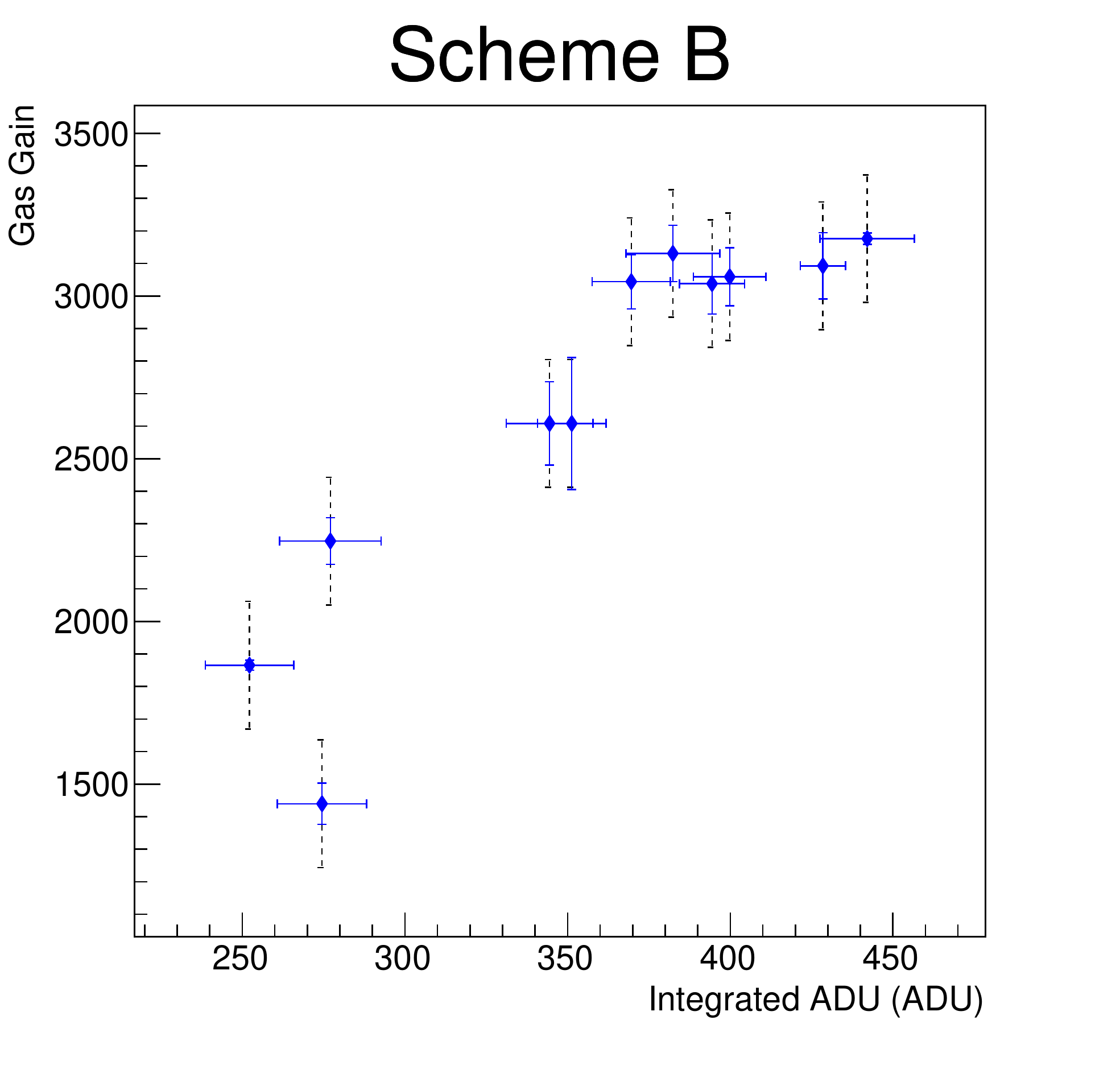}}\\ 
\subfloat[]{\label{hptpcPaper:sec:ChargeLight:fig:schemeC_ChargeLight}
\includegraphics[width=0.49\columnwidth, trim = 0 0 0 0, clip=true]{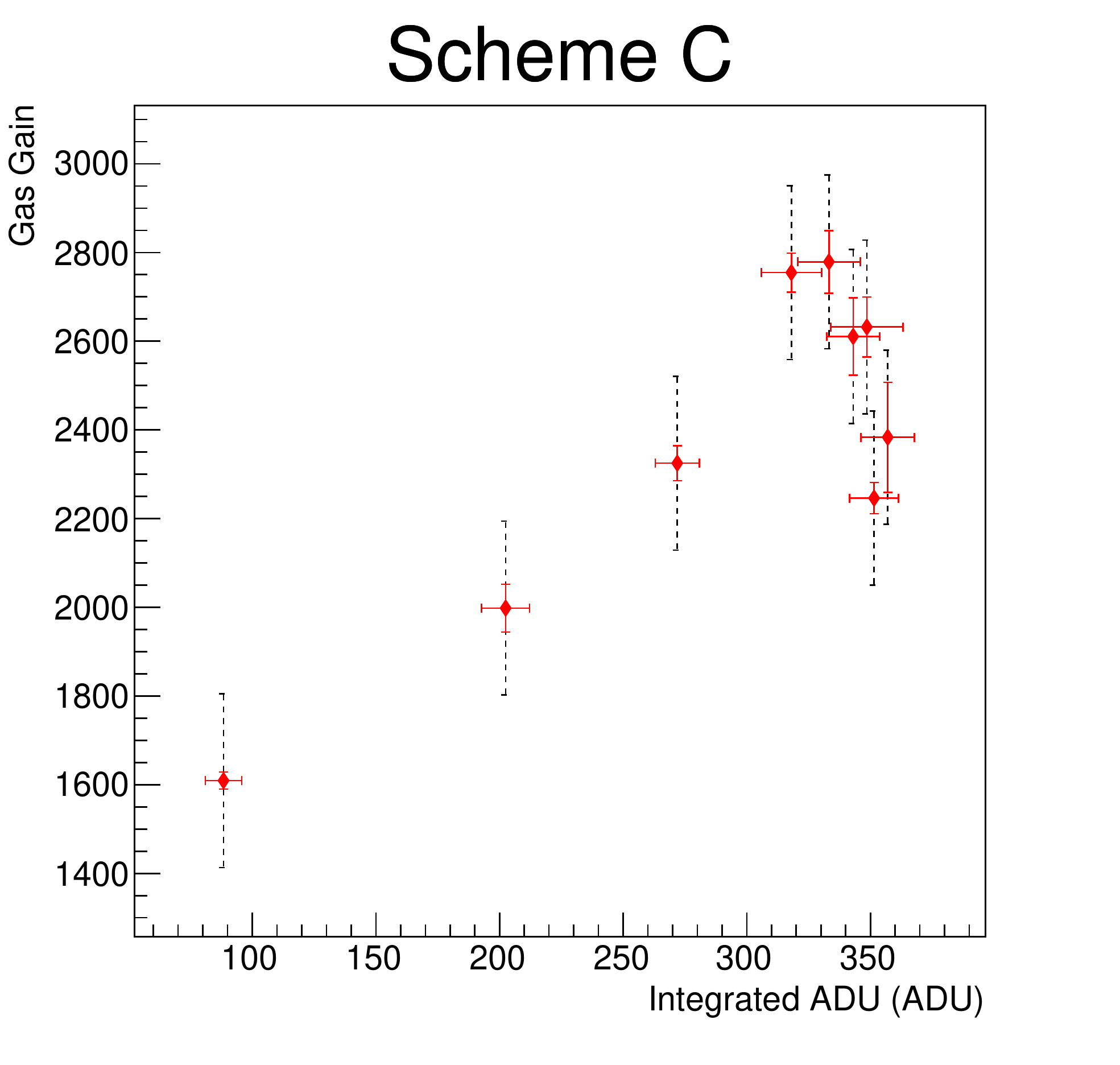}}
\caption{\label{hptpcPaper:sec:ChargeLight:fig:schemeABC_ChargeLight}Measured light intensity (Integrated \si{ADU}) (\figref{hptpcPaper:sec:opticalgain:fig:schemeABC_source}) plotted against the gas gain measured in the charge readout on anode 3 (\figref{hptpcPaper:subsec:hptpcPerformance:chargeGain:fig:GasGainVsVoltageSchemes}, right column) for Schemes A, B and C.}
\end{figure}
\begin{figure}
\centering
\subfloat[]{\label{hptpcPaper:sec:ChargeLight:fig:schemeA_ChargeLight_Ratio}
\includegraphics[width=0.49\columnwidth, trim = 0 0 0 0, clip=true]{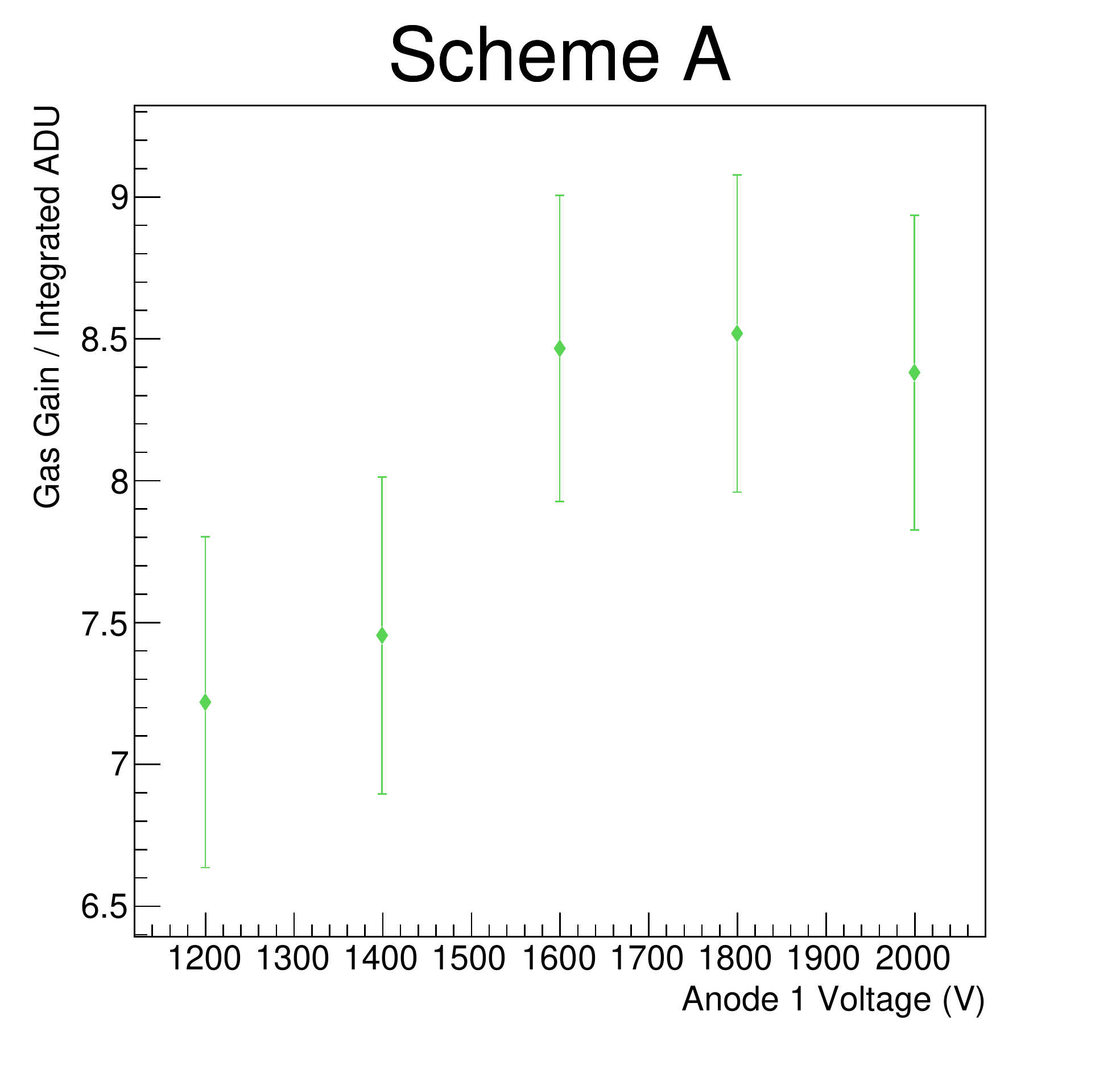}}
\subfloat[]{\label{hptpcPaper:sec:ChargeLight:fig:schemeB_ChargeLight_Ratio}
\includegraphics[width=0.49\columnwidth, trim = 0 0 0 0, clip=true]{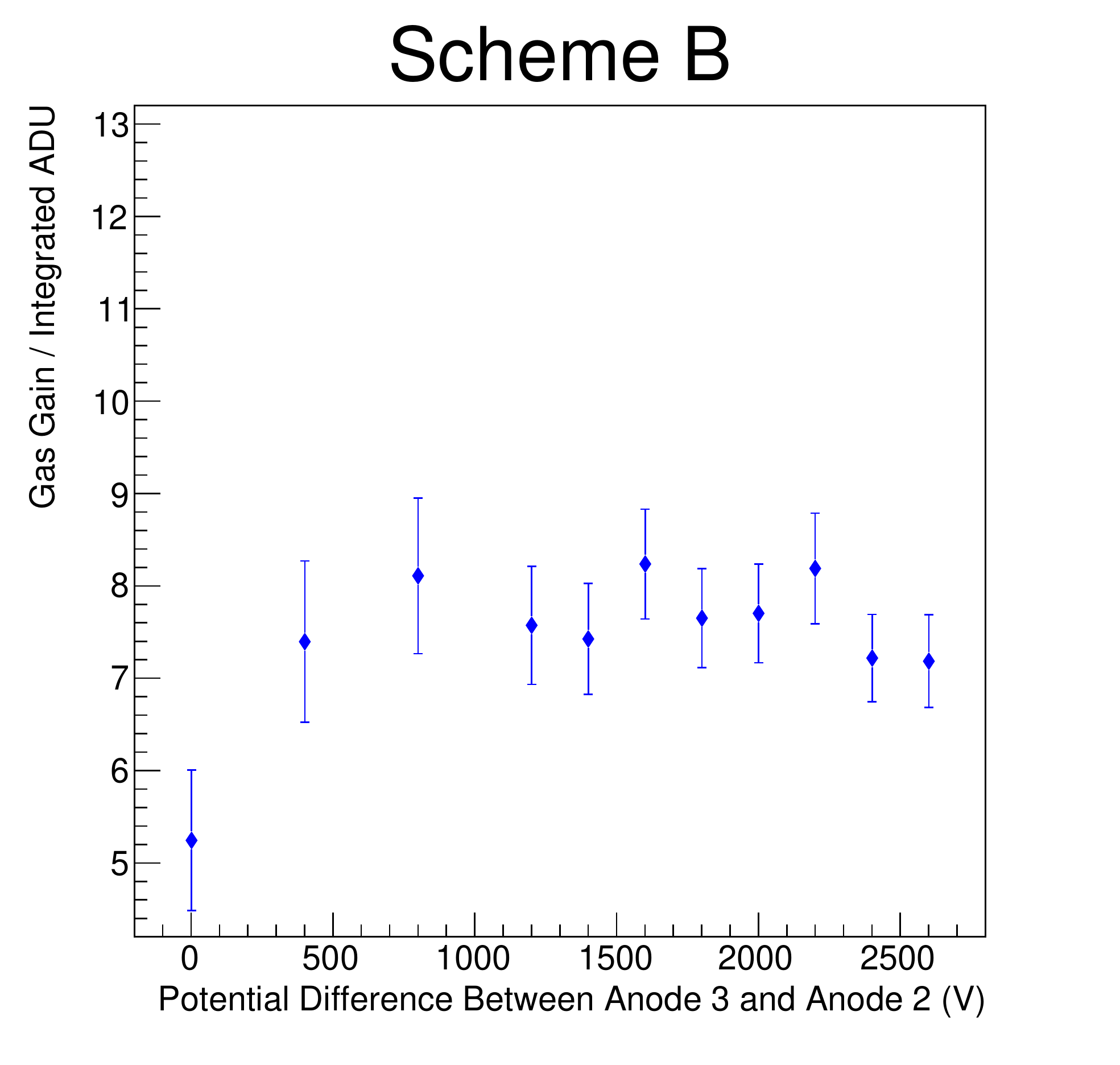}}\\ 
\subfloat[]{\label{hptpcPaper:sec:ChargeLight:fig:schemeC_ChargeLight_Ratio}
\includegraphics[width=0.49\columnwidth, trim = 0 0 0 0, clip=true]{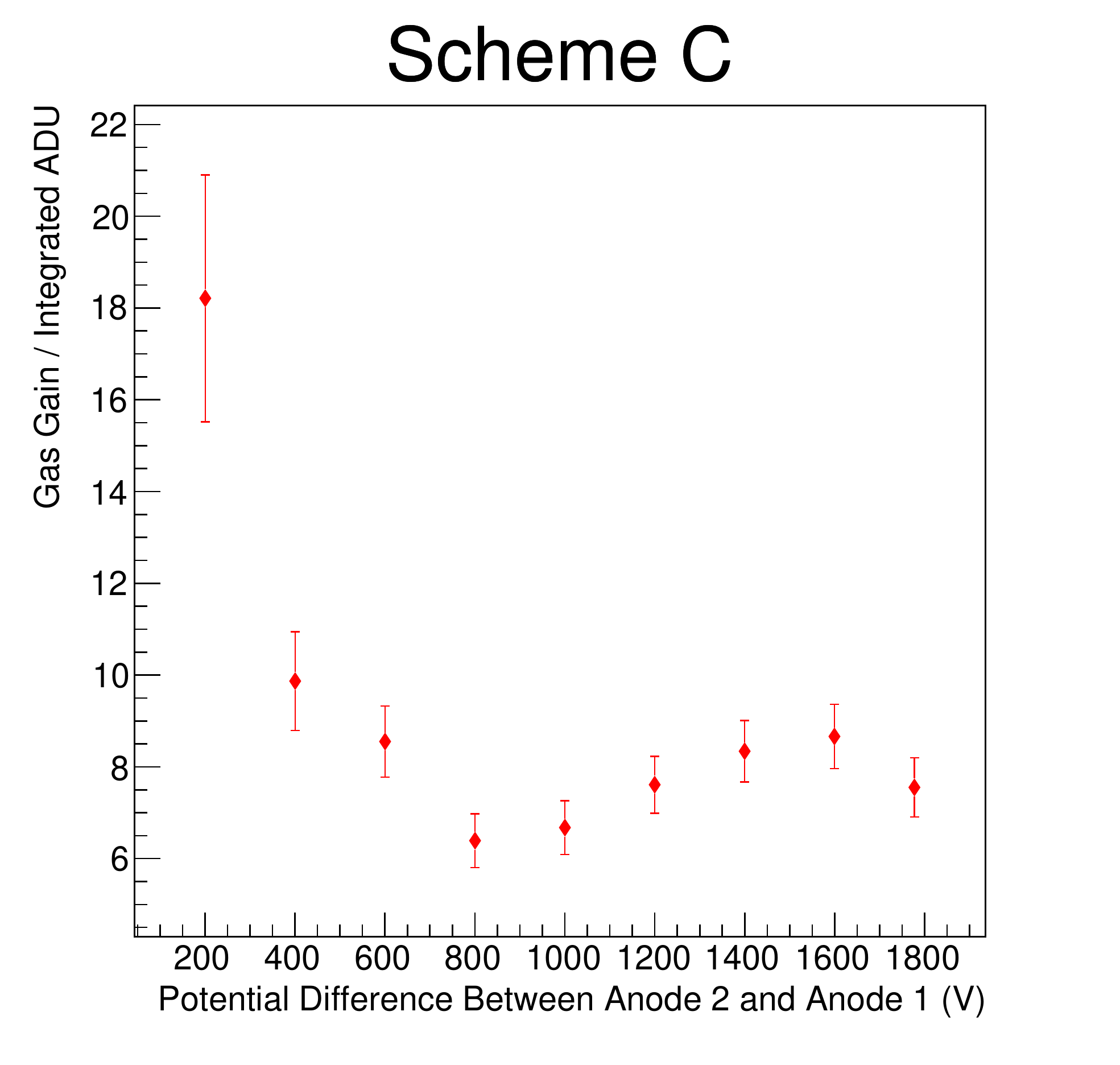}}
\caption{\label{hptpcPaper:sec:ChargeLight:fig:schemeABC_ChargeLight_Ratio}Ratio of gas gain measured in the amplification region at anode 3 (\figref{hptpcPaper:subsec:hptpcPerformance:chargeGain:fig:GasGainVsVoltageSchemes}, right column) to the measured intensity (integrated \si{ADU}) (\figref{hptpcPaper:sec:opticalgain:fig:schemeABC_source}) vs \protect\subref{hptpcPaper:sec:ChargeLight:fig:schemeA_ChargeLight_Ratio} anode 1 voltage ($V_{a1}$) where the voltage differences between the meshes is always $\Delta V_{a12}=\Delta V_{a23}=\SI{1200}{\volt}$ \protect\subref{hptpcPaper:sec:ChargeLight:fig:schemeB_ChargeLight_Ratio} potential difference between anode 2 and anode 3 ($V_{a23}$) while the anode 1 and 2 voltages are kept constant \protect\subref{hptpcPaper:sec:ChargeLight:fig:schemeC_ChargeLight_Ratio} potential difference between anodes 1 and 2 ($\Delta V_{a12}$) while $V_{a1}$ is kept constant and $\Delta V_{a23}$ is maintained at \SI{1200}{\volt}}
\end{figure}
In this section we present the results of the combined optical and charge gain analysis. The optical and charge gain analyses described in \secref{hptpcPaper:sec:ccdAnalysis} and \secref{hptpcPaper:sec:chargeAnalysis} were performed on data taken simultaneously with both readout systems. 
We investigate the correlation between optical gain and charge gain in \figref{hptpcPaper:sec:opticalgain:fig:schemeABC_source} and \figref{hptpcPaper:subsec:hptpcPerformance:chargeGain:fig:GasGainVsVoltageSchemes}. Plots of the optical gain against the charge gain for Schemes A, B and C are shown in \figref{hptpcPaper:sec:ChargeLight:fig:schemeABC_ChargeLight}. \Figref{hptpcPaper:sec:ChargeLight:fig:schemeABC_ChargeLight_Ratio} shows the ratio of the charge gain to the measured light intensity in \si{ADU} as a function of the relevant voltage in the respective voltage scheme. The larger of the two charge gain error bars (shown dotted in \figref{hptpcPaper:sec:ChargeLight:fig:schemeABC_ChargeLight}) has been propagated through to produce the error bars seen in \figref{hptpcPaper:sec:ChargeLight:fig:schemeABC_ChargeLight_Ratio}.

To measure the correlation factor between the two gain measurements we use the Pearson correlation coefficient, which takes values between -1 and 1 for fully negative and positive correlated data. The coefficient is zero for uncorrelated data. In order to take the uncertainties of our measurements into account we take every measured value as the centre of a normal distribution and its uncertainty as the distribution's standard deviation. From these distributions a 1000 random data series are drawn for each voltage scheme with the same number of points as in the original data series in \figref{hptpcPaper:sec:ChargeLight:fig:schemeA_ChargeLight_Ratio}, \figref{hptpcPaper:sec:ChargeLight:fig:schemeB_ChargeLight_Ratio} and \figref{hptpcPaper:sec:ChargeLight:fig:schemeC_ChargeLight_Ratio} and the correlation factor is calculated for each of them. The correlation factors quoted in the following are the mean of these 1000 correlation factors and their standard deviation.

Both the optical and charge analysis found Scheme A to be consistent with no change in gain. The figures in this section show continued support for this case as expected. The correlation factor of the data sample in \figref{hptpcPaper:sec:ChargeLight:fig:schemeA_ChargeLight} is $0.50\pm0.37$. \Figref{hptpcPaper:sec:ChargeLight:fig:schemeB_ChargeLight} and \figref{hptpcPaper:sec:ChargeLight:fig:schemeC_ChargeLight} have a positive correlation between optical and charge gain in Schemes B and C of $0.85\pm0.06$ and $0.75\pm0.11$. Measuring a correlation between the electron and the photon yield in the amplification region suggests that the measured light is produced within the avalanches and thus the light yield increases with the charge gain. In Scheme C, we see a defined saturation of the optical gain above $\Delta V_{a12}=\SI{800}{\volt}$ (\figrefbra{hptpcPaper:sec:opticalgain:fig:schemeC_source}). This effect is not clearly visible in the charge gain analysis (\figrefbra{hptpcPaper:subsec:hptpcPerformance:chargeGain:fig:GasGainVsVoltageSchemeCAnode3}). However, given the size of the gas gain error bars it is not possible to confidently exclude this as a possibility. 

\Figref{hptpcPaper:sec:ChargeLight:fig:schemeABC_ChargeLight_Ratio} shows a largely consistent ratio of gas gain to the measured light gain in integrated ADU of around 8 for all schemes. The only deviation from this ratio occurs at the lowest voltage settings for Schemes B and C.
In \secref{hptpcPaper:subsec:hptpcPerfomrance:lightGain:subsubsec:results:electroncount} we found that there are $(2.2 \pm 0.5) \times 10^{-3}$ photons in the amplification region per primary electron in the drift volume, when analysing the voltage setting with the highest light yield. The authors of \cite{Fraga2001} examine pure argon at a pressure of \SI{3}{bar} absolute, too. For this gas they measure, albeit with a much smaller detector and a two mesh amplification region with \SI{4}{\milli\meter} distance, a maximal value of $\sim\!\!0.5$ photons per primary electron in the drift volume. This is two orders of magnitude higher than the value we observe. Furthermore, they observe this photon yield at a charge gain of about 10 while their measured light gain as function of charge gain saturates somewhere in-between charge gains of 5 and 10. In \cite{Fraga2001} the measurement is performed with a x-ray tube as radiation source and a photo-diode mounted close to the amplification region. This set-up allows the authors to operate in a low charge gain regime, where the cross-section for excitations can be higher than at larger charge gains where the ionisation cross-section dominates.\\
To improve the concept of an optical HPTPC with a mesh based amplification region it could be considered to have an amplification stage followed by a region of lower field in which the amplified electrons predominately excite gas atoms or molecules. The difficulty with such a combination of amplification and scintillation regions is that the electron transmissions between two meshes depends on the ratio of the fields on either side -- therefore only a fraction of the electrons from the high-field amplification gap(s) will reach the low-field scintillation gap.

%% file: sec_8_summary.tex
\section{Summary}
\label{hptpcPaper:sec:summary}

In order to reduce neutrino interaction related systematic uncertainties in future neutrino oscillation experiments, a key measurement is proton-nucleus scattering. Hadronic interactions as particles produced in neutrino interactions exit the nucleus and obfuscate the secondary particle multiplicity and kinematics, causing event migrations between data samples and introducing biases in neutrino event reconstruction. Measurements of protons interacting with nuclei can constrain these hadronic interactions and thereby reduce these biases. A HPTPC prototype detector with a three mesh amplification region has been constructed and operated at RHUL and CERN as a first step in the development of a HPTPC capable of performing these measurements.\\
In this work, for the first time, we demonstrate the successful combined optical and charge readout of a hybrid high pressure gaseous TPC with an active volume of $\sim\!\!\SI{0.5}{\meter\cubed}$. The optical readout utilises CCD cameras, most sensitive in the visible part of the wavelength spectrum.  Using cameras with an increased range of sensitivity to the VUV or wavelength shifting filters could be beneficial in future research.  In a series of pilot measurements we identified pure argon at an absolute pressure of \SI{3}{bar} as the gas best suited to perform in depth tests of the optical readout performance with a high pressure gas. Our measurements were done using the $\alpha$ particles emitted by an $\text{Am}^{241}$ source. When using the HPTPC with argon at \SI{3}{bar} we were not able to image tracks on an event by event basis with the optical readout. This may be due to the large diffusion in pure argon. Integrating over many exposures we measure an increasing light yield when increasing the electric field between mesh 2 and mesh 3 ($E_{a23}$), where the mesh number increases for anodes further away from drift region. An increase in light yield is also measured when increasing the field between mesh 1 and mesh 2 ($E_{a12}$). In this case the light yield reaches a plateau when $E_{a12}\sim E_{a23}$. At the maximum light yield measured, we find that there are $(2.2 \pm 0.5) \times 10^{-3}$ photons in the amplification region per primary electron in the drift volume.\\
The analysis of the charge signals reveals that light gain and charge gain are correlated and that the gas gain at the voltage settings of the maximal light yield is 3000. The first mesh in the cascade contributes the largest fraction of the amplification stages gain of $\sim\!\!70$ whilst the following meshes contribute another factor of about $8$ and $5.5$, respectively.